\documentclass{aa}  

\usepackage{graphicx}
\usepackage{float}
\usepackage{txfonts}
\usepackage{ulem}
\usepackage{placeins}
\RequirePackage{siunitx}					
\newcommand{\kms}{\,km\,s$^{-1}$} 

\newcommand{\hi}{\ion{H}{i}}
\newcommand{\Ha}{H$\alpha$\,}
\newcommand{\ha}{H$\alpha$\,}

\newcommand{\vaz}{$V_\theta$}
\newcommand{\vrad}{$V_R$}

\newcommand{\vlos}{$V_{\rm los}$}
\newcommand{\slos}{$\sigma_{\rm los}$}

\begin{document}

   \title{Asymmetries in random motions of neutral Hydrogen gas in spiral galaxies}

   \author{P. Adamczyk
          \inst{1,2}\fnmsep\thanks{\email{p.adamczyk@free.fr, philippe.amram@lam.fr, astro.chemin@gmail.com, benoit.epinat@lam.fr}}
    \and
        P. Amram\inst{2}
    \and
        L. Chemin\inst{3}
    \and
        B. Epinat\inst{2,4}
    \and
        J. Braine\inst{5}
    \and
        F. Combes\inst{6}
    \and
        W. J. G. de Blok\inst{7, 8, 9}
          }

   \institute
            {Centro de Astronomia, Universidad de Antofagasta, Avda. U. de Antofagasta, 02800 Antofagasta, Chile
         \and Aix Marseille Univ, CNRS, CNES, LAM, 38 rue Frédéric Joliot Curie, 13338 France
         \and Instituto de Astrofisica, Universidad Andres Bello, Fernandez Concha 700, Las Condes, Santiago RM, Chile
         \and  Canada-France-Hawaii Telescope, 65-1238 Mamalahoa Highway, Kamuela, HI 96743, USA
         \and Laboratoire d’Astrophysique de Bordeaux, Univ. Bordeaux, CNRS, B18N, Allée Geoffroy Saint-Hilaire, 33615 Pessac, France
         \and Observatoire de Paris, LERMA, Collège de France, CNRS, PSL University, Sorbonne University, 75014 Paris
         \and ASTRON, Netherlands Institute for Radio Astronomy, Oude Hoogeveensedijk 4, 7991 PD Dwingeloo, The Netherlands
         \and Kapteyn Astronomical Institute, University of Groningen, Landleven 12, 9747 AD Groningen, The Netherlands
         \and Department of Astronomy, University of Cape Town, Private Bag X3, Rondebosch 7701, South Africa
             }

   \date{Received 27 April 2023 / Accepted 8 June 2023}

 
  \abstract
   {The velocity dispersion ellipsoid of gas in galactic discs is usually assumed isotropic. Under this approximation, no projection effect occurs in the random motions of gas, as traced by the line-of-sight velocity dispersion.
   However, it has been recently shown that random motions of the neutral Hydrogen gas of the Triangulum  galaxy (M33) exhibit a bisymmetric perturbation which is aligned with the minor axis of the galaxy, suggesting a projection effect.}
   {To investigate if perturbations in the velocity dispersion of nearby discs are comparable to those of M33, the sample is extended to 32 galaxies from The \hi  \ Nearby Galaxy Survey (THINGS) and the Westerbork \hi\ Survey of Spiral and Irregular Galaxies (WHISP).}
   {We study velocity asymmetries in the disc planes by performing Fourier transforms of high-resolution \hi\ velocity dispersion maps corrected for beam smearing effects, and measure the amplitudes and phase angles of the Fourier harmonics.}
   {In all velocity dispersion maps, we find strong perturbations of first, second and fourth orders. The strongest asymmetry is the bisymmetry, which is predominantly associated with the presence of spiral arms. The first order asymmetry is generally orientated close to the disc major axis, and   the second and fourth order asymmetries are preferentially orientated along intermediate directions between the major and minor axes of the  discs. These results are evidence that strong projection effects shape the \hi\ velocity dispersion maps. The most likely source of systematic orientations is the anisotropy of velocities, through the projection of streaming motions stronger along one of the planar directions  in the discs. Moreover, systematic phase angles of asymmetries in the \hi\ velocity dispersion could arise from tilted velocity ellipsoids, that is when the velocities are correlated. We expect a larger incidence of correlation between the radial and tangential velocities of \hi\ gas  with $|\rho_{R\theta}| \sim 0.6$, which could be tested against the kinematics of the youngest stellar populations of the Milky Way. }
   {\hi\ velocity dispersions cannot be considered devoid of projection effects. The systematic orientations of asymmetries can be explained by the projection of  unresolved streaming motions arising mainly from spiral arms. Our methodology is a powerful tool to constrain the dominant direction of streaming motions and thus the shape of the velocity ellipsoid of \hi\ gas, which is de facto anisotropic  at the angular scales probed by the observations. 
   The next step is to study the shape of the velocity ellipsoids of molecular and ionised gas and their link with galaxy mass and/or morphology, in addition to extending the sample size.}

   \maketitle
%
\section{Introduction}
\label{Introduction}
 
In discs of galaxies, large-scale perturbations like bars, spiral arms, warps, or lopsidedness make the orbits non-circular. They generate streaming motions, i.e. asymmetric radial and  tangential velocities in the stellar and gaseous media in galactic discs \citep[e.g.][]{1980visser}.
This is beautifully illustrated from an observational viewpoint via the kinematics of millions of individual stars in the Large Magellanic Cloud, for which  the Gaia Collaboration used astrometry to deduce the velocity fields of both the radial and tangential components of a disc galaxy \citep{2016prusti, 2018brown, 2018helmi, 2021brown}, and from which significant asymmetries are observed in the velocity maps, as caused by the bar and  spiral arms of the LMC \citep{2021luri}.

The study of kinematic asymmetries has become relevant in galactic dynamics, as it impacts our knowledge of the  structure of dark matter halos and the comparison with simulations made in a cosmological context. Interstellar gas is thought to follow circular orbits as it undergoes dissipative collisions.  If asymmetries are important, then there is no guarantee that rotation curves trace perfectly the mass distribution of galaxies. Asymmetries could affect the inner slope of gaseous rotation curves and require a model including asymmetries to determine the scale parameters of dark matter halos \citep[]{2006hayashi,2019oman}.  
Strategies have thus been developed to quantify asymmetries in gaseous velocity fields in the last 25 years. 
One approach is to measure harmonic velocity components which overlap with the axisymmetric motions \citep{1997schoenmakers, 2006MNRAS.366..787K}. 
Early works focused on the large-scale kinematic lopsidedness in \hi\ velocity fields, a perturbation of first order which causes the kinematic centre to drift with radius.  In this case, rotation curves from  the approaching and receding halves are not consistent with each other. This could be due to a misalignment between the axes of the disc and the host dark matter halo \citep{1999swaters}.  \citet{2008trachternach} estimated asymmetries for a large sample of \hi\ velocity fields to study the elongation of the gravitational potential.
They found that the asymmetries do not alter the mass distribution inferred from the axisymmetric rotational motions. \citet{2007spekkens} measured bisymmetric perturbations in CO and \Ha velocity fields to highlight the importance of bisymmetric flows caused by a stellar bar on the inner shape of the rotation curve.
\citet{2016chemin} modelled the harmonics observed in the \ha\ velocity field of a grand design spiral through direct derivation of  gravitational potentials and assessed the impact of asymmetries on the structure of the host dark matter halo. 

While studying streaming motions in ordered velocity fields has become routine, little is known about asymmetries in maps of apparently random motions of interstellar gas, as traced by the line-of-sight velocity dispersion (\slos). 
The velocity dispersion can be used to assess the dynamical heating of discs, and to estimate the support due to pressure and how it compares with rotational motions \citep[e.g.][]{1997combes,2009koyama,2010bershady,2015oh}.
The main reason dispersion has not been studied extensively is that \slos\ mixes turbulence, asymmetric drift, shear, superposition of multiple elliptical orbits,  unresolved motions on relatively large scales.
The local velocity dispersion ellispoid of gas is also usually considered isotropic because the gaseous component dissipates energy through collisions between clouds in all directions. Within this hypothesis, no projection effect occurs in the random motions of gas, as traced by \slos, unlike that of collisionless stars.
Furthermore, the projection of the velocity ellipsoid can be degenerate if the ellipsoid is anisotropic. The finite velocity and spatial resolution further complicate the interpretation of maps of \slos. 

In a recent study  of  the Local Group spiral M33,  \citet{2020chemin} found perturbations in the random motions through Fourier transforms of \slos\ maps of 21-cm \hi\ data at various angular resolutions. Bisymmetry dominated in the outer half of the disc with a phase aligned with the minor axis of M33. Is this typical of spirals in general? Is it due to an anisotropic velocity ellipsoid harbouring a dominant radial component, as suggested by these authors?
The goal of this study is to determine whether what was observed in M33 is a general feature of spiral discs.

Our goal is to systematically search for asymmetries in velocity dispersion maps, identify their origin, and evaluate the consequences for galactic dynamics.
In this article, we present the first census of asymmetries in the velocity dispersion of \hi\ gas in local, massive discs. The  samples used for this analysis are described in Sect.~\ref{sec:sample}. The asymmetries are measured through Fast Fourier Transforms of velocity dispersion maps from interferometric data with a careful treatment of the effect of beam smearing (BS hereafter), as detailed in Sect.~\ref{sec:method}. The  properties of the asymmetries are presented in Sect.~\ref{sec:asym}, and the discussion of their origin is given in Sect.~\ref{sec:discussion}. Finally, we provide a synthesis and conclusions in Sect.~\ref{sec:conclusion}.

\section{Selection of a working sample}
\label{sec:sample}

\begin{table*}[t]
\centering
\begin{tabular}{ccccccccccc}
\hline
\hline
Galaxy  &  $\alpha$(J2000) &   $\delta$(J2000) & D & $V_{sys}$ & Incl & PA & $B_{maj}$ &  $B_{min}$ & $B_{pa}$ & $\Delta_V$ \\ 
name    &  (hh mm ss)      &  (dd mm ss)    & Mpc & \kms & $\mathrm{deg}$ & $\mathrm{deg}$ &  $\mathrm{arcsec}$/ $\mathrm{pc}$ &  $\mathrm{arcsec}$/ $\mathrm{pc}$ & $\mathrm{deg}$ & \kms \\ 
(1) & (2) & (3) & (4) & (5) & (6) & (7) & (8) & (9) & (10) & (11)\\
\hline 
NGC925  & 02 27 16.5 & +33 34 44 & 9.2  & 546 & 66 & 287  &  5.9 / 263  &  5.7 / 254  & 31  & 2.6 \\  
NGC2403 & 07 36 51.1 & +65 36 03 & 3.2  & 133 & 63 & 124  &  8.8 / 136  &  7.7 / 119  & 25  & 5.2 \\  
NGC2841 & 09 22 02.6 & +50 58 35 & 14.1 & 634 & 74 & 153  &  11.1 / 756 &  9.4 / 641  & -12 & 5.2 \\  
NGC2903 & 09 32 10.1 & +21 30 04 & 8.9  & 556 & 65 & 204  &  15.3 / 659 &  13.3 / 573 & -51 & 5.2 \\  
NGC2976 & 09 47 15.3 & +67 55 00 & 3.6  & 1   & 65 & 335  &  7.4 / 129  &  6.4 / 112  & 72  & 5.2 \\  
NGC3031 & 09 55 33.1 & +69 03 55 & 3.6  & -40 & 59 & 330  &  12.9 / 225 &  12.4 / 216 & 80  & 2.6 \\  
NGC3198 & 10 19 55.0 & +45 32 59 & 13.8 & 661 & 72 & 215  &  13.0 / 867 &  11.6 / 774 & -59 & 5.2 \\  
NGC3521 & 11 05 48.6 & -00 02 09 & 10.7 & 804 & 73 & 340  &  14.1 / 730 &  11.2 / 580 & -62 & 5.2 \\  
NGC3621 & 11 18 16.5 & -32 48 51 & 6.6  & 729 & 65 & 345  &  15.9 / 508 &  10.2 / 326 & 4   & 5.2 \\  
NGC3627 & 11 20 15.0 & +12 59 30 & 9.3  & 708 & 62 & 173  &  10.6 / 477 &  8.9 / 400  & -48 & 5.2 \\  
NGC4736 & 12 50 53.0 & +41 07 13 & 4.7  & 307 & 41 & 296  &  10.2 / 232 &  9.1 / 207  & -23 & 5.2 \\  
NGC5055 & 13 15 49.2 & +42 01 45 & 10.1 & 497 & 59 & 102  &  10.1 / 493 &  8.7 / 425  & -40 & 5.2 \\  
NGC6946 & 20 34 52.2 & +60 09 14 & 5.9  & 44  & 33 & 243  &  6.0 / 171  &  5.6 / 160  & 7   & 1.3 \\  
NGC7331 & 22 37 04.1 & +34 24 57 & 14.7 & 818 & 76 & 168  &  6.1 / 433  &  5.6 / 398  & 34  & 5.2 \\  
NGC7793 & 23 57 49.7 & -32 35 28 & 3.9  & 226 & 50 & 290  &  15.6 / 295 &  10.9 / 206 & 11  & 2.6 \\  
\hline
UGC01256   & 01 47 53.9  &  +27 25 55 & 8.8  &   426  &    70 & 69   &  25.6 / 1090 &  11.6 / 494 & -1.0  & 16.5 \\                          
UGC01913   & 02 27 16.9  &  +33 34 44 & 10.8 &   553  &    59 & 287  &  16.7 / 872  &  8.6 / 449  &  0.0  & 4.1 \\                          
UGC02455   & 02 59 42.5  &  +25 14 19 & 7.5 &   373  &    42 & 208  &  27.4 / 1417 &  11.2 / 580 &  -1.0 & 2.1 \\                          
UGC04284   & 08 14 40.1  &  +49 03 42 & 10.7 &   547  &    60 & 170  &  13.1 / 855  &  10.2 / 666 &  0.0  & 4.1 \\                          
UGC04305   & 08 19 04.3  &  +70 43 18 & 5.3 &   158  &    51 & 195  &  12.3 / 607  &  11.6 / 572 &  -1.0 & 2.1 \\                          
UGC04325   & 08 19 20.5  &  +50 00 35 & 10.3  &   506  &    68 & 60   &  15.1 / 490  &  11.8 / 383 &  -1.0 & 4.1 \\                          
UGC04499   & 08 37 41.5  &  +51 39 09 & 13.5 &   687  &    81 & 151  &  14.8 / 937  &  11.7 / 741 &  -1.0 & 4.1 \\                          
UGC05414   & 10 03 57.2  &  +40 45 27 & 10.2  &   604  &    54 & 216  &  17.7 / 583  &  11.4 / 375 &  -1.0 & 2.1 \\                          
UGC05721   & 10 32 17.2  &  +27 40 08 & 6.6  &   532  &    62 & 273  &  29.7 / 1236 &  13.0 / 540 &  1.2  & 4.1 \\                          
UGC05789   & 10 39 09.5  &  +41 41 13 & 13.1  &   738  &    63 & 38   &  18.4 / 863  &  11.9 / 558 &  0.0  & 4.1 \\                          
UGC07323   & 12 17 30.2  &  +45 37 09 & 8.6  &   516  &    52 & 23   &  17.5 / 728  &  12.7 / 528 &  0.0  & 4.1 \\                          
UGC07766   & 12 35 57.7  &  +27 57 35 & 9.7  &   814  &    65 & 328  &  22.1 / 589  &  9.3 / 283  &  0.0  & 4.1 \\                          
UGC07831   & 12 39 59.3  &  +61 36 33 & 4.4 &   146  &    70 & 302  &  13.0 / 773  &  8.6 / 511  &  0.0  & 4.1 \\                          
UGC07853   & 12 41 32.9  &  +41 09 04 & 8.6 &   538  &    58 & 212  &  19.8 / 1006 &  12.9 / 655 &  0.0  & 4.1 \\                          
UGC08490   & 13 29 36.6  &  +58 25 14 & 5.5  &   202  &    59 & 185  &  13.5 / 490  &  11.3 / 410 &  -1.0 & 4.1 \\                         
UGC11891   & 22 03 33.7  &  +43 44 56 & 12.3 &   461  &    43 & 130  &  16.5 / 854  &  11.4 / 590 &  -1.0 & 4.1 \\                         
UGC12632   & 23 29 58.7  &  +40 59 25 & 10.5  &   422  &    37 & 20   &  18.4 / 472  &  12.2 / 313 &  0.0  & 4.1 \\   
\hline
\end{tabular}
\caption{Parameters of the galaxy sample. Sources from THINGS are the NGC names, those from WHISP the UGC names. (1) Name of the object; (2)-(3) Right ascension and Declination (J2000) of the galactic centre; (4) Galaxy distance  from \cite{2008AJ....136.2563W} for THINGS, and deduced from the systemic velocity taken in NED corrected from Virgo infall, assuming $H_0 = 67.8$~\kms~Mpc$^{-1}$ for WHISP; (5)-(6)-(7) Systemic velocity, mean inclination and position angle from \cite{2008AJ....136.2648D}; (8)-(9) Major and minor axes of the synthesised beam; (10) Beam position angle; (11) Spectral sampling.}
\label{tab:galaxysample}
\end{table*}

Several factors limit the modelling of spatially resolved velocity dispersions. The smearing induced by the finite telescope beam can produce systematic asymmetries in velocity dispersion maps  (see Sect.~\ref{sec:method}). For a given instrumental configuration of spatial and spectral resolution and sampling, the BS effect increases with the distance and inclination of the galaxies.

For these reasons, our sample is from the high-resolution \hi  \ Nearby Galaxy Survey \citep[THINGS,][]{2008AJ....136.2563W}. 
With galaxy  distances between 2 and 15 Mpc, THINGS yields linear resolutions up to $\sim 900$ pc. The velocity resolution is 2.6 or 5.2 \kms, and the pixel size  1.5\arcsec, except for the galaxy NGC2403 (1\arcsec). We have used the three moment maps of the integrated \hi\ emission (0th moment), line-of-sight velocity  and velocity dispersion (1st and 2nd moments, respectively), which are made available by the THINGS collaboration\footnote{\url{https://www2.mpia-hd.mpg.de/THINGS/Data.html}}. 
In particular, we used the data obtained with natural weighting. We however verified that the conclusions of this analysis are not changed by using higher resolution kinematics, from maps obtained with robust weighting. A discussion of the impact of the angular resolution can be found in Sect.~\ref{sec:bs-ha}.  

 To perform the Fourier analysis of velocity dispersion fields, robust constraints on the geometry of the discs are necessary, and particularly the variation of the inclination and the position angle of the discs as a function of galactocentric radius. The disc warping parameters for 19 of the 34 THINGS galaxies have been measured with tilted-ring models of the velocity fields by \cite{2008AJ....136.2648D}. Among several methods applied to obtain velocity fields from the \hi\ data cubes, these authors adopted results from Hermite h3 polynomials fittings, due to their stability at low signal-to-noise ratio (SNR). They  explained that two masks were successively applied to filter low-quality regions  from the velocity fields. The first one consisted in rejecting \hi\ profiles (i) for which the fitted maximum intensity was lower than $3\sigma_{ch}$, with $\sigma_{ch}$ being the average noise in the profile outside of the emission line, and (ii) for which the dispersion of the fitted function was lower than the channel separation. The second mask  consisted in a sigma-clipping on the \hi\ column density maps to suppress noise pixels and to exclude regions inside $r_{min}$ and outside $r_{max}$, which are respectively the innermost and outermost radii of the tilted-ring models fitted of the final velocity fields. They finally measured the warp parameters and rotation curves,  adopting a sampling of two points per synthesised beam size. 

We masked the moment maps following \cite{2008AJ....136.2648D} and adopted the parameters of their tilted-ring models. This makes our gaseous and kinematic distributions, and the galaxy cylindrical frames fully consistent with their analysis. A minor difference with their study is that we considered sampling the radial profiles with one data-point per beam only, to minimize the correlation between adjacent rings. Finally, among the 19 galaxies, we selected the 15 massive, regular galaxies, excluding dwarfs or strongly disturbed galaxies (NGC 2366, DDO 154, IC 2574 and NGC 4826). 

\cite{2020chemin} showed that the angular resolution did not affect the detection of the bisymmetry/anisotropy
observed in  \hi\ velocity dispersion maps of M33, as it was seen at 70, 100 and 490 pc resolution. To increase the size of our sample, we included sources from the Westerbork \hi\ Survey of Spiral and Irregular Galaxies (WHISP) \citep{2001ASPC..240..451V}. 
The reduction pipeline of the WHISP observations provides data at resolutions of $14\arcsec \times\, 14\arcsec/\sin{\delta}$, where $\delta$ is the galaxy declination, $30\arcsec \times\, 30\arcsec$, and $60\arcsec \times\, 60\arcsec$, for a spectral resolution of 2.06 to 16.5 \kms. 
At distances out to more than 100 Mpc, the implied linear resolution for many WHISP galaxies is beyond the kiloparsec scale. 
Considering the balance between spatial resolution and sensitivity, and, due to the necessity to work with an appropriate spatial resolution, we only selected the sources with a physical resolution better than 1.5 kpc \citep{adamczyk}. This corresponds to lower SNR measurements than for THINGS targets, and consequently to a smaller extent of the gas distribution and kinematics.
We  further required the neutral gas to cover seven times the size of the beam \citep[as suggested in][to derive rotation curves]{1978PhDT.......195B}, in the same inclination range as  THINGS, and with no obvious signs of tidal interactions.  In the end, 17 WHISP galaxies were added to the THINGS sample.
We produced moment maps for them by means of CAMEL, a  Python tool\footnote{CAMEL stands for Cube Analysis: Moment maps of Emission Lines, see \url{https://gitlab.lam.fr/bepinat/CAMEL}} described in \citet{Epinat+2012}. 
We then performed  tilted-ring models of WHISP data cubes using the package 3D-Based Analysis of Rotating Object via Line Observations (3D-Barolo, \citealp{2015MNRAS.451.3021D}). This software creates a model data cube from input  centre of the rings, inclination and position angle, systemic velocity, scale height of the disc, and velocity dispersion, which is fitted to the observations.  
To  determine the kinematic centre, we first kept the inclination and position angle fixed, with the centre of radial rings and the systemic velocity as free parameters. Then, we fixed the centre and let the inclination and position angles vary to get the warp parameters.  

The final sample of galaxies is presented in Tab.~\ref{tab:galaxysample} which lists the coordinates, distance, systemic velocity, the mean inclination and position angle of the discs, and the properties of the synthesised beam.
With mean source distances of 8 and 9 Mpc for THINGS and WHISP respectively, the mean beam $B_{maj}$ of 11\arcsec\  and 18\arcsec\ correspond to average linear resolutions of 470 pc and 800 pc, respectively.
The THINGS sample is the main reference  and the WHISP sample is  used to check the observed trends are not linked to a particular telescope. Section~\ref{sec:asym} shows that the WHISP data fully support the trends found with THINGS data only.

\section{Methodology}
\label{sec:method}

 Observations yield the integrated flux of gas emission lines, the line-of-sight velocity and velocity dispersion. 
 This observed dispersion cannot be modelled directly as it results from the convolution of the  random motion in the plane of galaxies ($\sigma$) with the Line Spread Function of the instrumental device ($\sigma_{\rm LSF}$) and the broadening from  thermal processes occurring in the gas ($\sigma_{\rm T}$).  The observed dispersion is then expressed by:
\begin{equation}
\sigma_{\rm los}=\left(\sigma^2+\sigma_{\rm LSF}^2+\sigma_{\rm T}^2\right)^{1/2}
\label{eq:sigmaobs} 
\end{equation}

Additionally, due to the finite angular resolution of observations, $\sigma$  is affected by a smearing effect. This comes from unresolved velocity gradients in the line-of-sight velocity, which broadens emission lines.

Below we describe how we modelled the BS effect, and how a corrected velocity dispersion has been inferred. Then, we describe how Fourier transforms  constrain the amplitudes and phases of asymmetries in the velocity dispersion maps.  

\begin{figure*}[th]
    \includegraphics[height=4.8cm,trim={0.7cm 0 1cm 0},clip]{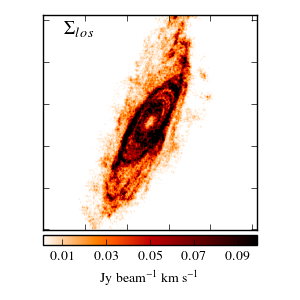}
    \includegraphics[height=4.8cm,trim={0.95cm 0 1.05cm 0},clip]{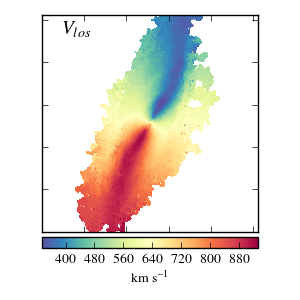}
    \includegraphics[height=4.8cm,trim={0.95cm 0 1.05cm 0},clip]{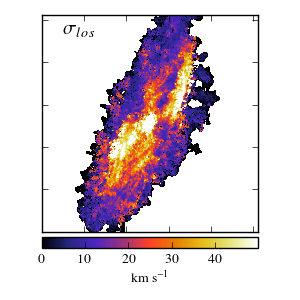}
    \includegraphics[height=4.8cm,trim={0.95cm 0 1.05cm 0},clip]{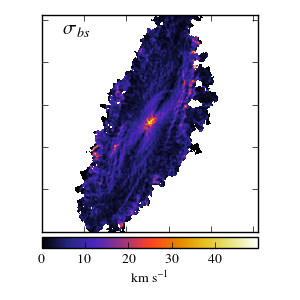}
    \includegraphics[height=4.8cm,trim={0.95cm 0 0.85cm 0},clip]{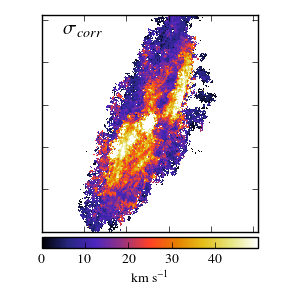}
    \caption{Density and velocity maps of NGC2841. From left to right we show : the observed flux density map, the observed velocity field, the observed velocity dispersion, the beam smearing model, and the velocity dispersion map corrected from the beam smearing effect.}
    \label{fig:n2841-ex}
\end{figure*}  

\subsection{Modelling the effect of beam smearing in a velocity dispersion map}
\label{sec:beamsmearinggaussian}

Beam smearing impacts the flux, velocity field, and velocity dispersion maps extracted from datacubes in a different way. Its main effect on velocity fields is to weaken the inner gradient of the rotation curve by mixing data from adjacent regions in the disc.
\cite{2008AJ....136.2648D} showed that BS does not have a significant impact on THINGS galaxies rotation velocities. They built mock datacubes, smoothed the cubes to the THINGS resolution, and derived the velocity fields and rotation curves from the mock datacubes. Compared to the input used to make the datacube, deviations were less than $\sim 1$ \kms.  

For velocity dispersion maps, one expects BS to be more prominent where large variations of radial velocities are observed locally, for instance due to rotation curve inner gradients and to variations of $\cos(\theta)$, $\theta$ being the azimuth angle (see Eq. \ref{eq:vlos_disc}). We need to model the impact of BS on velocity dispersion maps by taking into account the variations of velocities in the two spatial dimensions. We use the tool MocKinG\footnote{MocKinG stands for modelling Kinematics of Galaxies, see \url{https://gitlab.lam.fr/bepinat/MocKinG}} based on an analytical formula presented in \cite{2010MNRAS.401.2113E} who derived the observed velocity dispersion in a galaxy following:
\begin{equation}
\sigma^2=\sigma_{\rm corr}^2+\sigma_{\rm bs}^2 \text{~,}
\label{eq:bs1}
\end{equation}
with 
\begin{equation}
\sigma_{\rm corr}^2 = \frac{\iint_{\rm pix} \left[ \sigma_{\rm loc}^2 M\right] \otimes \mathrm{PSF}~ds}{\iint_{\rm pix} M \otimes \mathrm{PSF}~ds} \text{ ,}
\label{eq:bs2}
\end{equation}
and 
\begin{equation}
\sigma_{\rm bs}^2=\frac{\iint_{\rm pix}\left[V_{\rm los}^2 M\right] \otimes \mathrm{PSF}~ds}{\iint_{\rm pix} M \otimes \mathrm{PSF}~ds}
-\left( \frac{\iint_{\rm pix} \left[ V_{\rm los} M\right] \otimes \mathrm{PSF}~ds}{\iint_{\rm pix} M \otimes \mathrm{PSF}~ds} \right)^2 \text{~,}
\label{eq:bs3}
\end{equation}
where $\sigma_{\rm loc}$ is the local velocity dispersion map, $V_{\rm los}$ is the line-of-sight velocity field, $M$ is the line flux map, $\otimes \mathrm{PSF}$ represents the two-dimension convolution by the PSF, and $\iint_{\rm pix} ds$ integrates over the surface of the pixel. 
Equation~\ref{eq:bs2} shows that $\sigma_{\rm corr}$ is impacted by BS on the local velocity dispersion $\sigma_{\rm loc}$ whereas Eq.~\ref{eq:bs3} shows that unresolved velocity shears in the first moment map create an artificial line broadening. Our goal is to study $\sigma_{\rm corr}$ by correcting the observed velocity dispersion $\sigma$ for the term $\sigma_{\rm bs}$ of Eq.~\ref{eq:bs3}.
The maps of first two moments ($M$ and $V_{\rm los}$) in this equation should be at high-resolution and free from BS to properly account for velocity variations inside the beam and pixels. In practice, such high-resolution maps do not exist so the observed flux is usually used for $M$, and either a high-resolution model \citep[see e.g.][]{2010MNRAS.401.2113E} or an observed velocity field are used for $V_{\rm los}$. We used MocKinG with the observed \hi\ flux maps and velocity fields to compute the BS correction.
It is thus asymmetric by construction and accounts for both circular and non-circular motions from the kinematics.
This leads to a fair correction, although not perfect, and is discussed in Sect.~\ref{sec:bs-ha}, \ref{sec:BS_residuals}, \ref{sec:orderedvelocityeffect}, and App.~\ref{app:toymodels}.
These BS dispersion maps were modelled assuming a bi-dimensional Gaussian synthesised beam (see Tab.~\ref{tab:galaxysample}). 
The effect of the observed dirty beams, which are not entirely elliptical, is addressed in Sect.~\ref{sec:dirty_beam}. 

An example of BS modelling with the galaxy NGC2841 is shown in Fig. \ref{fig:n2841-ex}. This fast-rotating galaxy (second panel) has an inclination of $\sim 74\degr$. The fourth panel shows the corresponding $\sigma_{\rm bs}$, which, once subtracted to the observed dispersion (middle panel), yields  the corrected velocity dispersion map (right panel) from which  asymmetries are measured. The general behaviour of the smearing effect is thus that $\sigma_{\rm bs}$ is larger at low radius and exhibits a typical X-shape pattern. On average,  $\sigma_{\rm bs}$ decreases with radius, but non-negligible values are also observed near non-axisymmetric density and velocity features (see Sect.~\ref{sec:bs_properties}), underlining the benefits of modelling the smearing from 2D data. 

\begin{figure}[t]
    \includegraphics[width=9cm,trim={0 0 1cm 0.8cm},clip]{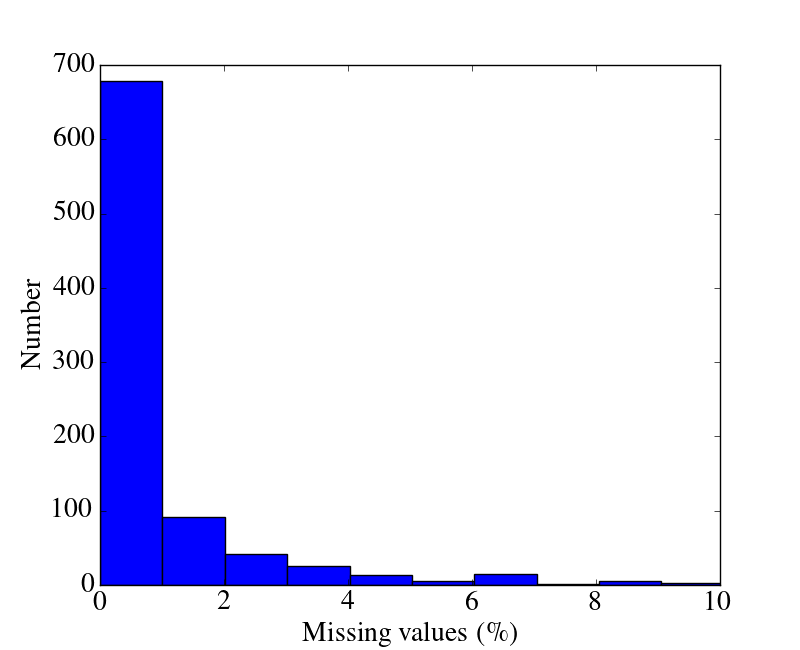}
    \caption{Distribution of the fraction of missing values for the radial bins in the THINGS galaxies. }
    \label{fig:frac_nan}
\end{figure}

\subsection{Fourier series modelling of velocity dispersion maps}
\label{sec:modfft}

In order to quantify asymmetries, we performed Fast Fourier Transforms (FFT) of the maps of \hi\ random motions. 
Discrete FFTs of the corrected $\sigma_{\rm los}^2$ 
maps were performed using the FFTPACK package of the SciPy libraries in Python (see App.~\ref{app:biaises}). 
In cylindrical coordinates, the discrete FFT of the velocity dispersion is: 
\begin{equation}
\sigma^2_{\mathrm{asym}} = \sum_{k=0}^{N-1} \sigma_k^2 \cos (k(\theta - \phi_k)) \text{ ,}
\label{eq:dft}
\end{equation}
where $\sigma_k$ and $\phi_k$ are the amplitude and phase angle of an asymmetry of order $k$,  $N$ is the number of orders in the decomposition, equivalent to the number of elements along the considered ring, and $\theta$ is the azimuthal angle in the galaxy plane measured from the semi-major axis of the receding disc half.
With this FFT formalism, we make the choice of studying the azimuthal asymmetry, i.e. we cannot study the asymmetry with respect to the disc plane in the vertical direction. Nevertheless, the present study includes radial, azimuthal and vertical components projected on the sky plane, assuming either that contributions along the direction orthogonal to the disc plane are averaged across the line of sight, which is in practice more valid for face-on than for edge-on galaxies, or that discs are infinitely thin. 
We use the squared velocity dispersion because $\sigma^2_{\rm los}$ is a simple sum of quadratic terms (Eqs.~\ref{eq:sigmaobs} and~\ref{eq:bs1}, see also App.~\ref{app:anisotropy}, Eq.~\ref{eq:slos_disc_simpl}). Consistent results are obtained when FFTs of the linear dispersion are calculated.

In practice, we decomposed a galactic disc into a series of concentric rings whose geometry was defined by the tilted-ring models (Sect.~\ref{sec:sample}). 
Given the different angular sizes of the galaxies, the adopted ring width of one full width at half beam power leads to differing numbers of rings per galaxy, from 7 for NGC3627 to more than 140 for NGC2403. Such a non-uniformity in the number of radial bins among galaxies has no visible consequence on the results (see Sect.~\ref{sec:asym}).

For each dispersion map, we first subtracted $\sigma^2_{\rm LSF}$ and $\sigma^2_{\rm T}$ from the  squared observed dispersion, as well as the corresponding 2D map of $\sigma^2_{\rm bs}$ described in Sect.~\ref{sec:beamsmearinggaussian}. 
All pixels with resulting negative quadratic velocity dispersion after these subtractions were discarded from the maps at this stage of the process, because such values are unphysical. 
The instrumental dispersion is $\sigma_{\rm LSF} = \Delta_V/2.35$ and $\sigma_{\rm LSF} = 1.2\Delta_V/2.35$ for the THINGS and WHISP galaxies, respectively, where $\Delta_V$ is the velocity channel width listed in Tab.~\ref{tab:galaxysample}. 
\hi\ is a mixture of cool ($\sim 100$ K) and warm ($\sim 5000-8000$ K) gas.  In M33, the \hi\ velocity dispersion was sometimes narrower than $\sigma_T$ if the warm gas was assumed to dominate \citep{2020chemin}.  Hence, we do not consider $\sigma_T$ here. We note  that this has no consequence hereafter, because considering gas as a warm neutral medium is equivalent to subtracting quadratically $\sigma_{\rm T} \sim 6$ \kms\ from the axisymmetric term $\sigma_0$.  
This latter term, measured as the mean dispersion of a given ring, is then subtracted quadratically from all pixel values inside the considered ring. Therefore,  the observed $\sigma^2_{\mathrm{asym}}-\sigma^2_0$ are centred on 0, and can be negative (see Eq.~\ref{eq:dft}). 
We also point out that $\sigma_{\rm T}$ could be asymmetric and vary over the disc. In such a case, fluctuations of $\sigma_{\rm T}$ as a function of the position have been accounted in the measurements of $\sigma_{\mathrm{asym}}$, though it is not possible to disentangle these effects from those arising from other local motions of, e.g., gravitational origins, without being able to measure locally the gas temperature.

We then sorted the squared velocity dispersions with increasing values of azimuth and apply the FFT, leading to a harmonic decomposition with $N/2$ terms, where $N$ is the number of pixels in the considered radial bin. Incomplete coverage of azimuths as caused by missing pixels (not-a-number values) could seen by the FFT as artificial perturbations. In the THINGS sample, about 77\%  of the 883 available tilted rings show less than 1\% of missing values of all available pixels, and 95\% less than 4\% of missing azimuthal angles (Fig.~\ref{fig:frac_nan}). The missing factor is thus low, and we verified that it has no impact in the analysis. Within WHISP data, we rejected 30\% of the  initial 203  rings  because they had more than 40\% of missing values. Within the 144 remaining rings, $15 \%$ of the pixels have missing values. Even though no impact was detected, we replaced the missing values by the azimuthally  averaged dispersion before the modelling. 

\begin{figure*}[t]
\noindent
\centering
    \includegraphics[height = 3.5cm]{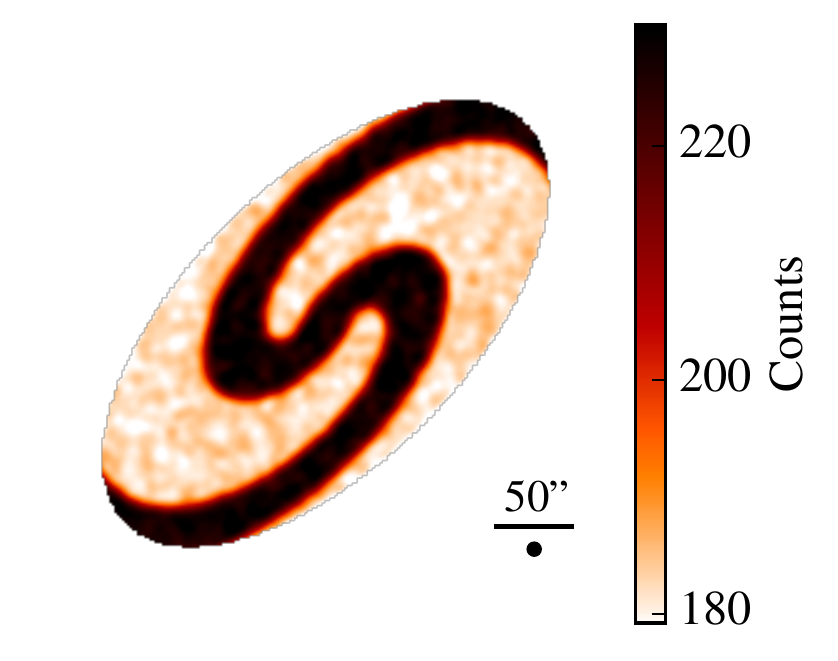}
    \includegraphics[height = 3.5cm]{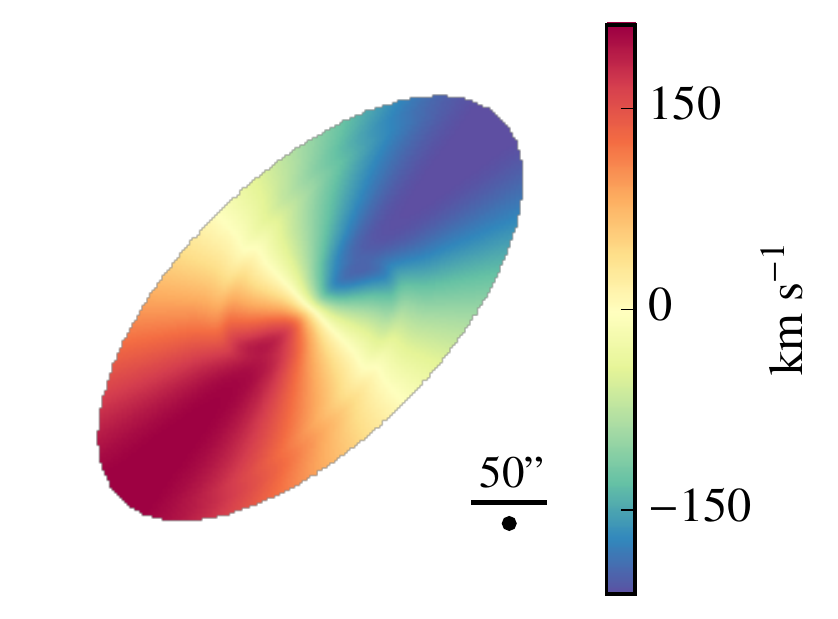}
    \includegraphics[height = 3.5cm]{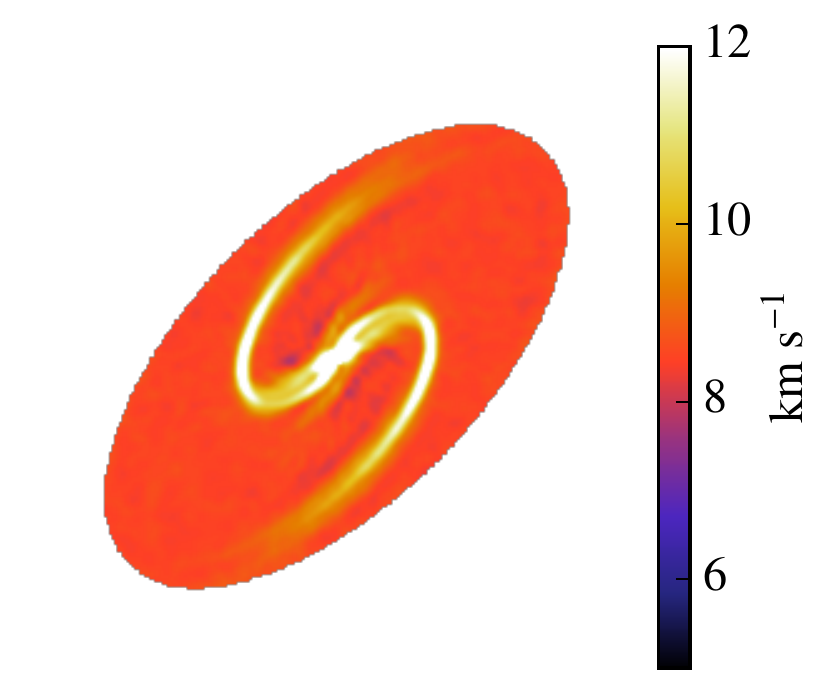}
    \includegraphics[height = 3.5cm]{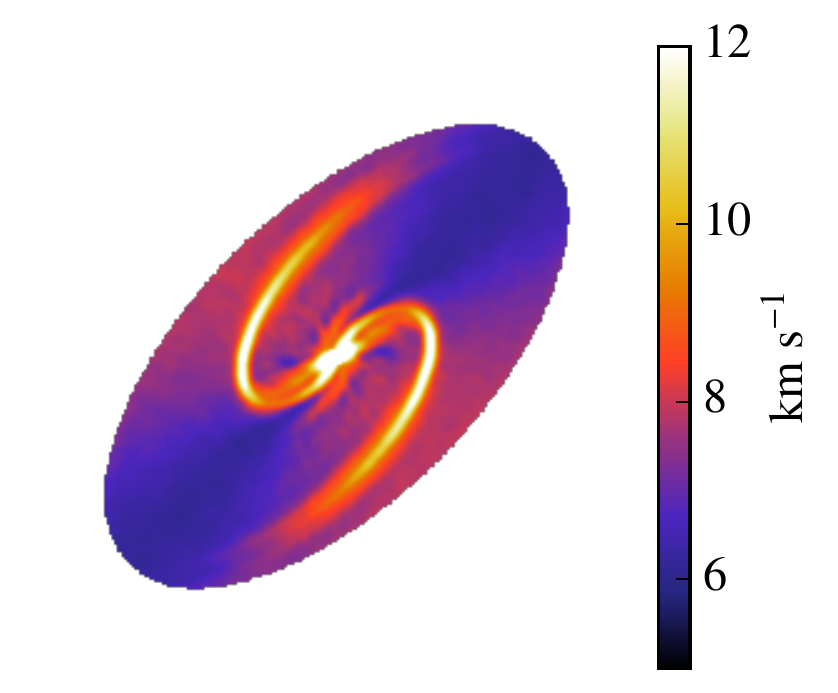} \\
    \includegraphics[height = 4cm]{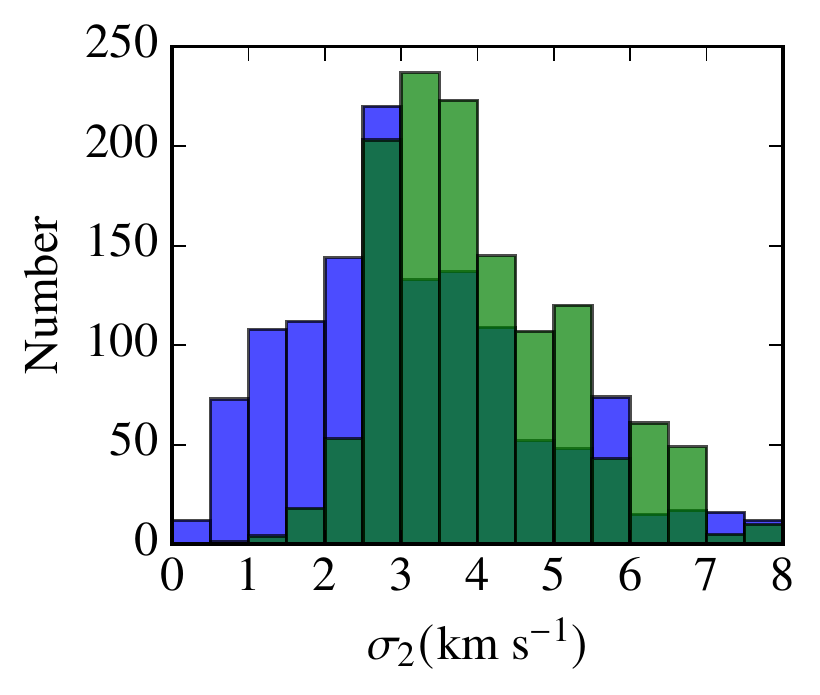}
    \includegraphics[height = 4cm, trim = 0.6cm 0 0 0 , clip]{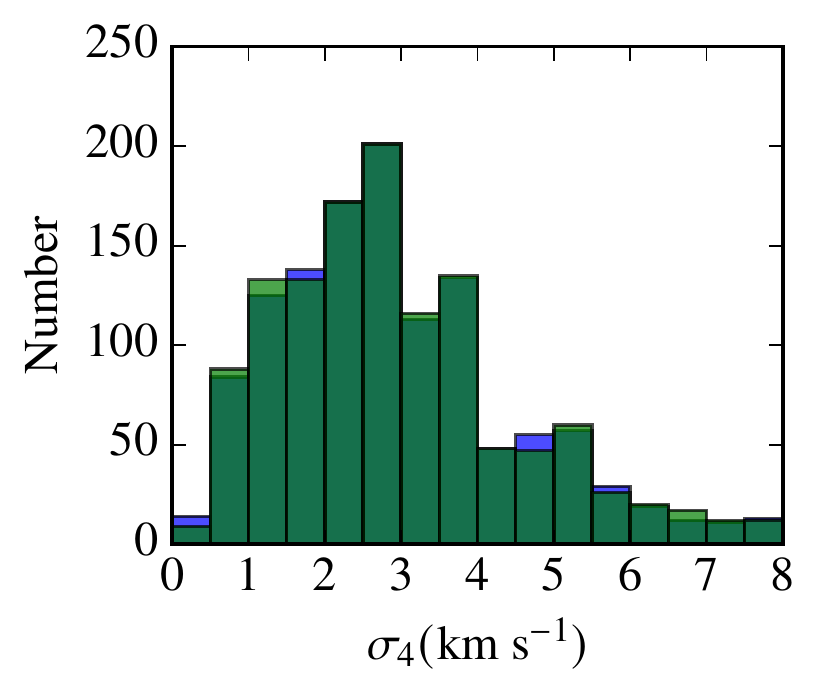}
    \includegraphics[height = 4cm, trim = 0.6cm 0 0 0 , clip]{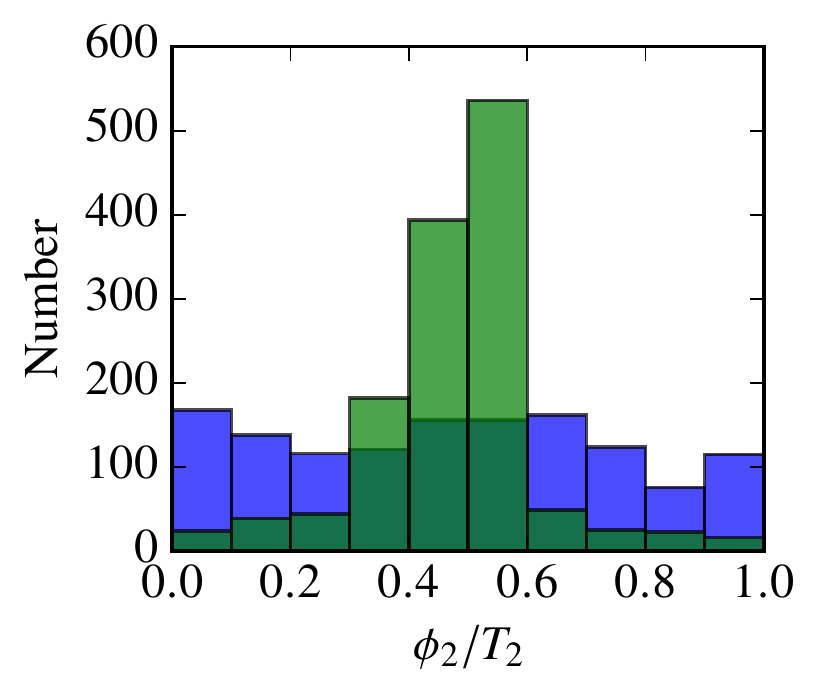}
    \includegraphics[height = 4cm, trim = 0.6cm 0 0 0 , clip]{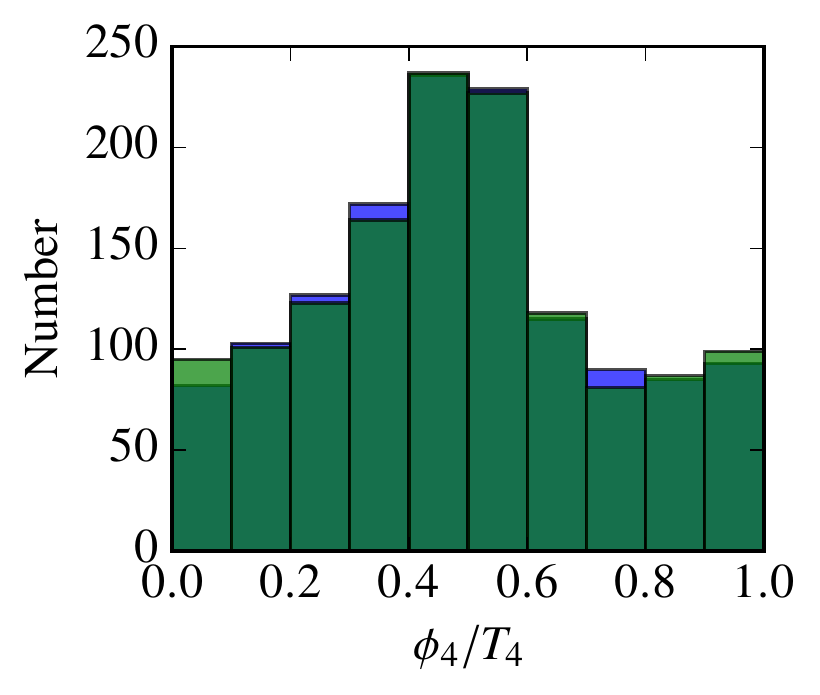}
    \caption{Maps and histograms of toy models with a steep rotation curve ($v_0=250$ \kms, $r_s=20\arcsec$ and $\gamma = 2$) with a spiral perturbation.
   We show (top), from left to right, the density map, the velocity field, and the BS corrected dispersion maps in the isotropic and anisotropic cases, obtained with an inclination of 60\degr\ and with $\phi_{sp}=135$\degr.
   The bottom panels show amplitude and phase histograms of orders $k = 2$ and $k = 4$. Blue and green histograms correspond to the cases with  local uniform isotropy with large-scale asymmetries, and  local uniform anisotropy with large-scale asymmetries, respectively.
    }
    \label{fig:modele_spivrvphi}
\end{figure*} 

\subsection{Generating toy models to study the impact of unresolved ordered velocity variations}
\label{sec:toy_models}
As discussed in Sect. \ref{sec:beamsmearinggaussian}, BS is present in our data and is corrected in this study using observed line flux maps and velocity fields, which limits the accuracy of the correction. In order to study the impact of residual BS effects, we perform toy models for cases with barely resolved motions due to large-scale axisymmetric rotation of various strengths, with and without additional asymmetric velocity perturbations on local scales, under both isotropic and anisotropic hypotheses.
We build toy models to produce data cubes, velocity, and \slos\ fields, and calculate the  FFT of \slos. 
This enables us to generate different configurations and identify the conditions for which systematic values in the distributions of $\sigma_k$ and/or $\phi_k$ could occur. Our toy models have no dynamical basis, yet they are very useful to assess the effects of beam smearing, anisotropy and streamings in \slos. The full description of the toy models is presented in App.~\ref{app:toymodels}. 

We produced mock datacubes of $400\times 400$  pixels (scale of 1\arcsec, or $\sim 50$ pc at a typical distance of 10 Mpc), with 200 spectral elements with a 3~\kms\ velocity sampling using $5\times 10^6$ uniformly distributed points. To first order, the galaxy is assumed to be an axisymmetric rotating disc to which velocity perturbations can be added. Two rotation curves, one with a weak velocity gradient and a moderate velocity plateau, the other with a steep inner gradient quickly reaching a high velocity plateau, were used. Sharp planar velocity perturbations are produced by a bisymmetric spiral pattern, with five possible inner angles. 
No lopsidedness ($k=1$ mode) is introduced for simplicity. 
Therefore, the perturbed models have intrinsically anisotropic velocities on large scales, as illustrated by the ellipsoid elongated in the radial dimension in App.~\ref{app:toymodels}.
We also synthesize velocity anisotropy in the toy models by modifying the  shape of the velocity ellipsoid locally, i.e. by choosing a uniform radial bias $\sigma_\theta=0.7\sigma_R$ with null covariance, with $\sigma_R=8$~\kms\ and $\sigma_z=5$~\kms, in opposition to isotropic velocity distributions ($\sigma_R=\sigma_\theta=\sigma_z =8$~\kms).  The choice of $\sigma_\theta=0.7\sigma_R$ comes from the radial bias seen of young stellar populations in the disc of the Milky Way \citep{2022gaiadrimmel}.
This is an additional effect to the streaming-driven velocity anisotropy, and switching it off and on in the toy models enables to assess   other specific sources of anisotropy not accounted for by the streaming  perturbations in the planar velocity components.

For each particle, velocities are drawn randomly with a mean velocity and a velocity dispersion for each component. The mean in radial, azimuthal and vertical velocities are computed from the rotation curve and the perturbation.  The velocities are then projected along the line-of-sight for three inclinations (45\degr, 60\degr\, and 75\degr). For each inclination, cubes were created for 12 orientations of the spiral perturbation.  Gaussian smoothing was then applied to mimic BS (8 pixels FWHM, corresponding to $\sim 400$~pc), before extracting moment maps. Beam smearing corrections were applied as for observed data.
Because of projection effects, a perturbation along the radial (azimuthal) direction is mainly seen along the minor (major) axis in the line-of-sight velocity fields and velocity dispersion maps (see Fig.~\ref{fig:toy_models}).
FFTs of $\sigma_{\rm los}^2$  were calculated for 35 independent rings, yielding profiles and distributions of $\sigma_k$ and $\phi_k$, as in Sect. \ref{sec:asym} using 
the 36 models (3 inclinations, 12 orientations of the perturbation). Figure~\ref{fig:modele_spivrvphi} shows a velocity perturbation (case where the velocity perturbation vector is orientated towards the closest position of the spiral) for the steep rotation curve. From the maps, we can see residual BS effects due to both large-scale rotation and to the unresolved perturbation. The velocity perturbation is more orientated azimuthally in the centre than in the outer parts. The effect of the radial bias is also clearly seen along the minor axis in both the velocity dispersion map and in the histograms of the second order phases.
Other cases are shown in App.~\ref{app:toymodels}. An analysis of BS residuals induced by large-scale rotation is provided in Sect.~\ref{sec:BS_residuals} and the analysis of projection effects from asymmetric perturbations in Sect.~\ref{sec:orderedvelocityeffect}.

\section{Asymmetries in velocity dispersion maps}
\label{sec:asym}

\begin{figure*}[t]
\noindent
\centering
    \includegraphics[height = 5.5cm, trim={2cm 0 2.2cm 0},clip]{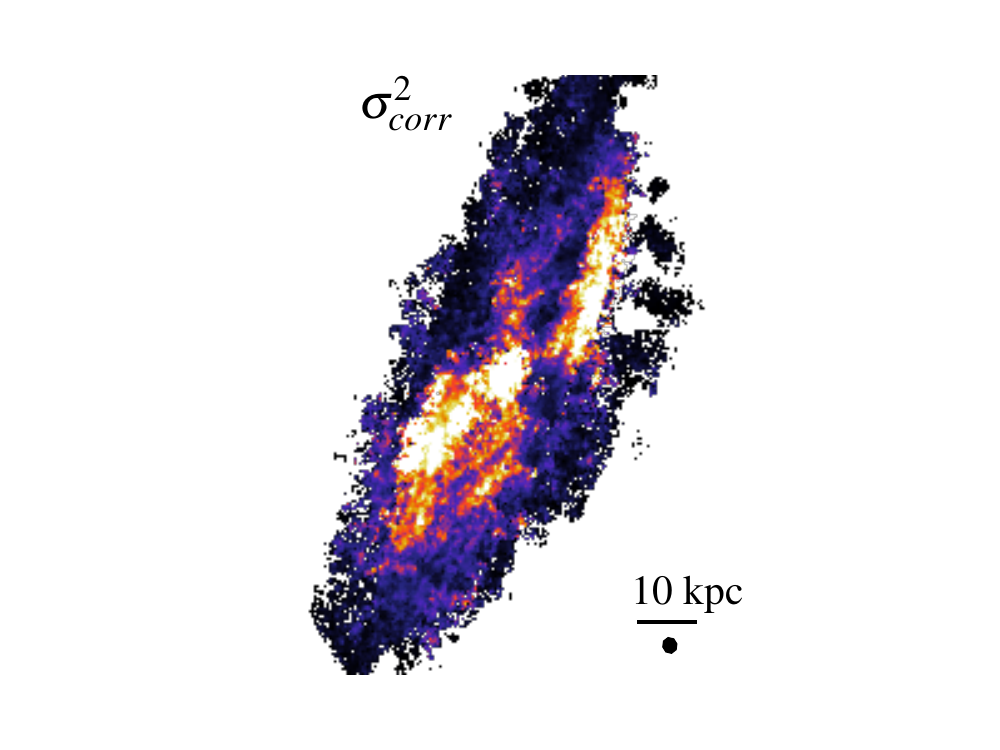}
    \includegraphics[height = 5.5cm, trim={2.3cm 0.5 0cm 0},clip]{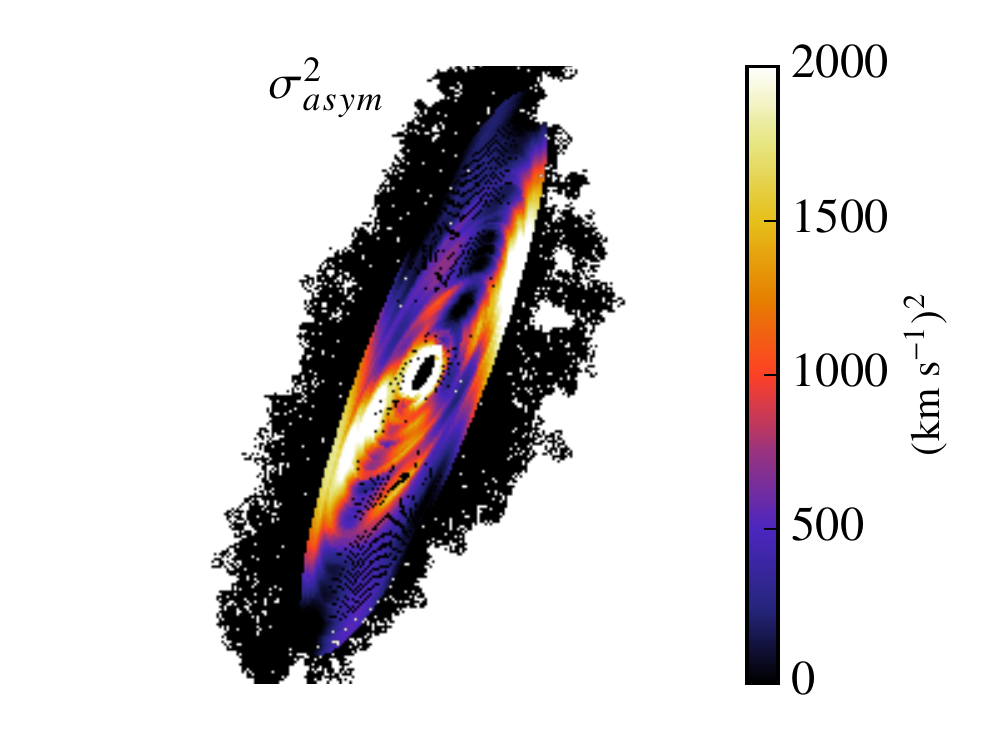}
    \includegraphics[height = 5.5cm, trim={2.5cm 0.5 0cm 0},clip]{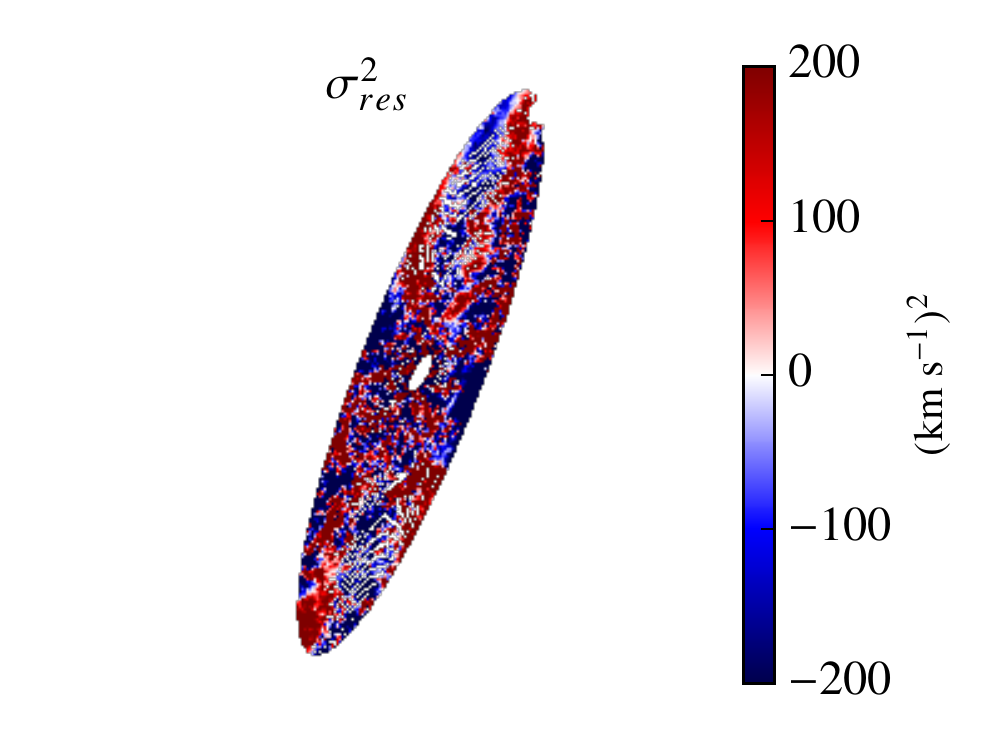} \\
    \includegraphics[height = 5.5cm, trim={3cm 0 3.8cm 0},clip]{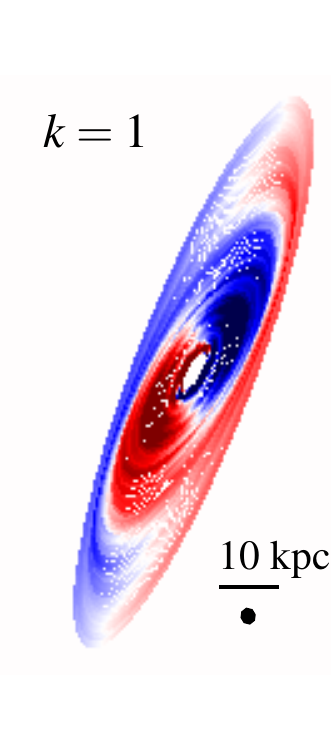}
    \includegraphics[height = 5.5cm, trim={3cm 0 4cm 0},clip]{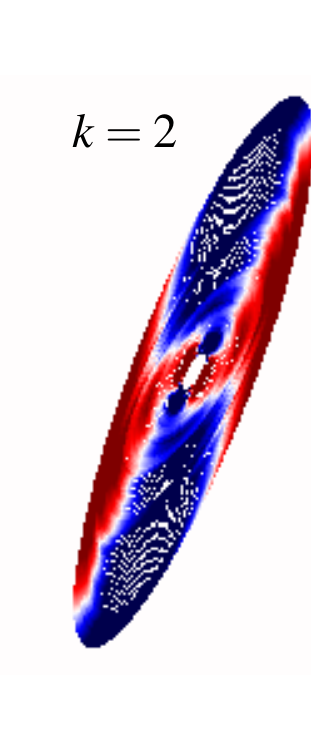}
    \includegraphics[height = 5.5cm, trim={3cm 0 4cm 0},clip]{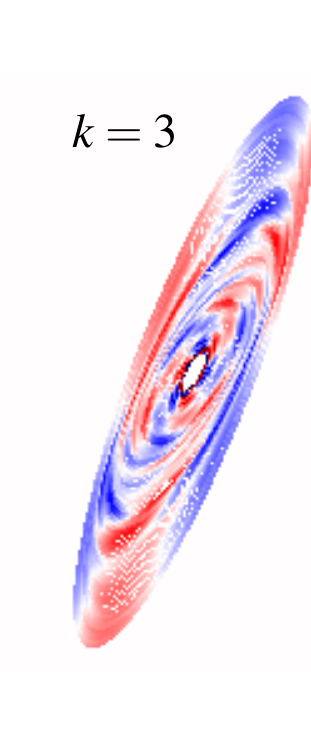}
    \includegraphics[height = 5.5cm, trim={2cm 0 0cm 0},clip]{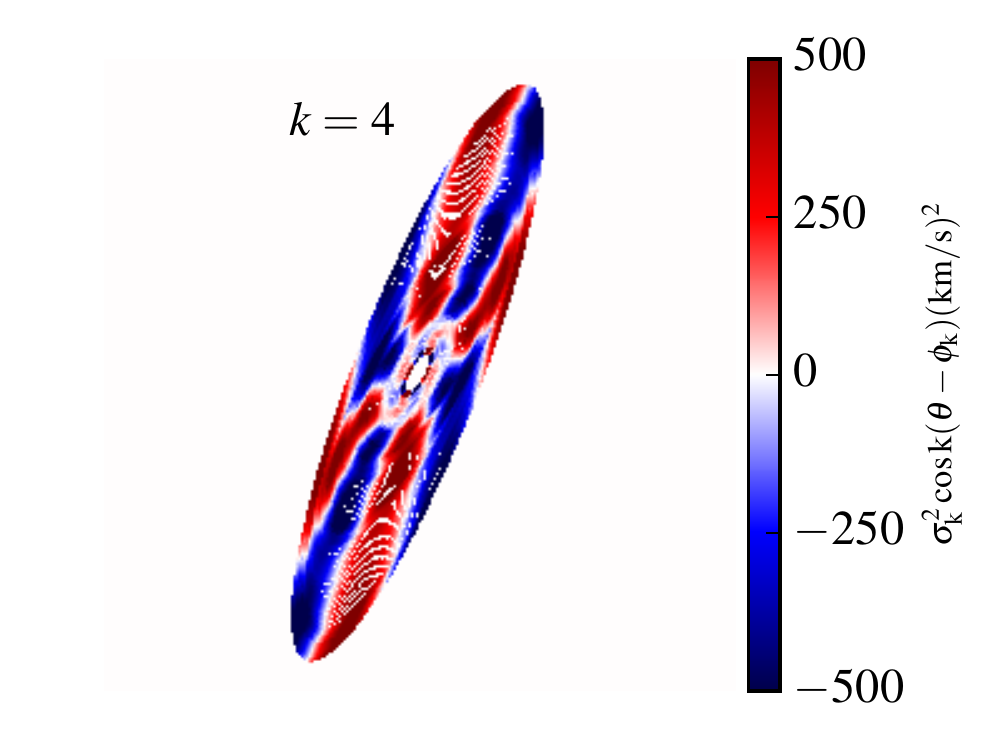} \\
    \includegraphics[height = 6cm, trim={0.7cm 0 0.5cm 0},clip]{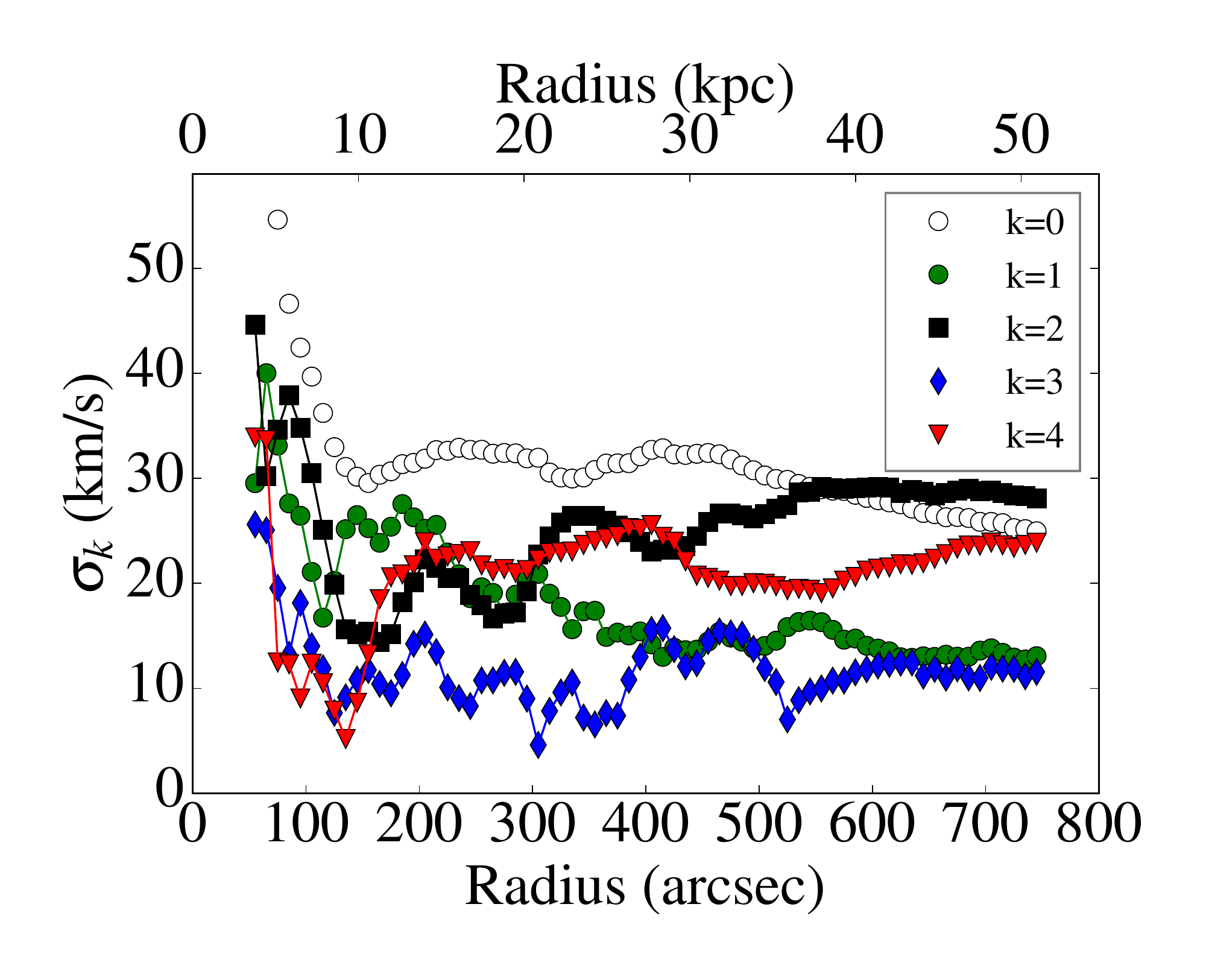} 
    \includegraphics[height = 6cm, trim={0.7cm 0 0cm 0},clip]{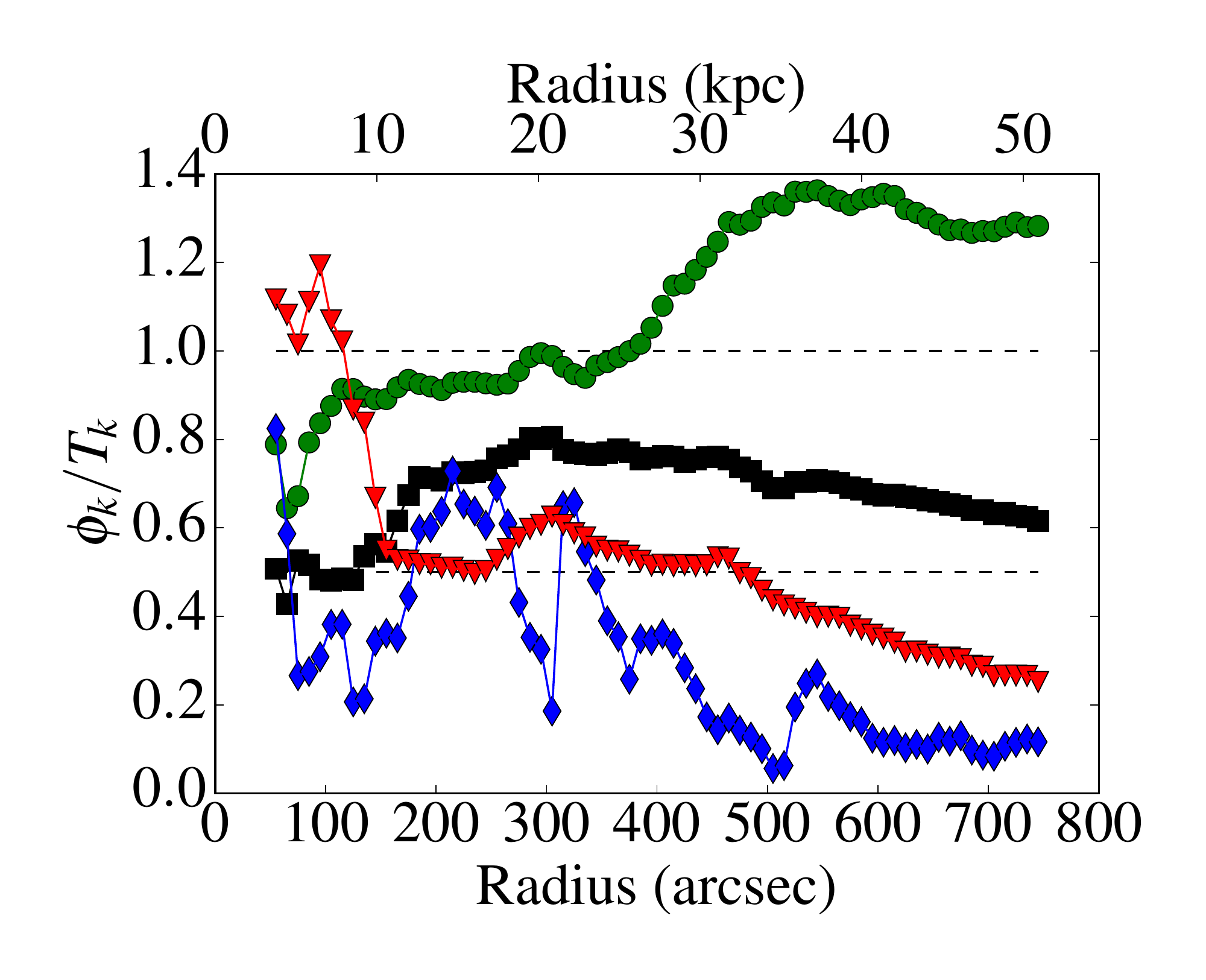} \\
    \caption{Examples of FFT results with the galaxy NGC2841. (Top) Observed squared velocity dispersion and its modelling through FFT up to the order 4, and their residuals. (Middle panel) Individual orders of the FFT projected in the plane of the galaxy. (Bottom) Amplitudes (left) and phase angles (right) of FFT coefficients as a function of the radius.}
    \label{fig:n2841-exfft0}
\end{figure*}  

Coefficients of the FFTs were calculated for the 883 and 144 radial rings of THINGS and WHISP sources, respectively.  
Of the first 20 orders, the dominant asymmetries are those up to the fourth harmonic, in agreement with \citet{2020chemin}, so we limit the analysis to $k\le 4$ hereafter.

\subsection{A case study: NGC 2841}
\label{sec:FFT-case-study}

The results of the Fourier analysis for all galaxies from the THINGS sample are presented in App.~\ref{sec:mapsgalaxies}. To illustrate examples of results, Fig.~\ref{fig:n2841-exfft0} shows the process and results for NGC2841, from $\sigma^2_{\rm corr}$ to the FFT and the individual FFT components.
The normalised phase angles $\phi_k/T_k$, where $T_k=2\pi/k$, are shown, with values of $0.5$ and 1 corresponding to half and a full period. The normalised phases $\phi_1/T_1=0, 0.5, 1$ and $\phi_k/T_k=0$ for other orders are aligned along the major axis of the galaxy, and  $\phi_2/T_2=0.5$ along the minor axis.  
In NGC2841, $\sigma^2_{\rm corr}$ is high along a cross shaped pattern, with strong $k=2$ and $k=4$ perturbations and the $k=2$ perturbation grows stronger with radius. The residual $\sigma^2_{\rm res}$ is weak, showing that the first 4 orders reproduce the structure in $\sigma^2_{\rm corr}$. 
The galaxy is lopsided at small radii \citep[e.g.][]{1980baldwin} and this is detected by the $k=1$ mode of the FFT. The phase angle of the bisymmetry does not vary much beyond $R \sim 170\arcsec$, at a value of $\sim 0.7$ times the period of the $k=2$ asymmetry.

\begin{figure*}[ht]
\centering
    \includegraphics[height=4.1cm,trim={0.4cm 0.3cm 0.5cm 0.4cm},clip]{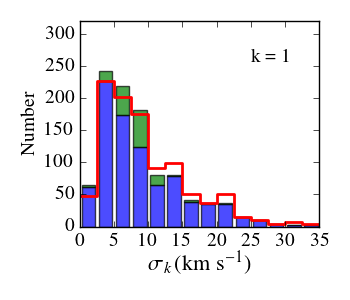}
    \includegraphics[height=4.1cm,trim={1.1cm 0.3cm 0.5cm 0.4cm},clip]{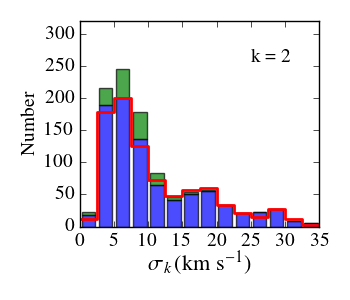}
    \includegraphics[height=4.1cm,trim={1.1cm 0.3cm 0.5cm 0.4cm},clip]{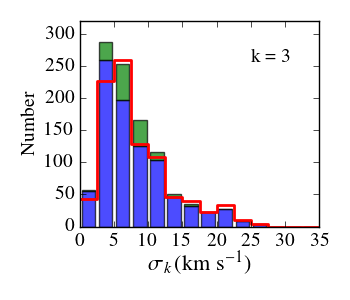}
    \includegraphics[height=4.1cm,trim={1.1cm 0.3cm 0.5cm 0.4cm},clip]{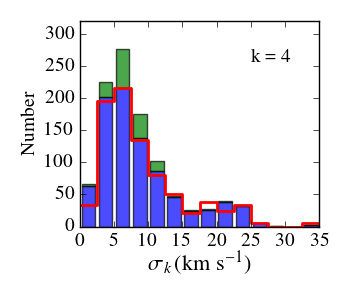} 
    \includegraphics[height=4.1cm,trim={0.4cm 0.3cm 0.4cm 0.4cm},clip]{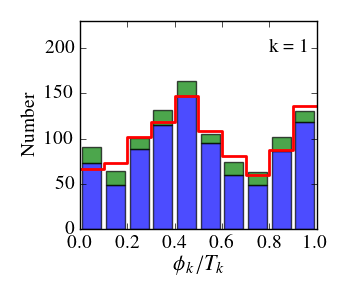}
    \includegraphics[height=4.1cm,trim={1.1cm 0.3cm 0.5cm 0.4cm},clip]{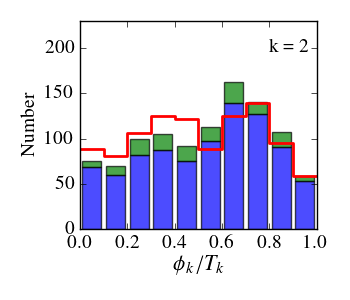}
    \includegraphics[height=4.1cm,trim={1.1cm 0.3cm 0.5cm 0.4cm},clip]{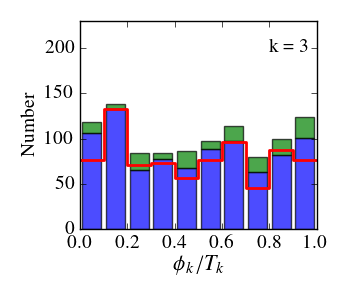}
    \includegraphics[height=4.1cm,trim={1.1cm 0.3cm 0.5cm 0.4cm},clip]{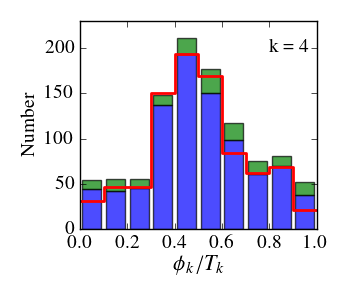}   
    \caption{Results of the FFT measurements of the \hi\ velocity dispersion maps. The histograms show the number of rings as a function of amplitude (top row) and normalised phase angle (bottom row) of the FFT harmonics of the THINGS sample (883 rings, in blue), the interpolated distribution with 20 rings per galaxy for THINGS (280 rings, in red, normalised to the maximum values of the blue histograms), and the WHISP sample (144 rings, in green). Phase angles are normalised to the period $T_k$ of each order.} 
    \label{fig:distrib}
\end{figure*} 

\subsection{Census of asymmetries in \hi\ random motions}
\subsubsection{General trends}
\label{sec:general_trend}
 
Figure~\ref{fig:distrib} presents the distributions of the amplitudes and normalised phase angles of the Fourier modes for our reference THINGS sample (blue histograms), and the WHISP sample appended to the THINGS sources  (green histograms).  Table~\ref{tab:ampl_distrib} lists the mean, median and standard deviation of the amplitudes for the 883 rings of the THINGS sample. 
Globally, the distributions of amplitudes have comparable shapes.  With an average of 11 \kms, the strongest Fourier mode is of second order, and the weakest is of third order (8 \kms).  
Most $k=2$ amplitudes are between 2.5 and 7.5 \kms\ with a few rings below 2.5 \kms, possibly due to noise.

For $k = 1$, we see in Fig.~\ref{fig:distrib} (bottom) that the phase angle peaks at $\phi_1/T_1=0.45$ and $0.95$ with dips at $0.15$ and $0.75$. 
The bisymmetric mode is maximal around $\phi_2/T_2=0.65$, and is minimum at 0. 
The $k = 3$ distribution is relatively flat with a maximum located at $\phi_3/T_3 = 0.15$. The $k=4$ histogram is strongly peaked at $\phi_4/T_4=0.5$. The uncertainties are approximately the bin size of the distributions ($0.1 T_k$). 

The distributions of Fig.~\ref{fig:distrib} may be biased by a few galaxies with more radial rings than others. To assess the impact of different numbers of rings, we measured new profiles of amplitudes and phase angles using 20 equally spaced rings per galaxy as all galaxies but NGC3627 have at least 20 independent rings.  The initial profiles were interpolated at the 20 new radii. This yields new distributions built on 280 radii for 14 galaxies (NGC3627 was excluded for this assessment). These histograms are shown as red lines in Fig.~\ref{fig:distrib}.  The agreement between the blue and red histograms shows that the unequal number of radial bins has little impact on the results.

We can estimate the significance of the dips and peaks of the phase distributions. Assuming that the gas velocity ellipsoids are isotropic, the velocity dispersion maps should  exhibit asymmetries randomly distributed over the plane of the sky. The 883 rings observed for the THINGS sample should thus yield, on average, $N=88$ counts in each of the 10 bins of the histograms.  If we quantify a confidence level as $\zeta = \sqrt{N}$, then the uncertainty is $\zeta \sim 9$.  
Keeping only the bins showing  an excess of at least $3\zeta$,  the peaks observed at $0.45$ and $0.95$ for $k=1$  are  detected at a level of $6.3\zeta$ and $3.3\zeta$ (147 and 119 counts, respectively),   $5.5\zeta$  at  $0.65$ for $k=2$  (140 counts),   $4.8\zeta$  at  $0.15$ for $k=3$ (133 counts), and up to $11\zeta$ for the bin at $0.45$ of $k=4$ (194 counts). Now keeping only the bins showing  a deficit of at least $3\zeta$,  the minima observed at $0.15$ and $0.75$ for $k=1$  correspond to a confidence level of  $4.2\zeta$ (49 counts),   $3.7\zeta$ at  $0.95$ for $k=2$ (53 counts), and   4.5 to $5.3\zeta$ at  $0.05, 0.15, 0.25$ and $0.95$ for $k=4$ (45, 42, 46 and 38 counts, respectively). Therefore, the systematic phase angles  are  not consistent with a random fluctuation of the orientations of asymmetries in the galaxies. We note that in the case of $k=3$ no trend is seen around the unique bin exceeding $3\zeta$.  It may be that this peak is caused by chance due to the limited statistics, and not by a systematic effect, unlike the  peaks or dips seen in the other asymmetries. 
With the interpolated profiles at 20 bins for each galaxy (red histograms of Fig.~\ref{fig:distrib}), we find less significant peaks and dips than with the 883 rings due to lower statistics, but the trends are preserved.

\begin{table}[!h]
\centering
\begin{tabular}{ccccc}
\hline
\hline
Order amplitude &  Mean &  Median  & Standard deviation \\ 
   & \kms   & \kms   &  \kms \\  
\hline              
$\sigma_1$       &  9.3 &  7.1  & 7.1  \\
$\sigma_2$       &  10.9 &  8.0  & 7.7  \\
$\sigma_3$       &  7.9 &  6.6  & 5.1  \\
$\sigma_4$       &  8.8 &  7.0  & 6.1  \\
\hline
\end{tabular}
\caption{Properties of Fourier amplitudes for the THINGS galaxies, as measured from 883 tilted rings.}
\label{tab:ampl_distrib}
\end{table}    

\subsubsection{Correlations between the Fourier modes}

\begin{figure*}[h]
\centering
    \includegraphics[width=5.3cm,trim={0.4cm 0.3cm 2.6cm 0.4cm},clip]{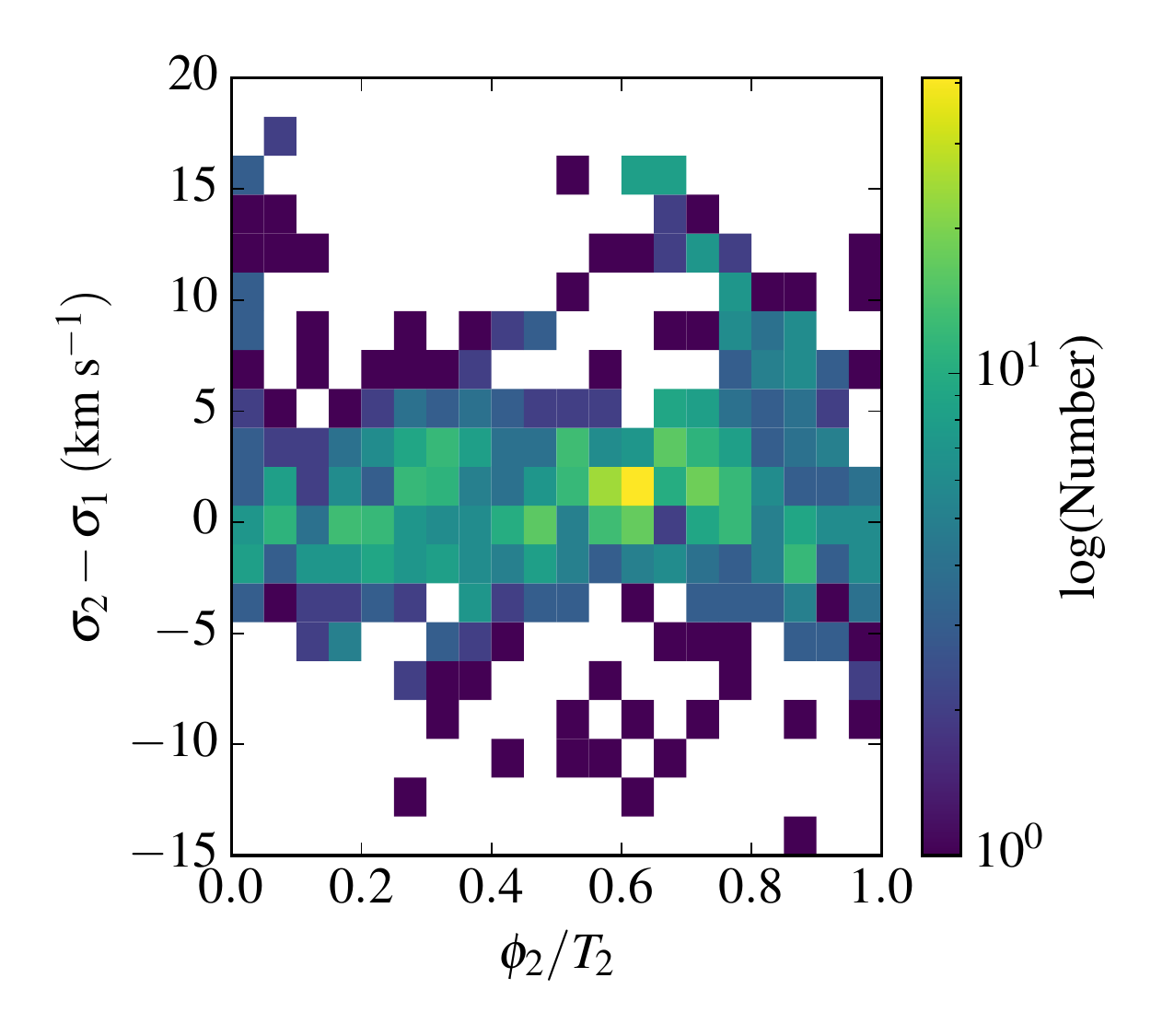}
    \includegraphics[width=5.3cm,trim={0.4cm 0.3cm 2.6cm 0.4cm},clip]{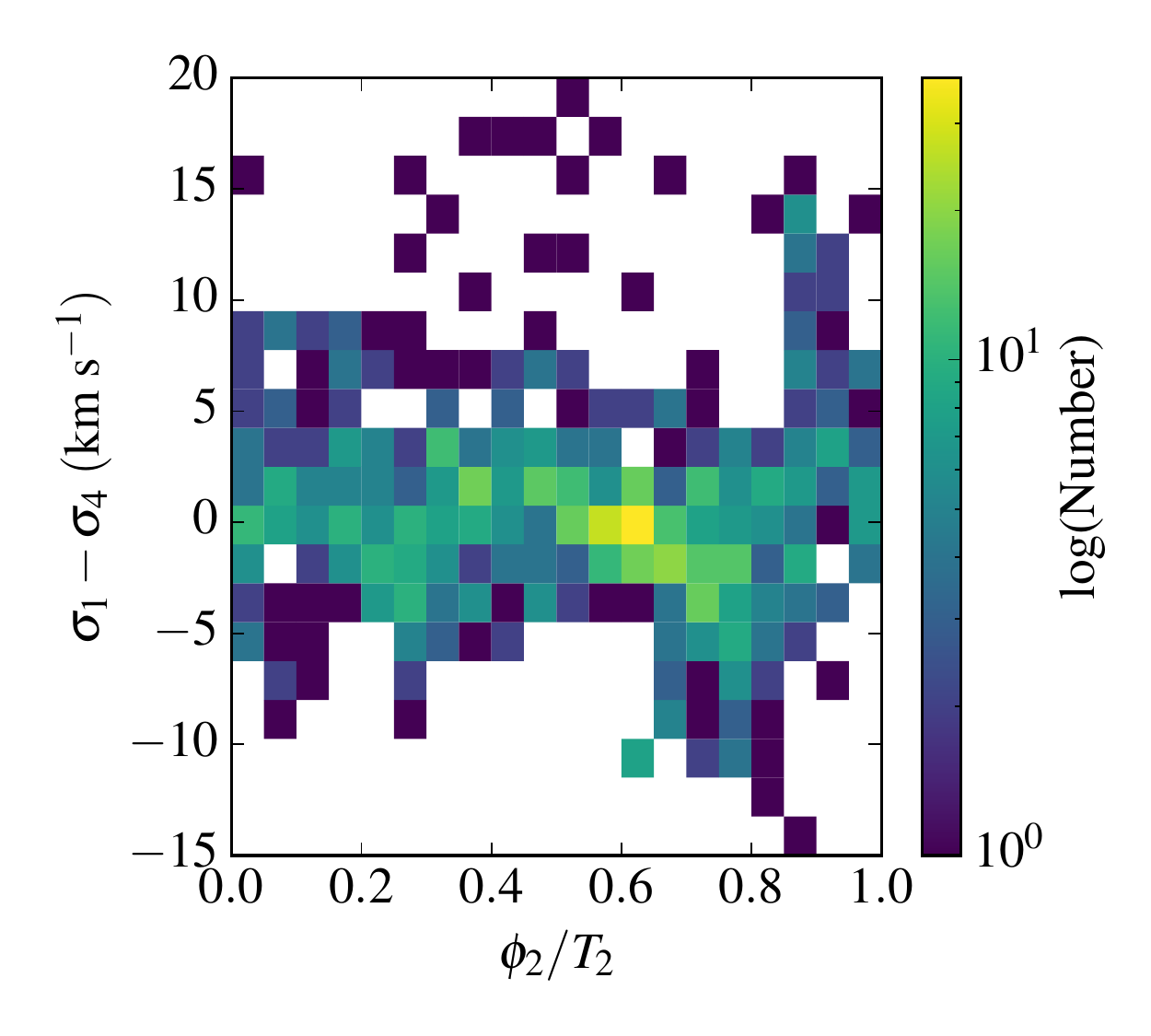}
    \includegraphics[height=5.8cm,trim={0.4cm 0.3cm 0.4cm 0.4cm},clip]{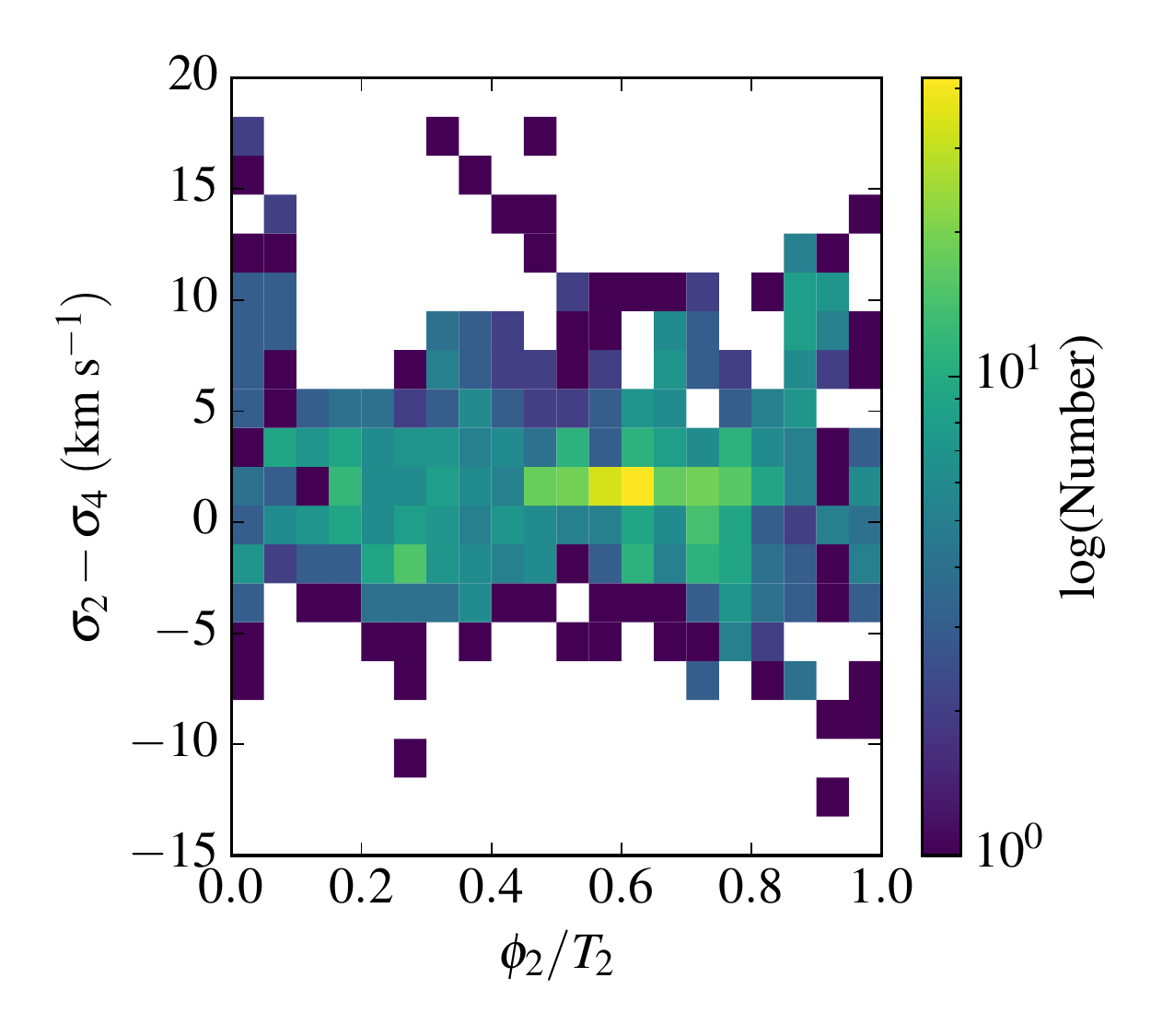} \\
    \includegraphics[height=5.2cm,trim={0.5cm 0.3cm 0.3cm 0.4cm},clip]{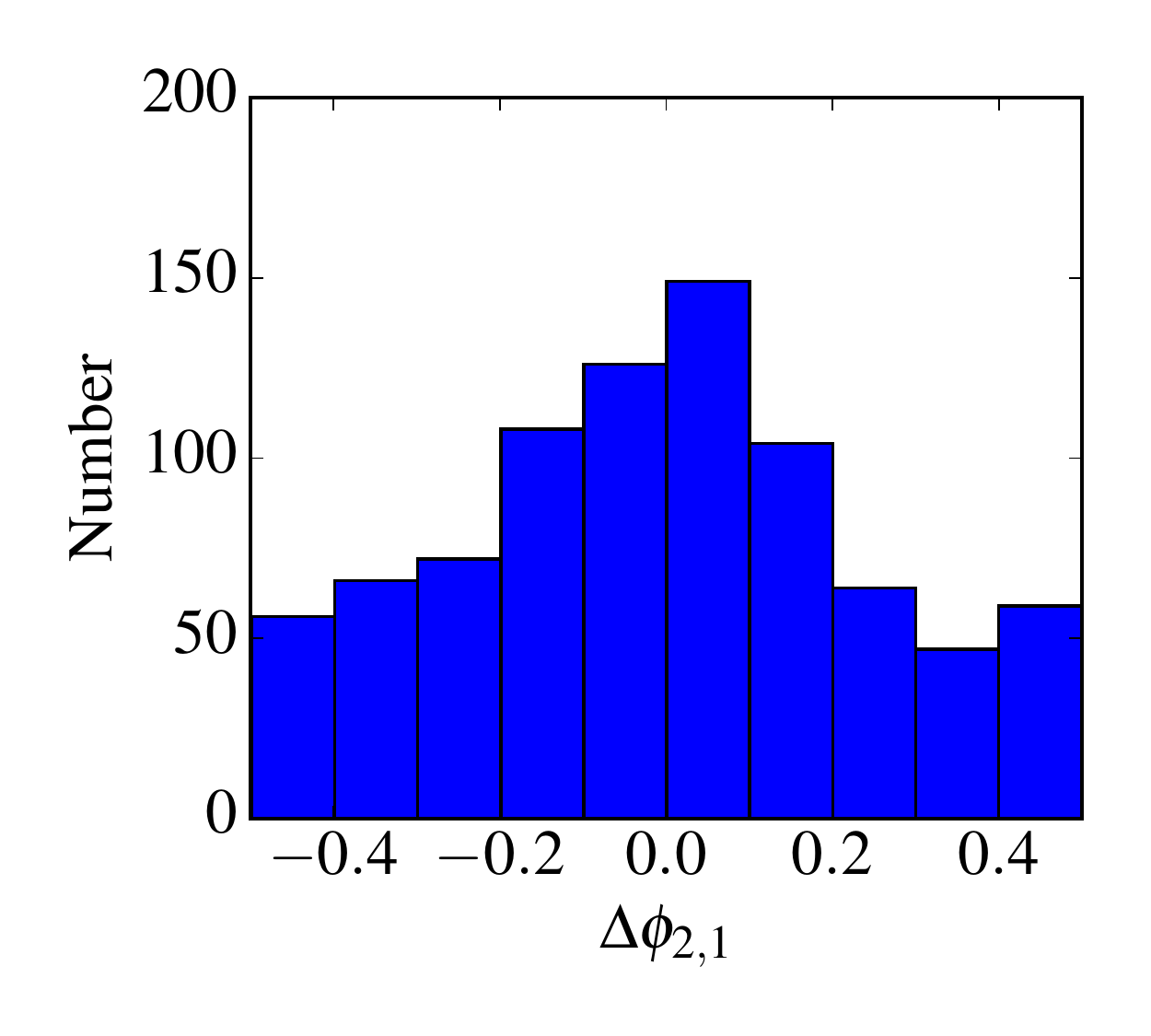}
    \includegraphics[height=5.2cm,trim={1.5cm 0.3cm 0.3cm 0.4cm},clip]{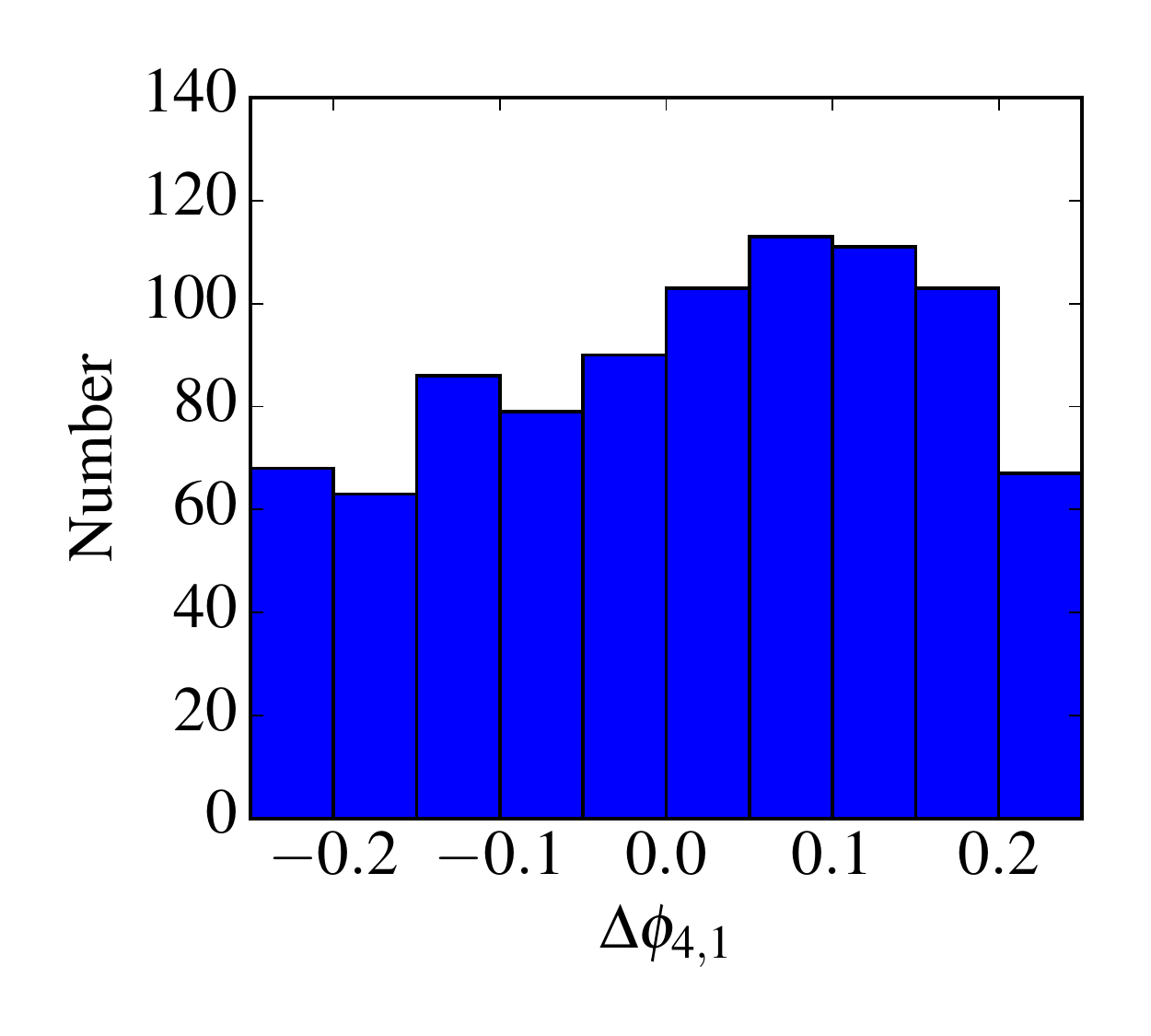}
    \includegraphics[height=5.2cm,trim={1.5cm 0.3cm 0.3cm 0.4cm},clip]{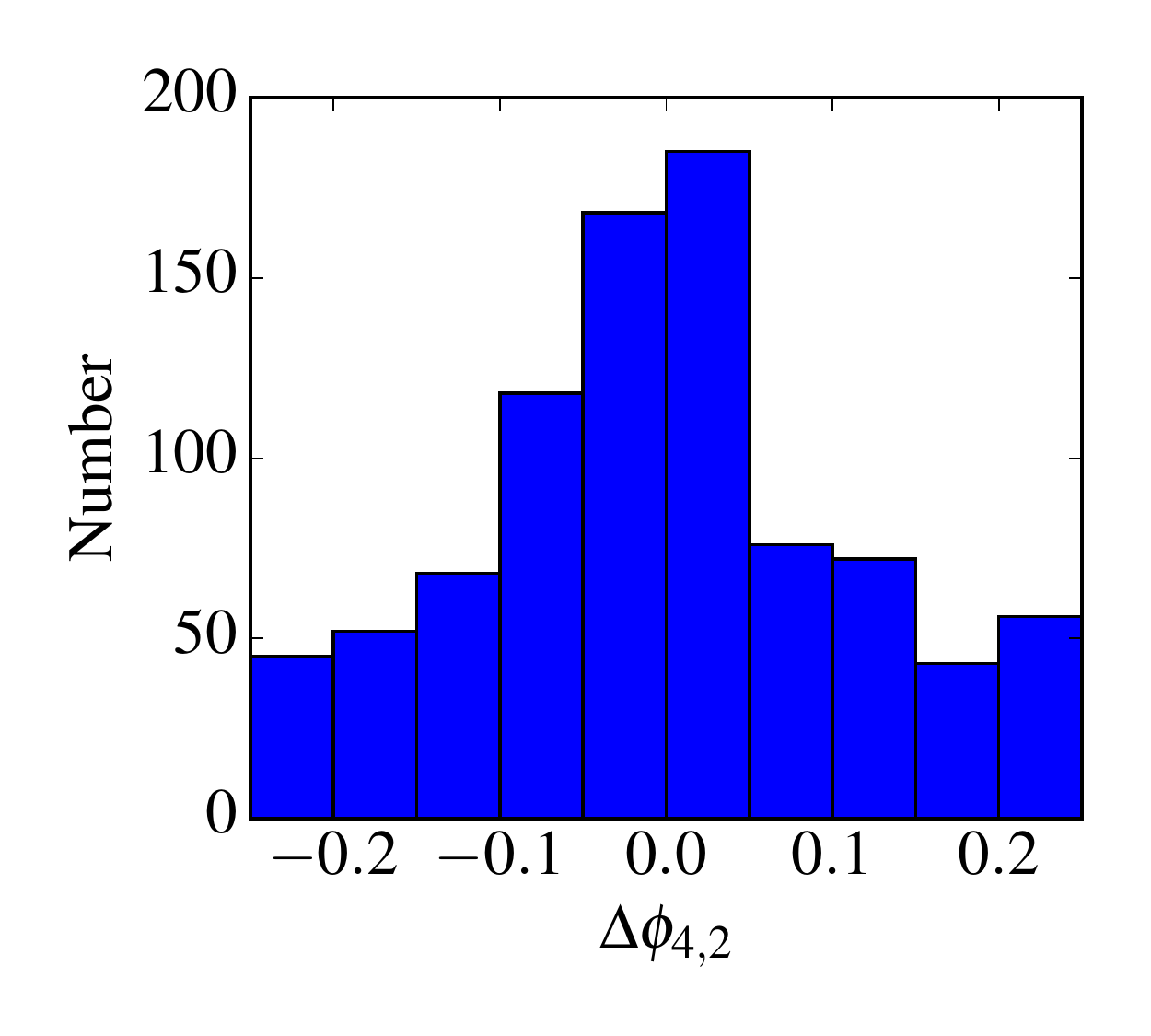}
    \caption{Comparisons of amplitudes and phase angles between the Fourier $k=1,2$ and 4 orders, shown as differences of amplitudes versus $\phi_2/T_2$ (upper row) and histograms of phase angles difference normalised to $\pi$ (bottom row). The comparisons between the orders $k = 1$ and $k = 2$,  $k = 1$ and $k = 4$, and $k = 2$ and $k = 4$  are shown in the left, middle, and right columns, respectively. In the upper row, the amplitude difference is colour coded with the logarithmic number of the tilted rings.}
    \label{fig:correlation}
\end{figure*}

The upper row of Fig.~\ref{fig:correlation} compares the Fourier amplitudes  
as a function of the phase angle $\phi_2/T_2$ for the strongest orders ($k=1,2,4$) for THINGS galaxies. The colour code is the number of tilted rings within bin widths of 0.05 for $\phi_2/T_2$, and 1.75 for the amplitude difference.
The density of tilted rings is highest around a difference of 0 \kms\ (in agreement with values given in Tab.~\ref{tab:ampl_distrib}), irrespective of the value of $\phi_2/T_2$, which implies correlated amplitudes. We measure a Pearson correlation coefficient of $0.8\pm 0.1$ between the amplitudes of orders 2 and 4.

The bottom row of Fig.~\ref{fig:correlation} compares the phase angle differences between orders $\Delta \phi_{m,n} = \phi_m - \phi_n$, and shows the differences within one period of the highest order, that is within $\pm T_m / 2$, with  $m > n$. 
The distributions of $\Delta \phi_{2,1}$  and $\Delta \phi_{4,2}$  are peaked and symmetric around zero, implying a correlation between the $k=1$ and $k=2$ modes and between $k=2$ and $k=4$.  These correlations suggest that $k=4$ is a harmonic of the second order perturbation.

\subsection{Correlations with the \hi\ flux density}

Several processes may induce correlations or anti-correlations between gas velocity dispersion and gas density.
A large velocity dispersion can arise from unresolved motions, such as insufficient spatial resolution or the presence of unresolved multiple peak profiles. These unresolved velocity gradients may result from various physical processes, including steep density gradients, gas compression in density waves, instabilities associated with spiral-like features, or starburst outflows.
The faintest regions may also display a high velocity dispersion because the signal-to-noise ratio (SNR) affects the profile widths, noisier profiles appearing broader.
Leaving aside the aforementioned observational caveats (low resolution, low SNR) and focussing on the resolved regions, correlations between dispersion and density can have different origins. 
Large-scale star formation is clearly linked to \hi\ content but this is not true for small scales \citep[e.g.][]{Zhou+18}.
As long as the \hi\ density is below the density necessary to gravitationally collapse and form molecular hydrogen, the \hi\ gas clouds have no reason to cool and the velocity dispersion to decrease. 
In some galaxies \citep[e.g. NGC 4214,][]{Wilcots+01}, the largest \ha\ velocity dispersion is observed in the diffuse ionised gas regions which often has a low \hi\ column density. Around  massive star forming regions, the large velocity gradients might be explained by outflows. In low \hi\ column density regions far from massive stars \citep[e.g. NGC 2366,][]{Hunter+01}, in the absence of a local source, the velocity width might be induced by long-range turbulent pressure.

Using NGC2841 as an example, Fig.~\ref{fig:n2841-ex} shows that the gas density ($\Sigma_{\rm los}$) is principally distributed in bright rings and outer spiral arms,
whereas larger velocity dispersion ($\sigma_{\rm los}$) is along a centred cross-shaped structure. In other words, the gas density is either anti-correlated or correlated with velocity dispersion, sometimes fainter in large velocity dispersion regions (e.g. in the galaxy centre), brighter at low velocity dispersion (e.g. in the NE and SW quadrants), but also brighter at larger velocity dispersion (e.g. NW and SE quadrants).
Across the sample, similar features are observed, and velocity dispersion seems correlated to spiral arms. Therefore, due to the wide diversity of processes, a galaxy-by-galaxy or a pixel-by-pixel comparison between the flux density and the velocity dispersion is beyond the scope of this paper. We rather aim at identifying global trends with respect to our analysis of the velocity dispersion.

\begin{figure*}[h]
\centering
    \includegraphics[width=6cm]{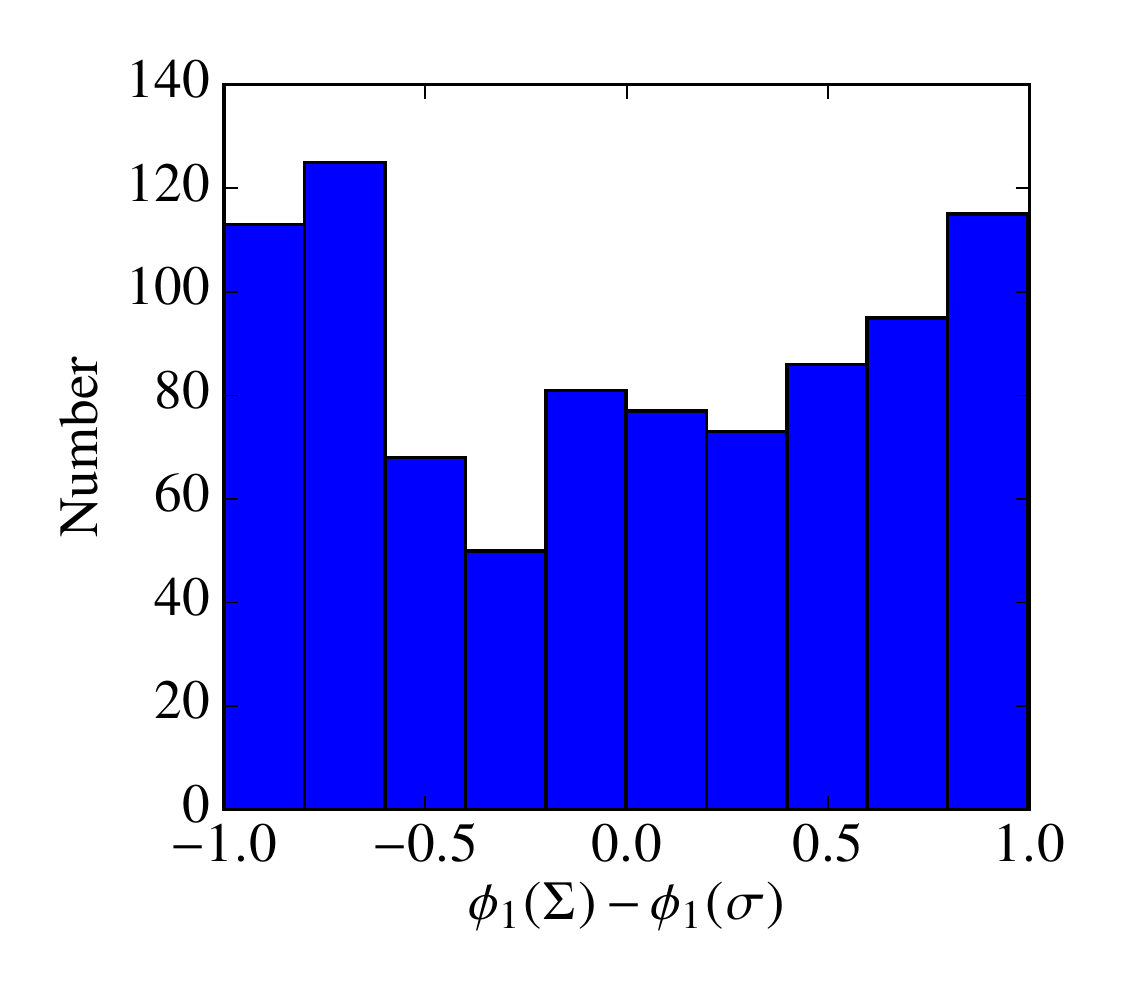}\includegraphics[width=6cm]{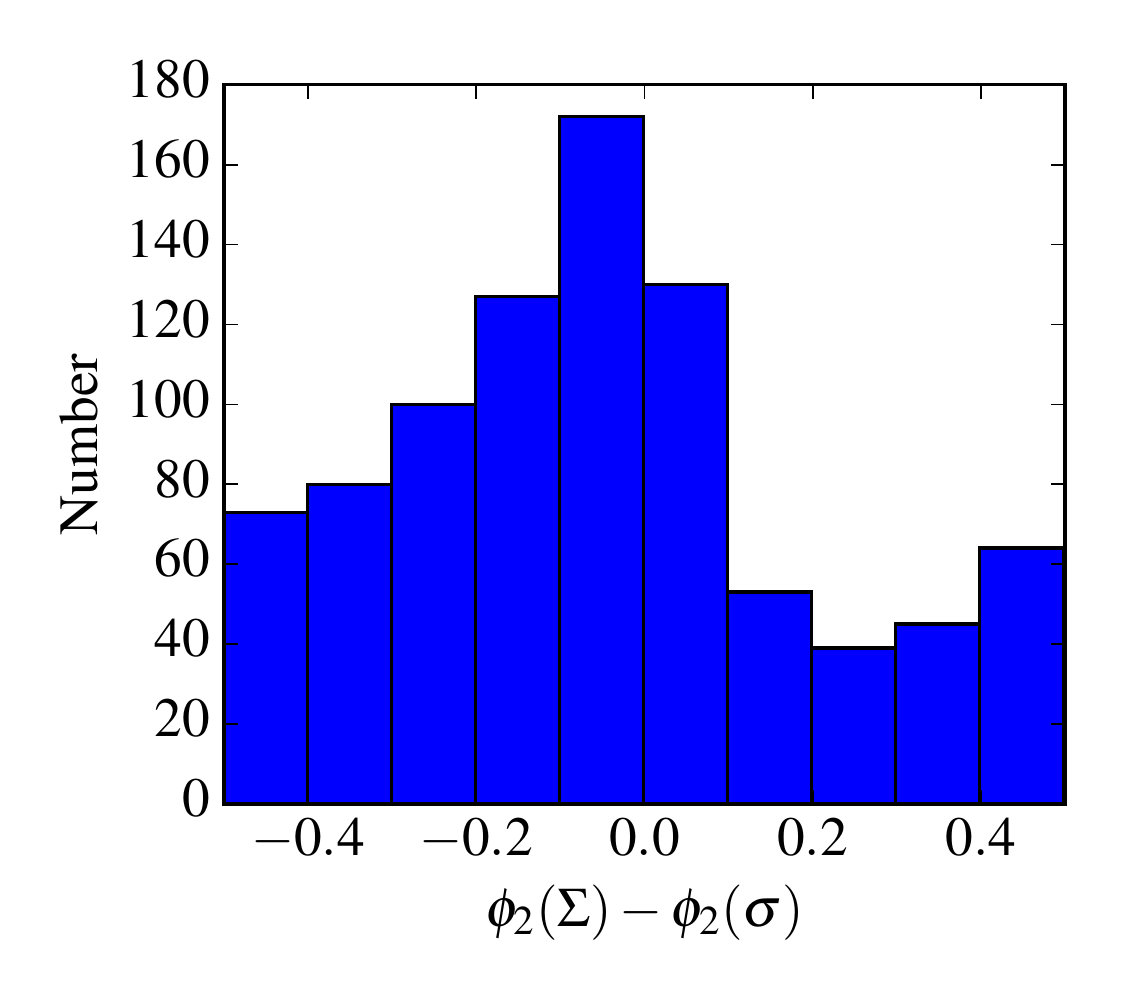}\includegraphics[width=6cm]{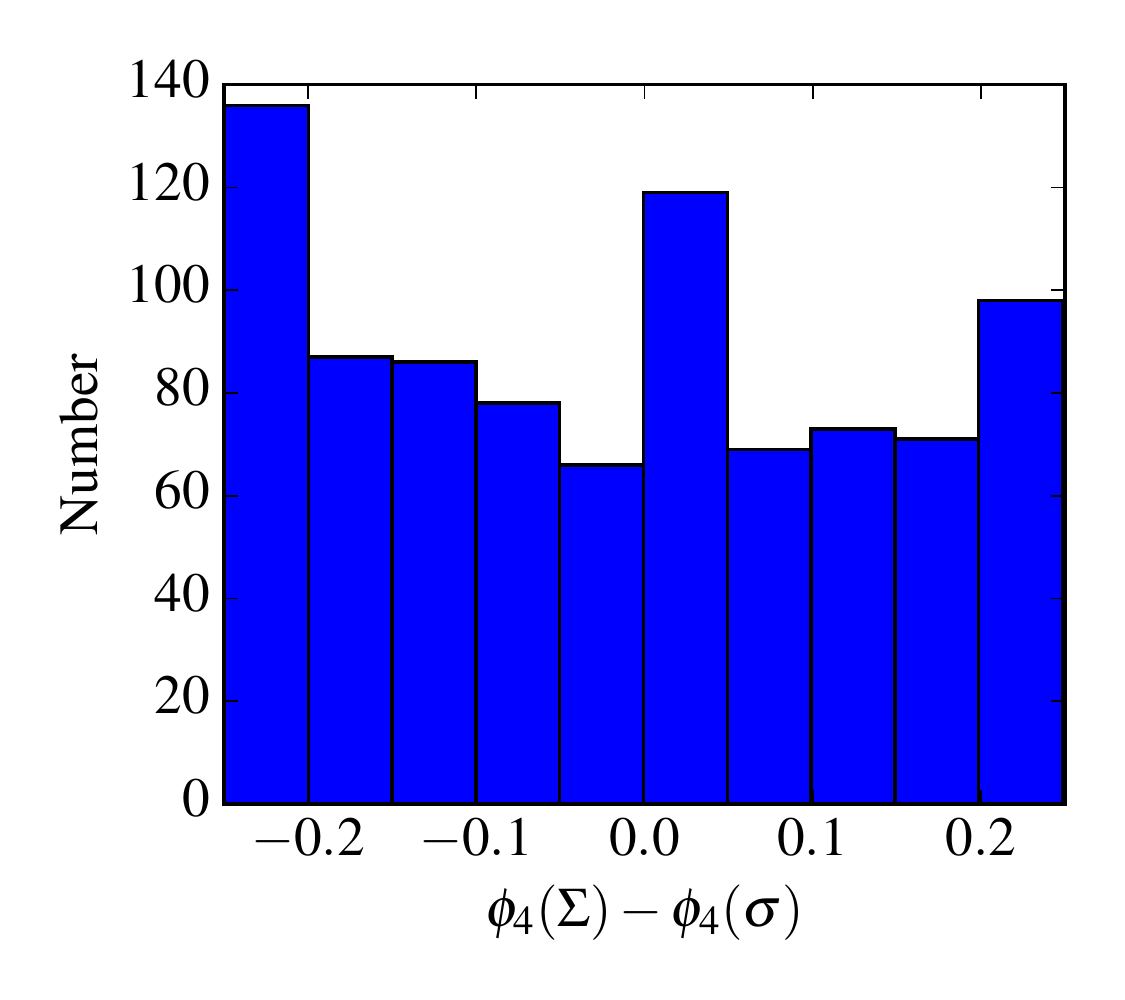}    
    \caption{Histograms of phase angles (normalised by $\pi$) difference between gas density and velocity dispersion $\phi_k (\Sigma) - \phi_k (\sigma)$ for $k=1, 2$ and 4 (left, middle, and right panels, respectively).}
    \label{fig:flvsdisp}
\end{figure*}

The correlation between gas density and velocity dispersion has therefore been investigated for the whole sample using FFTs of the density maps, in the same way as for the velocity dispersion. The phase angle differences $\Delta \phi_k \equiv \phi_k(\Sigma) - \phi_k(\sigma)$ are shown in Fig.~\ref{fig:flvsdisp}. 
As in NGC2841, we observe that all orders except $k=4$ exhibit correlated (peaks) and uncorrelated (troughs) phases. 
An anti-correlation is observed for $k=1$ (dip at $\Delta \phi_k \sim 0$), and a correlation   for $k=2$ (peak at $\Delta \phi_k \sim 0$).
These findings suggest that the asymmetries observed in the velocity dispersion are related to those of the gas distribution that are mainly induced by spiral arms, bars, warps and lopsidedness.

\subsection{FFT results versus galaxy properties}
\label{sec:phasevsproperties}

Our sample is somewhat biased by the fact that the smallest galaxies, which are usually the latest type and faintest ones, are also the closest\footnote{It is a well-known observational merit factor bias which consists in ``filling the field-of-view'' of the instrument as much as possible.}.
To look for an effect, we divided the sample into two subsamples of 7 galaxies each around the median value of each parameters: morphological type, optical radius, absolute magnitude, and metallicity. Class I (II) is the sub-sample with the more distant (nearest), or the brightest (faintest), or the earliest (latest) type, or  the largest (smallest) galaxies. 

\begin{figure*}[t]
    \centering
    \includegraphics[height=5cm]{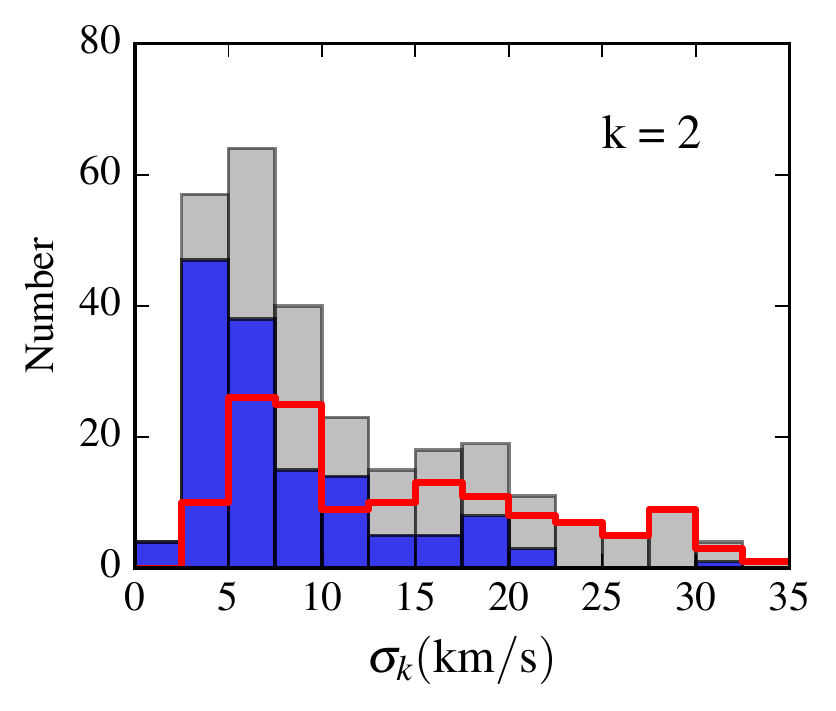}    
    \includegraphics[height=5cm]{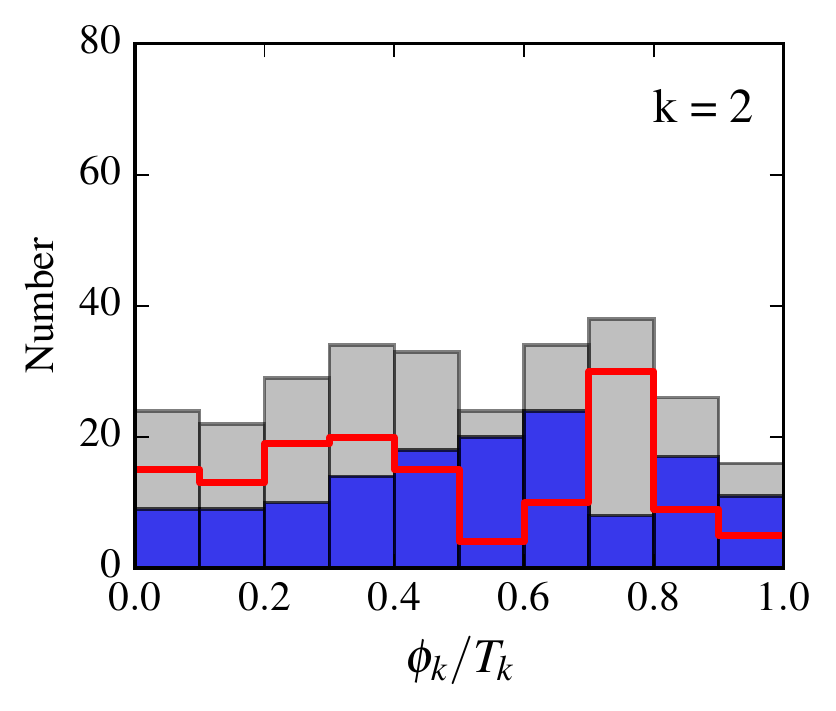}
    \includegraphics[height=5cm]{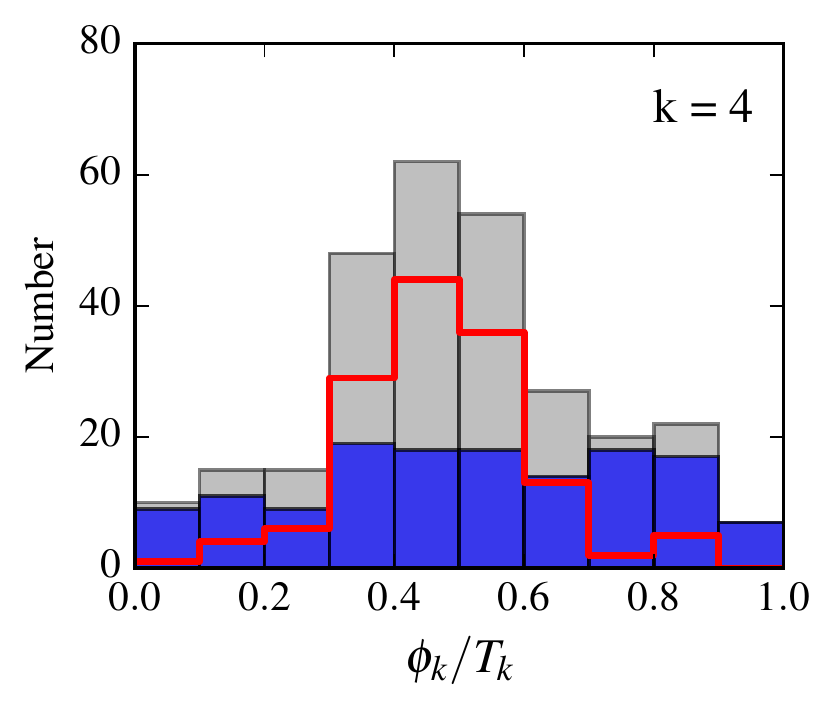}
    \caption{Histograms of the velocity dispersion amplitudes $\sigma_k$ ($k=2$, left panel) and of the normalised phase angles $\phi_k/T_k$ ($k=2$, middle panel and $4$, right panel) for the 7 galaxies brighter and fainter than $M=-20.7$ mag (red and blue colours, respectively) for the THINGS sample. The subsamples are the same as when splitting between high (red) and low (blue) projected maximum velocity (see Sect.~\ref{sec:BS_residuals}). The grey histograms are the total distributions from Fig.~\ref{fig:distrib}.}. 
    \label{fig:fftvsmabs}
\end{figure*}

In Fig.~\ref{fig:fftvsmabs}, we show histograms of $\sigma_2$, $\phi_2$, and $\phi_4$ for classes I (red) and II (blue) of absolute magnitude, which is the parameter providing the largest difference between classes I and II.
As expected, $\sigma_2$ is lower for small galaxies. 
For $k=2$, the incidence of $\phi_2/T_2 \sim 0.5$ is low (high, respectively) for class I galaxies (class II).
$\phi_2/T_2$ in class II galaxies agrees with the observations of M33 \citep{2020chemin}, which is a class II galaxy.
For orders $k=1$ and $3$, the phase does not depend significantly on the class.
$\phi_4/T_4$ has a bigger peak for class I than for class II but the difference is smaller for other parameters than the absolute magnitudes.
The $\phi_4/T_4$ distribution is almost flat for faint galaxies and peaked for  bright ones.  This could be due to residual BS (see Sect.~\ref{sec:BS_residuals}) or because higher gas density contrasts are expected in earlier type spirals, inducing larger perturbations (see Sect.~\ref{sec:orderedvelocityeffect}).  

\section{Discussion}
\label{sec:discussion}

Our FFT analysis of the velocity dispersion of the \hi\ gas shows that common values for $\phi_k/T_k$ are 
(1) $\phi_1/T_1 \approx 0.45$ or 0.95, (2) $\phi_2/T_2 \approx 0.35$ and $0.65$ and (3) $\phi_4/T_4 \approx 0.45$, with no major structure in the $k=3$ mode.  These peaks correspond to major axis of the 1st order asymmetry and half period of the 4th order asymmetry. Interestingly, finding $k=2$ asymmetries projected preferentially near the major axis is significantly less likely.

In the absence of instrumental effects such as BS at any scale, and assuming gas is isotropic, we expect the principal axes of the galaxies to be randomly orientated with respect to the spiral pattern or bar, so the distributions of $\phi_k/T_k$ should be uniform.
In this section, we discuss  the  origin of the alignment of phases around  particular angles.

\subsection{Systematic effects in the FFT measurements and the tilted-ring models} 
 
\subsubsection{Significance of FFT coefficients}
\label{sec:significance}

\begin{figure*}[t]
    \includegraphics[height=4.1cm,trim={0.4cm 0.3cm 0.4cm 0.2cm},clip]{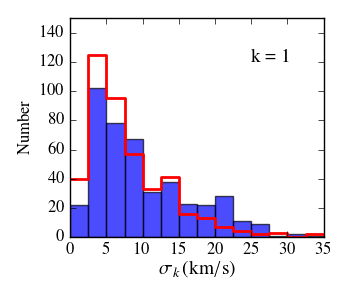}
    \includegraphics[height=4.1cm,trim={0.9cm 0.3cm 0.4cm 0.2cm},clip]{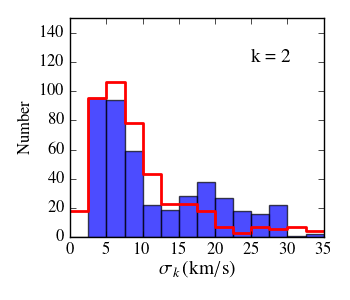}
    \includegraphics[height=4.1cm,trim={0.9cm 0.3cm 0.4cm 0.2cm},clip]{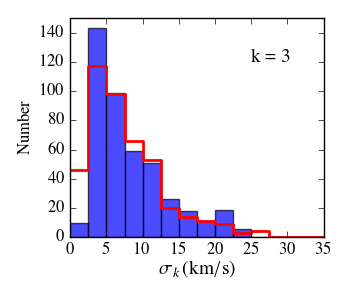}
    \includegraphics[height=4.1cm,trim={0.9cm 0.3cm 0.4cm 0.2cm},clip]{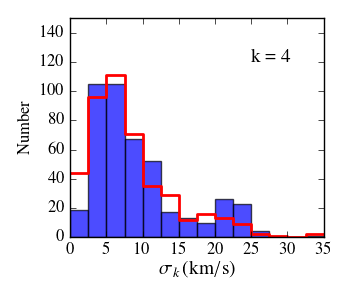}\\
    \includegraphics[height=4.1cm,trim={0.4cm 0.3cm 0.4cm 0.2cm},clip]{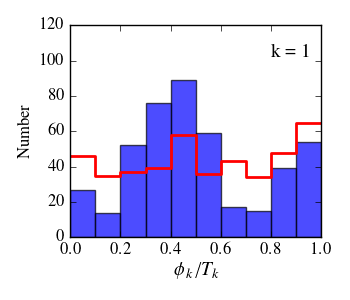}
    \includegraphics[height=4.1cm,trim={0.9cm 0.3cm 0.4cm 0.2cm},clip]{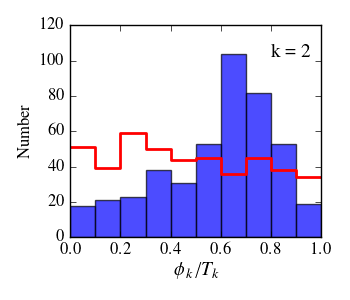}
    \includegraphics[height=4.1cm,trim={0.9cm 0.3cm 0.4cm 0.2cm},clip]{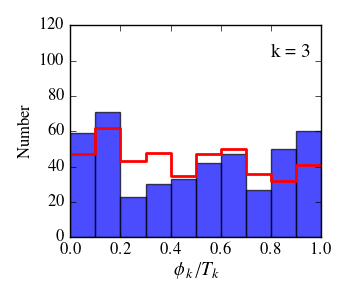}
    \includegraphics[height=4.1cm,trim={0.9cm 0.3cm 0.4cm 0.2cm},clip]{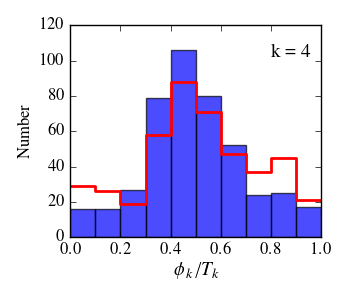}
    \caption{Histograms of amplitudes (upper row) and normalised phase angles (bottom row) resulting from the FFT analysis for the THINGS sample when splitting the sample in two, using the median significance of each order (see Sect.~\ref{sec:significance}). The blue and red distributions are for the rings with a significance respectively greater and smaller than the median value. }
    \label{fig:distrib-median}
\end{figure*}

In App.~\ref{app:biaises}, we describe the method used to infer the accuracy $d(\sigma^2_k)$ of a measured amplitude $\sigma_k^2$ (Eq.~\ref{uncertainty_approx}), and its significance $s_k=\sigma_k^2/d(\sigma^2_k)$.
For this exercise, we assume a reasonable uncertainty $d\sigma = 1$\kms\ on the standard deviation of the line spread function. 
By keeping only the rings for which the significance is greater than 3, we find that amongst the 883 rings of the THINGS sample,  
482 exceed this threshold for the order $k = 1$, 587 for $k = 2$, 431 for $k = 3$, and 536 for $k = 4$. 

We further divided the rings into high and low significance samples using the median significance of each order. The median values of the significance $s_k$ are 3.3, 5.2, 2.9, and 3.9 for $k=1,2,3$ and 4, respectively. 
The resulting  histograms are given in Fig.~\ref{fig:distrib-median}, with the low(high)-significance samples shown in red (blue, respectively).
A first result is that apart larger tails in the distributions of high-significance samples, the amplitudes are not much affected by the division into two subsamples.
Since some high amplitudes rings are located in the innermost part of galaxies where the number of pixels is much lower than in the outskirt, we observed that some high amplitudes rings (red tail) are less significant than some of the low amplitude ones (blue distribution at low amplitudes).
On the other hand, the low-significance samples show essentially random phases, except for order $k=4$, while the high-significance show the trends identified before ($\phi_1/T_1 \approx 0.45$, $\phi_2/T_2 \approx 0.65$ and $\phi_4/T_4 \approx 0.45$) but more clearly.
In other words, the rings of lower significance are consistent with having asymmetries randomly distributed in the sky plane (except for $k=4$), while those of higher significance seem to be more representative of the systematically orientated asymmetries in the velocity dispersion maps. Moreover, the fact that both sub-samples have $\phi_4$ centrally peaked around the half period is a hint that low amplitude residuals due to BS still affect the data, despite the correction applied before deriving the FFTs. We address this in Sect.~\ref{sec:bs}.

\subsubsection{Deviations in the tilted-ring model projection parameters}

We now test the robustness of our results with respect to variations of kinematic centre, inclination and position angle.
On the one hand, we changed the position of the kinematic centre of the rings by adding a random value selected from a Gaussian distribution of standard deviation of 1\arcsec\ and 3\arcsec, respectively for the right ascension and declination, as expected from their average uncertainties \citep{2008trachternach}.
On the other hand, we ignored warps by using the mean inclination and position angle for each galaxy, as given in \citet{2008AJ....136.2648D}.

The new distributions of phases are then compared to those of Fig.~\ref{fig:distrib}, by averaging the positive differences of histogram heights of all bins. 
Overall, the trends described in Sect.~\ref{sec:asym} are still present.  
Varying the kinematic centre leads to differences of 10\%, 7.2\%, 5.7\% and 7.7\%, for $k=1,2,3$ and $4$, respectively.
Not taking into account the disc warps has little impact as well, although the variations are more important, of 8\%, 15.6\%, 10\% and 22.8\%  for $k=1,2,3$ and $4$, respectively. 
We can thus exclude the possibility that the systematic phase angles of the asymmetries in the \hi\ velocity dispersion are caused by incorrect kinematic parameters.

\subsection{Systematic signatures associated to beam smearing}
\label{sec:bs}

Beam smearing  might provide residuals in the velocity dispersion beyond our correction process if locally the velocity gradient is not resolved as shown from Eq. \ref{eq:bs3}, or if the radio beam is not well represented by the 2D Gaussian model described in Sect.~\ref{sec:beamsmearinggaussian}.  An incomplete BS correction could contribute to or create velocity dispersion asymmetries.

\subsubsection{Properties of the beam smearing pattern}
\label{sec:bs_properties}

\begin{figure*}[t]
    \includegraphics[height=5.4cm,trim={0.5cm 0.3cm 3cm 0},clip]{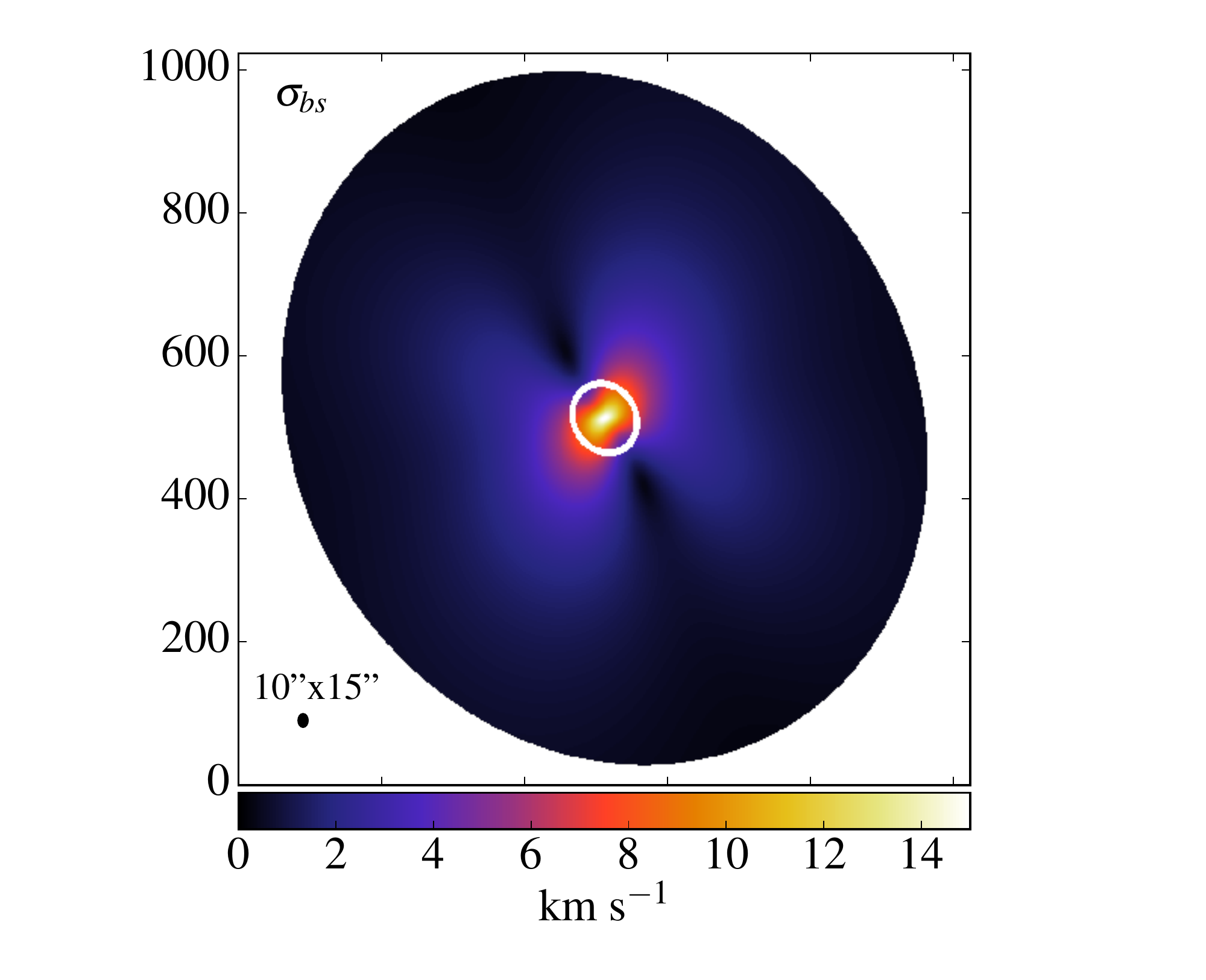}
    \includegraphics[height=5.4cm,trim={0.5cm 0 0.5cm 0},clip]{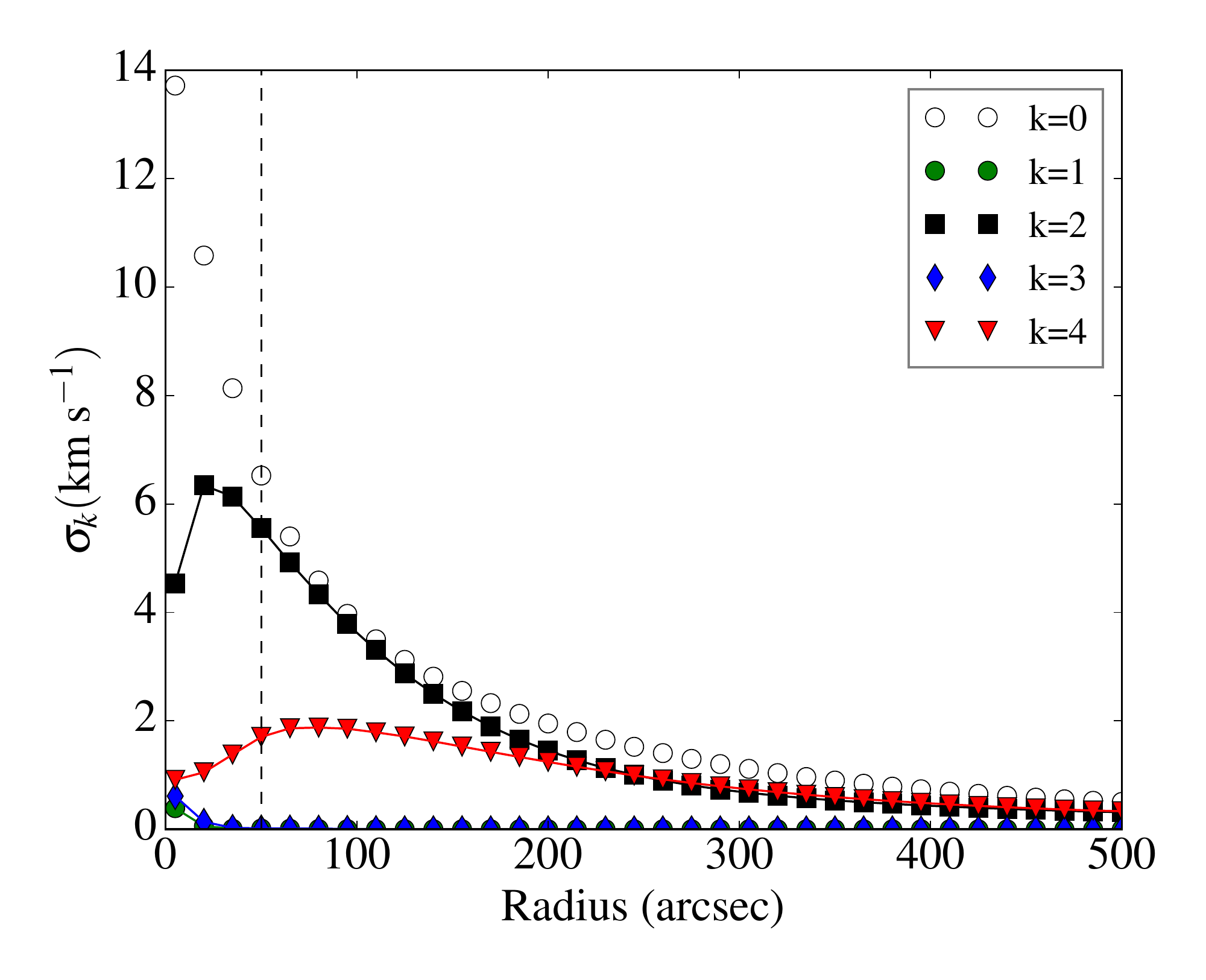}
    \includegraphics[height=5.4cm,trim={0.5cm 0 0 0},clip]{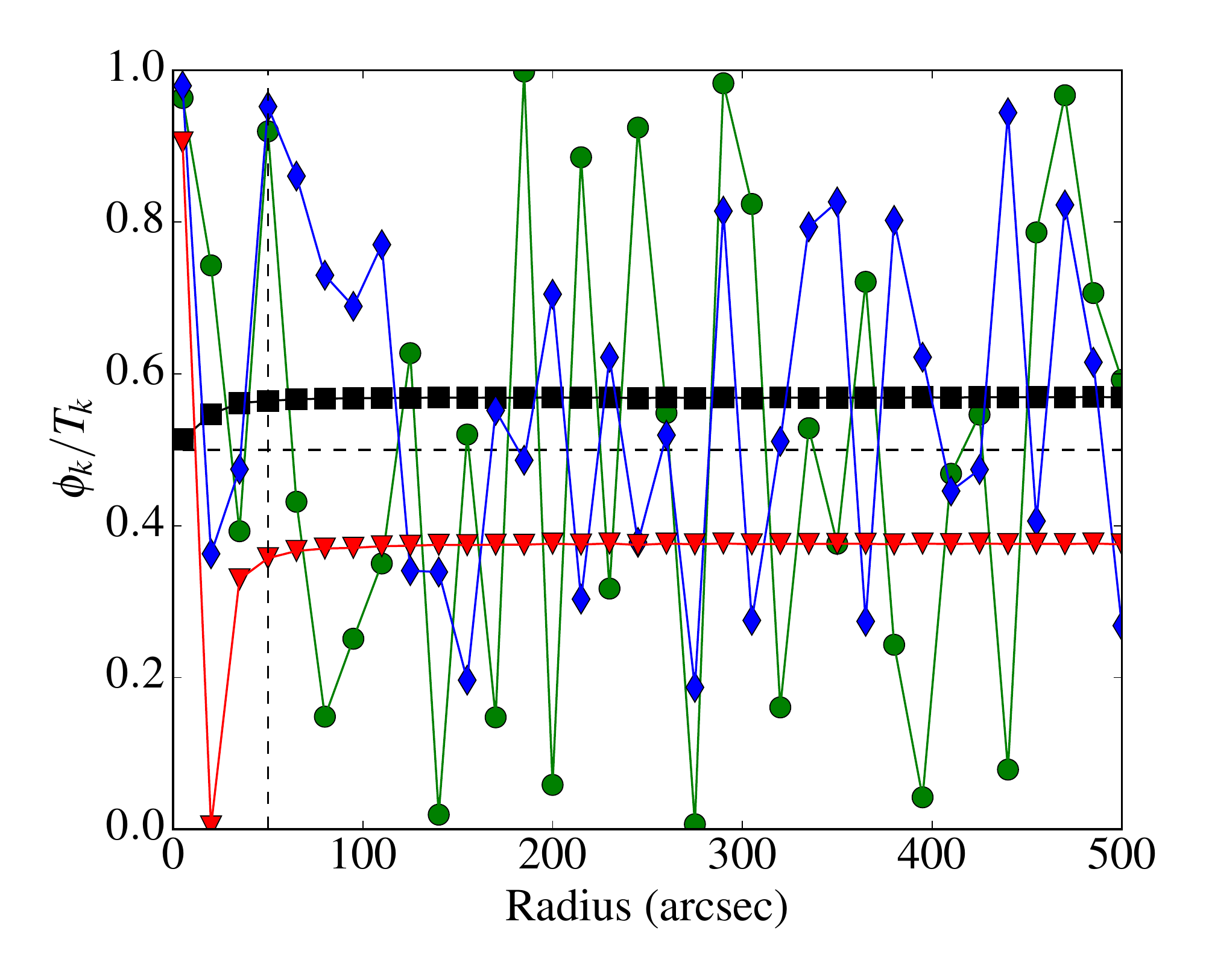} \\
    \includegraphics[height=5.4cm,trim={0.5cm 0.3cm 3cm 0},clip]{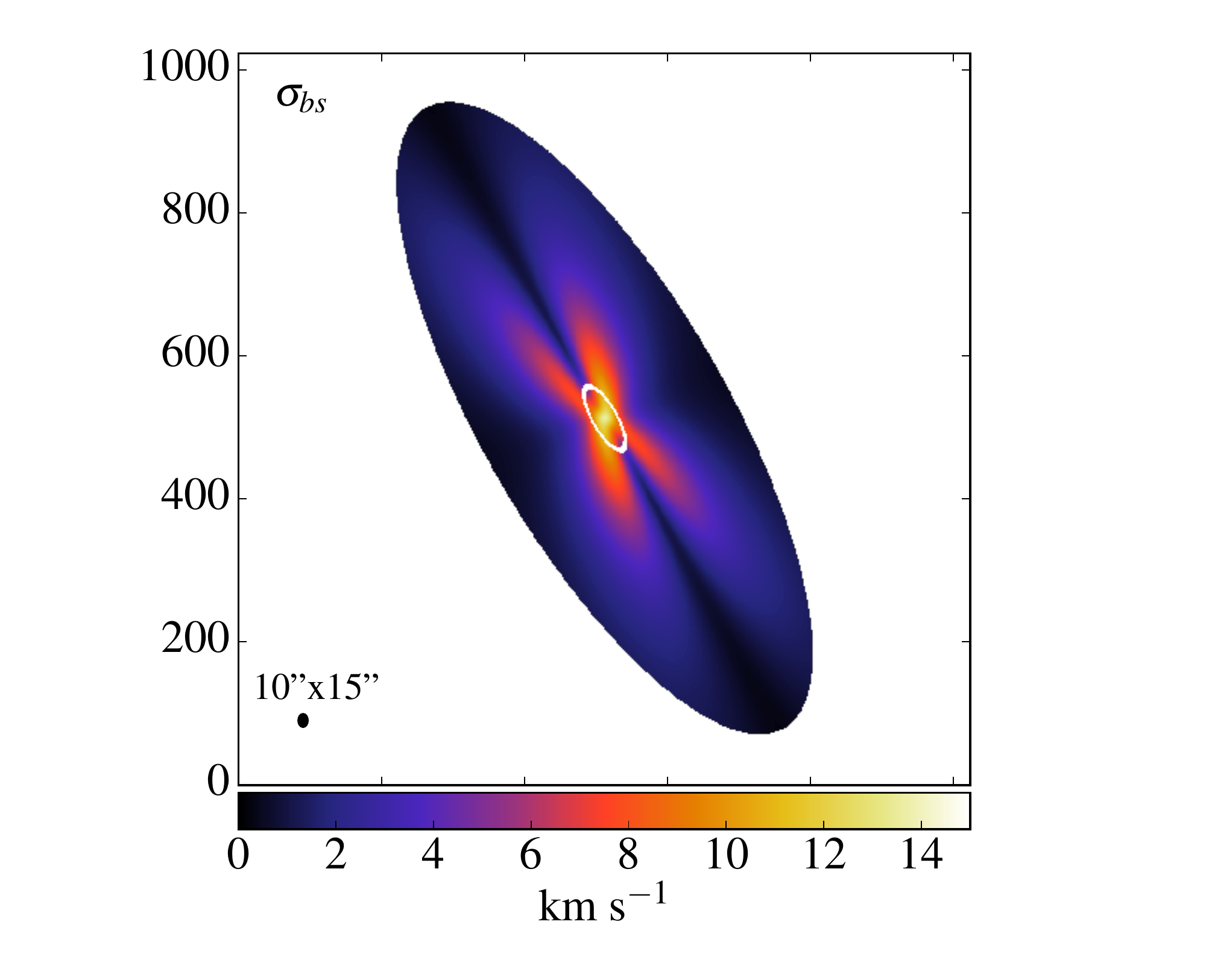}
    \includegraphics[height=5.4cm,trim={0.5cm 0 0.5cm 0},clip]{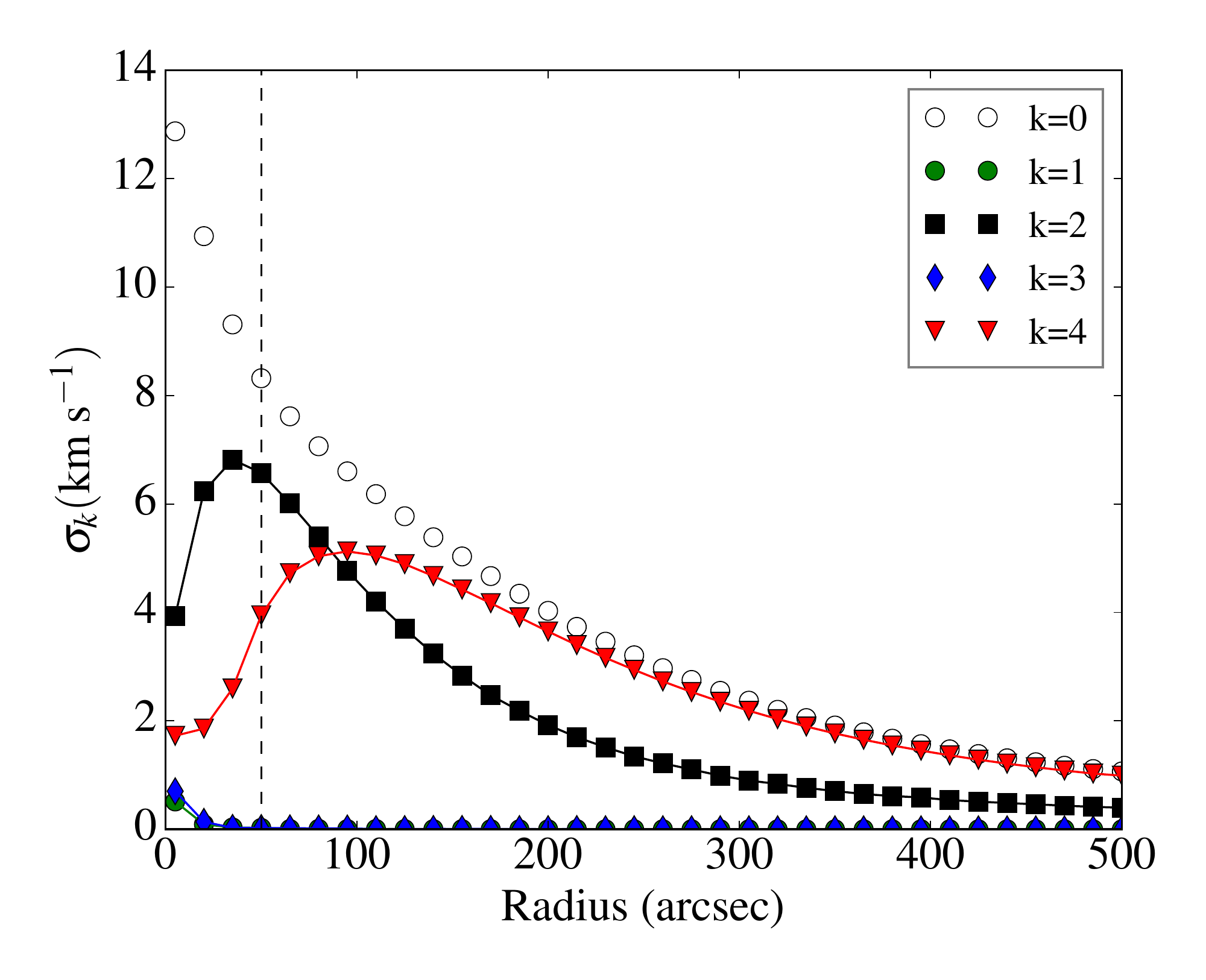}
    \includegraphics[height=5.4cm,trim={0.5cm 0 0.5cm 0},clip]{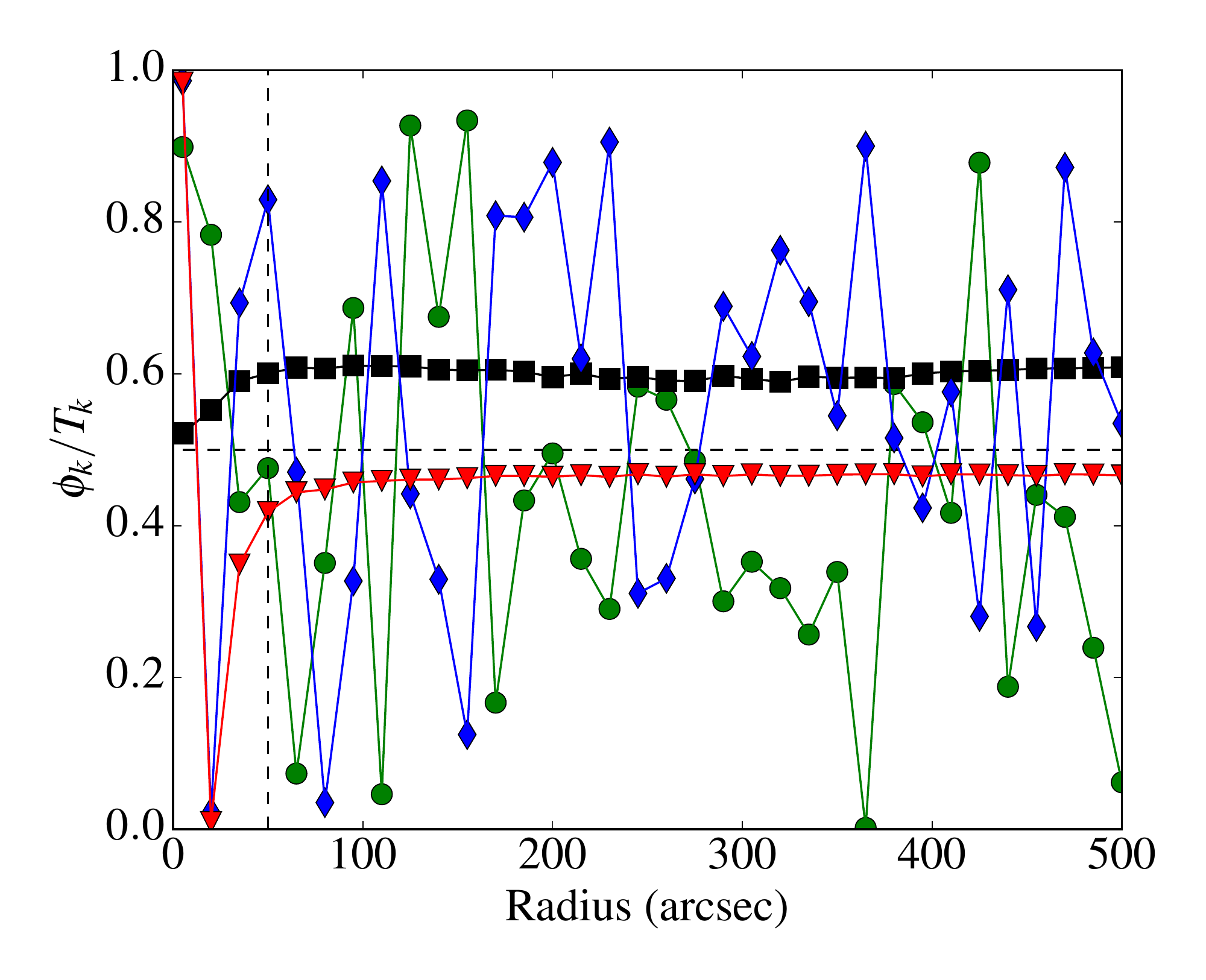}
    \caption{Results of the FFTs applied to the BS dispersion maps $\sigma_{\rm bs}$, as built from disc models projected at 30\degr\ (top row) and 70\degr\ (bottom row). \textit{Left panel:} Map of $\sigma_{\rm bs}$. The synthesised beam adopted in the toy model is shown on the bottom left corner. \textit{Middle panel:} Amplitudes of the Fourier modes for orders $k=0$ to $k=4$. \textit{Right panel:} Phase angles of the Fourier modes. A horizontal dashed line shows the half period of the Fourier modes.
    The disc scale-length is shown as a white contour (left) and as a vertical dashed line (middle and right).}
    \label{fig:bsvsinc}
\end{figure*}

To numerically explore the effect of BS, we first create a mock galaxy having both exponential flux and rotation curve models. 
In order to work with the same amplitude of the projected velocity field and thus with a similar BS induced velocity dispersion at zeroth order, we use a maximum projected velocity of 100\kms\ independent from the inclination.
The Gaussian beam width is set to $10\arcsec \times 15\arcsec$, and in order to sample correctly the exponential disc, we adopt a pixel size of 1\arcsec\ and a scale-length of 50\arcsec.
Figure~\ref{fig:bsvsinc} shows (left)  the BS  induced velocity dispersion maps at 30\degr\ (top) and  70\degr\ (bottom) inclinations. The central and right panels show the amplitudes and phase angles of the FFT decompositions. 
In our mock galaxy and PSF models, the BS pattern can be described using only orders $k = 2$ and $k = 4$. 
The ratio between the amplitude of these two orders vary with the inclination, with the $k = 4$ strength  being predominant at large inclination. This latter is maximum around the radius where the maximum rotation curve is reached.
The phases $\phi_2$ and $\phi_4$ are rather constant with radius, with their value depending on inclination. 
At fixed inclination, this constant value is related to the position angle of the galactic disc relative to that of the beam. In case both position angles match, or if the beam is circular, we have exactly $\phi_2 = 0.5 T_2$ and $\phi_4 = 0.5 T_4$.
For an inclination of 70\degr,  $\phi_2$ and $\phi_4$ are always in the range of $0.4 T_2 - 0.6 T_2$ and  $0.4 T_4 - 0.6 T_4$ respectively, with $\phi_4$ being close to its half period for most position angles. 

\begin{figure*}[t]
\centering
    \includegraphics[height=4.1cm]{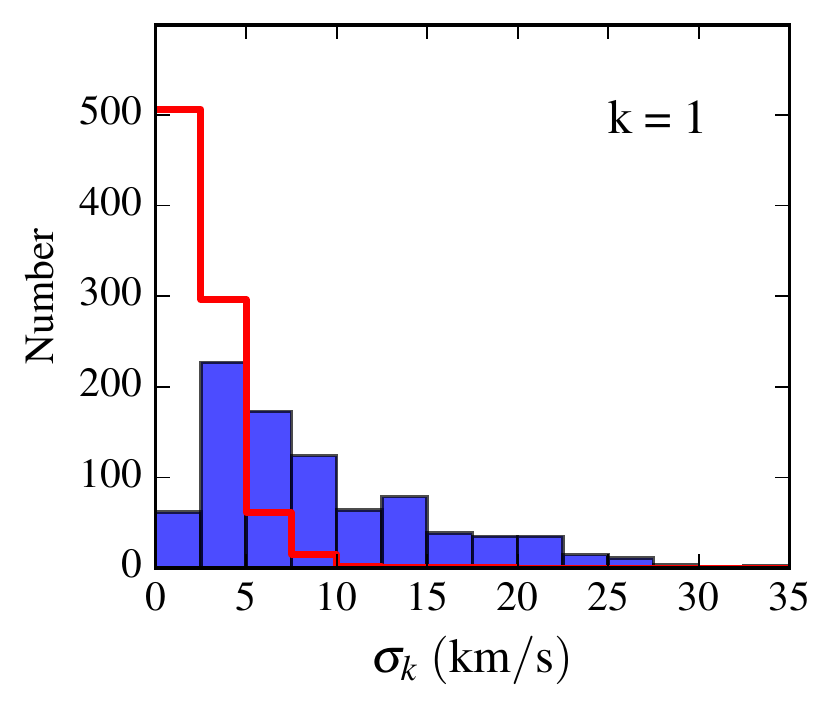}
    \includegraphics[height=4.1cm,trim={0.7cm 0cm 0cm 0cm},clip]{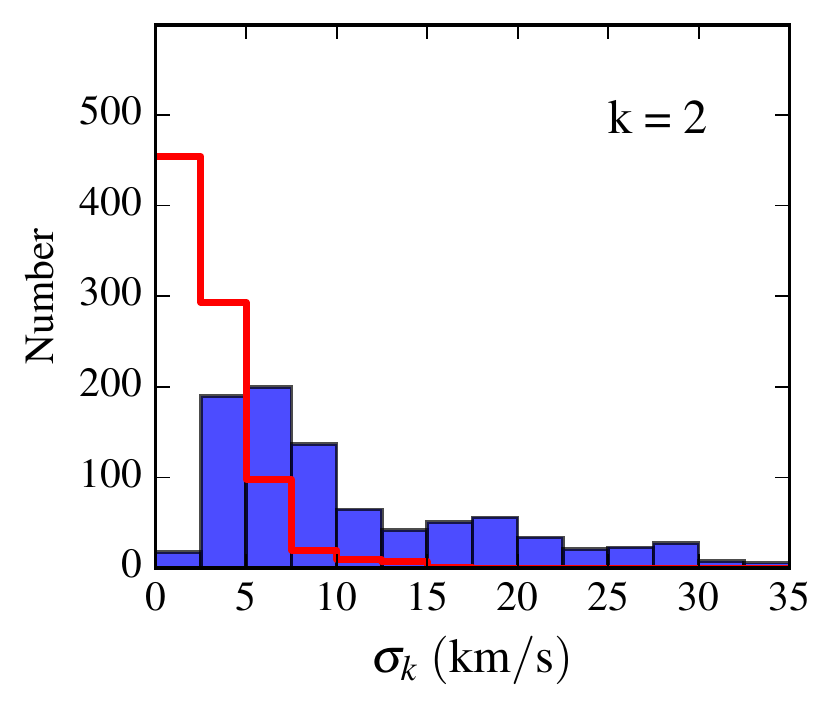}
    \includegraphics[height=4.1cm,trim={0.7cm 0cm 0cm 0cm},clip]{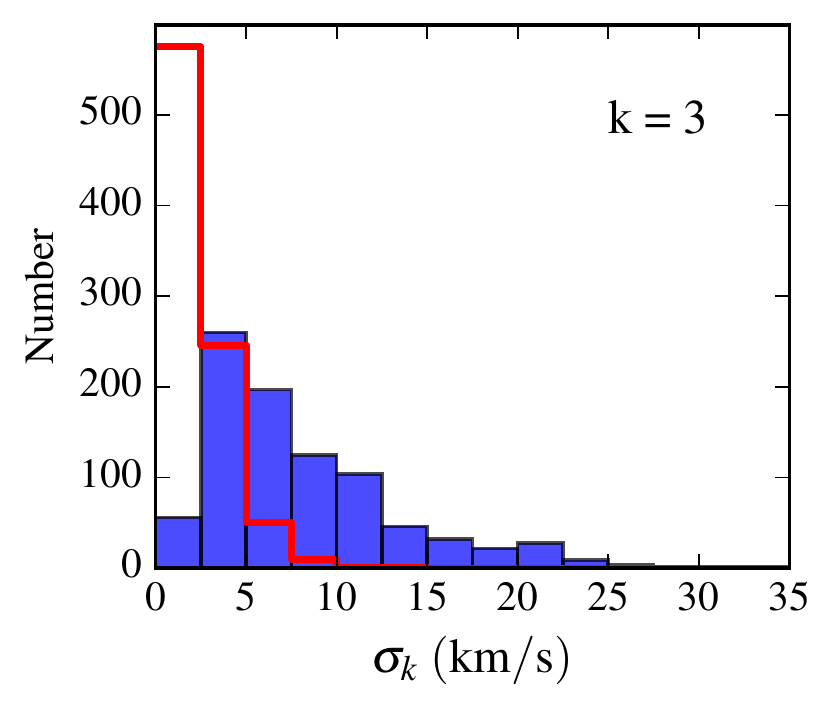}
    \includegraphics[height=4.1cm,trim={0.7cm 0cm 0cm 0cm},clip]{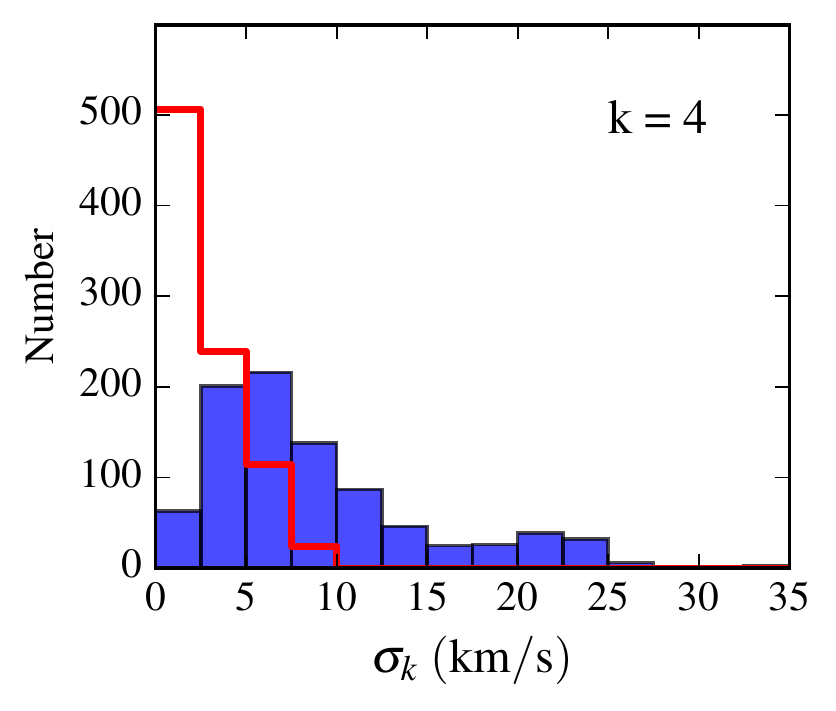}\\
    \includegraphics[height=4.1cm]{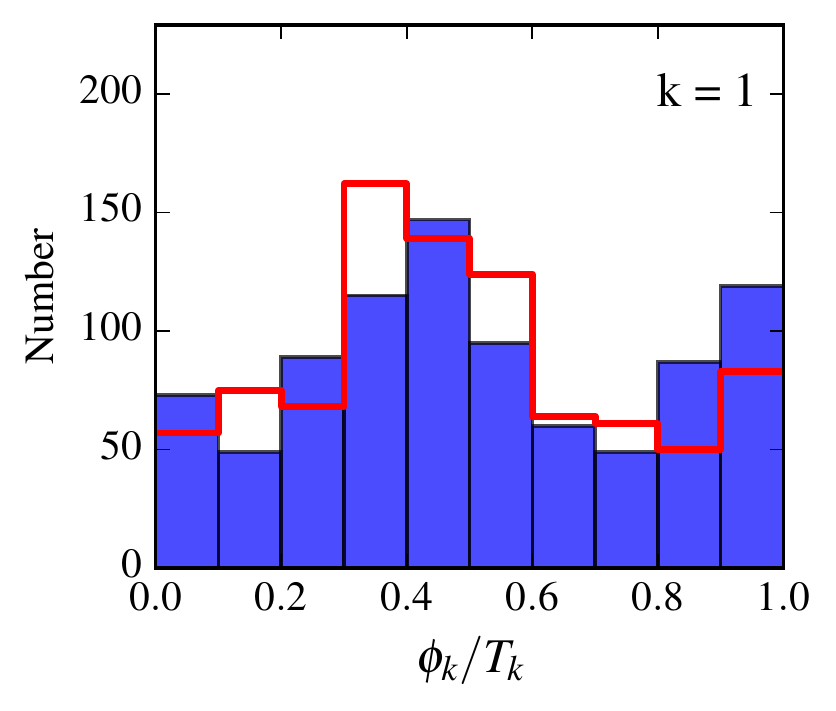}
    \includegraphics[height=4.1cm,trim={0.7cm 0cm 0cm 0cm},clip]{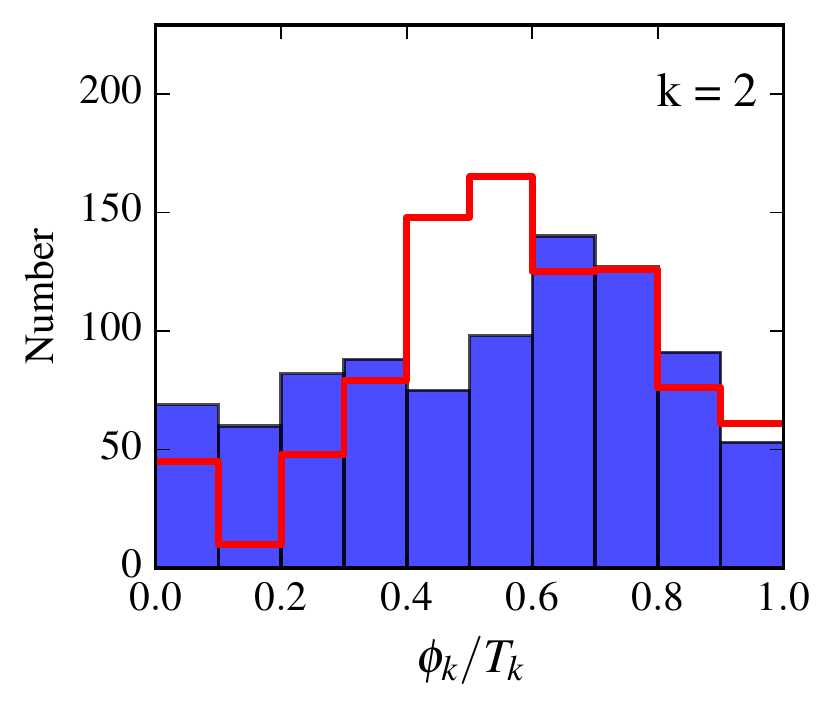}
    \includegraphics[height=4.1cm,trim={0.7cm 0cm 0cm 0cm},clip]{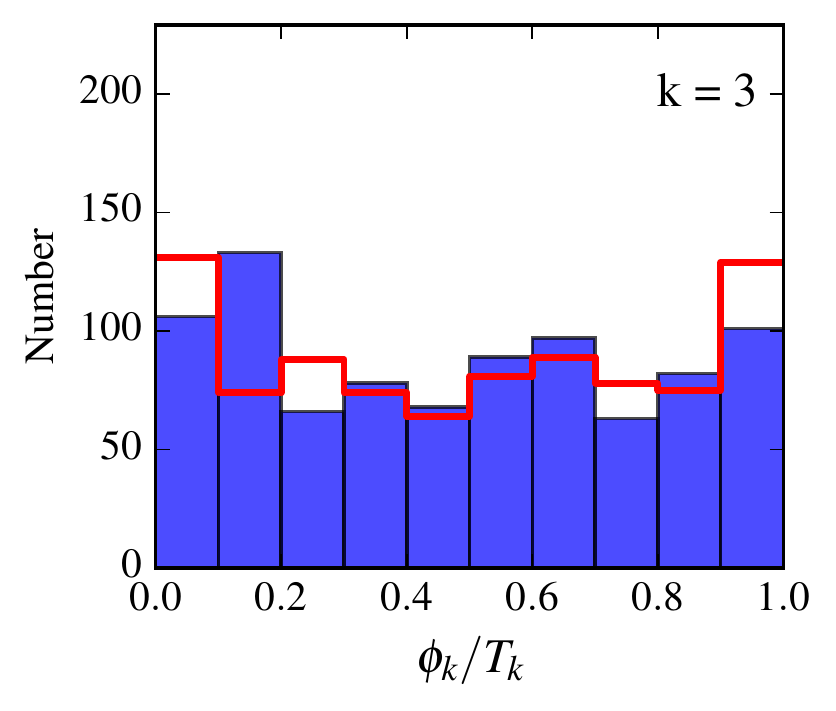}
    \includegraphics[height=4.1cm,trim={0.7cm 0cm 0cm 0cm},clip]{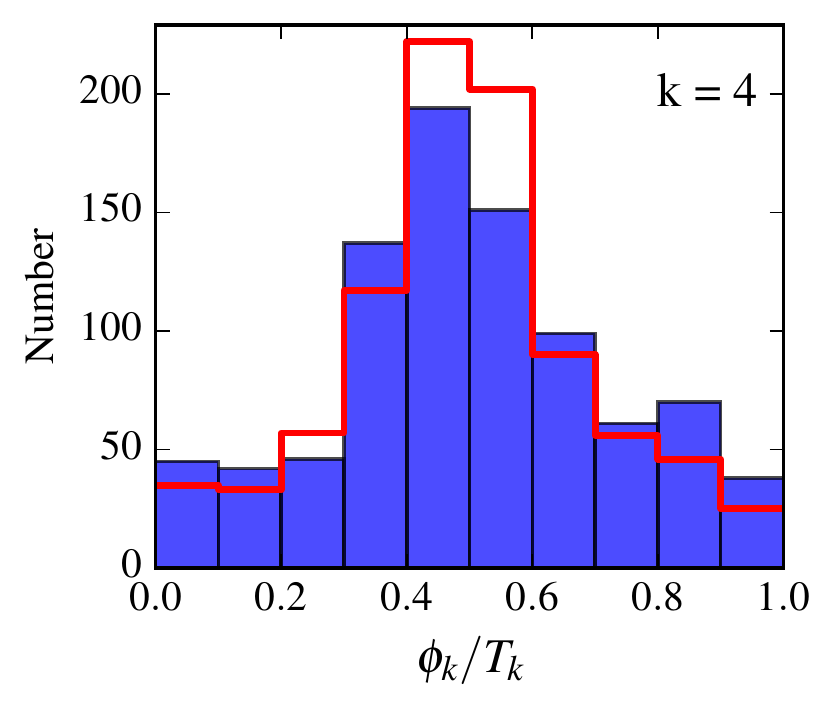}
    \caption{Histograms of amplitudes (upper row) and normalised phase angles (bottom row) resulting from the FFT analysis for the THINGS sample (in blue) and for the BS dispersion maps $\sigma_{\rm bs}$ (in red).}
    \label{fig:fftbs}
\end{figure*}

Figure~\ref{fig:fftbs} shows that asymmetries in  $\sigma_{\rm bs}$ are centrally peaked and  lower than the observed asymmetries. 
The phase angles of $\sigma_{\rm bs}$ are similar to those seen in the galaxies. Subtle differences are observed, like the peak of probability at $\phi_2/T_2 \sim 0.5$ not observed in galaxies, or the stronger and narrower peak at $\phi_4/T_4 = 0.5$.
Finally, we also compared the results of the FFTs obtained with and without BS correction, and found negligible differences. 
These findings indicate either that our data are sparsely affected by BS, or that the BS correction is underestimated. In Sect.~\ref{sec:bs-ha} and Sect.~\ref{sec:BS_residuals}, we further investigate the impact of using higher resolution data to infer BS correction and the amplitude of BS residuals respectively.

\subsubsection{Uncertainty on the beam shape}
\label{sec:dirty_beam}

The observed dirty beam of interferometric observations is complex.
In this section, we measure $\sigma_{\rm bs}$ using the observed dirty beam for 12 of the galaxies from THINGS. 
Maps of the dirty beam are mainly composed by a combination of two Gaussian functions, a narrow one to describe the PSF core, similar to the one of our model, and a second one with larger full width at half maximum to describe the PSF wings. 
The extent of dirty beam maps is often larger than those of the galaxies but the beam power at large angular distance from its centre is very low.
We apply a circular mask to the dirty beam map to deal mostly with positive PSF values, and get rid of small variations at large radius from the beam centre that could induce spurious effects in our analysis. 
We vary the radius of the masks from 5 to 50 pixels, in order to be consistent with our 2D Gaussian model which has an extent of 5 to 10 pixels within the THINGS sample. 
Using a dirty beam map truncated at the same size and scale as our 2D Gaussian model, the resulting $\sigma_{\rm bs}$ maps are very comparable to those obtained with the Gaussian beam. 
For example, within a cut of five pixels around the beam centroid, the  amplitude of $\sigma_{\rm bs}$ is $\sim 10$\% lower on average than for the Gaussian shape throughout the THINGS sample.
However, when the spatial extent of the dirty beam is larger, the patterns in  $\sigma_{\rm bs}$  are more prominent, leading to larger amplitudes. For instance, with a mask cut at a radius of 20 pixels, the amplitudes are twice larger than our model, on average. As for the phase angles, the minor difference between the two beam shapes is that the peaks are narrower and slightly larger for $k=2$ and $4$ in the case dirty beam.  
 
 \subsubsection{Kinematics at higher angular resolution}
\label{sec:bs-ha}

\cite{2006MNRAS.367..469D} and \cite{2008MNRAS.385..553D} published H$\alpha$ data cubes for respectively 28 and 37 galaxies from the SINGS sample \citep{Kennicutt2003}. Those observations were led with a Fabry-Perot interferometer at the  1.6m telescope of the Observatoire du Mont Megantic (OMM, Canada), and the 3.6m telescopes at the Canada-France-Hawaii Telescope and the European Southern Observatory (ESO, La Silla, Chile). In our sample, all galaxies were observed at OMM except NGC3621 and NGC7793, observed at ESO. 
In this section, we benefit from the higher angular resolution of the H$\alpha$  velocity fields to estimate, and correct from, the BS effects on \hi\ THINGS velocity fields. The angular resolution of the \ha\ data is seeing limited, $\sim 1\arcsec$ at ESO and $\lesssim 3\arcsec$ at OMM, which represents a gain of $\sim$ 2 to 13 with respect to THINGS data.  The underlying hypothesis is that the \ha\ kinematics have a similar behaviour as the \hi\ one. Nevertheless, the flux distribution of the ionised gas is more peaked due to the star forming regions. To model the \ha\ BS map, we used the \ha\ velocity field weighted by the flux distribution of the neutral gas. 
Since the kinematical maps available from \citet{2006MNRAS.367..469D} and \citet{2008MNRAS.385..553D} were obtained using adaptative smoothing, we compute new maps from the sky-removed data cubes. We first apply a 2-pixels Gaussian smoothing to increase the \ha\ SNR and then extract kinematical maps using the same barycentre method used in the two original papers. 
We also reproject the THINGS flux density maps on the same grid as \ha\ maps in order to match the positions and pixel size of both datasets. 
To compute the BS maps, we apply the same methodology as defined in Sect.~\ref{sec:beamsmearinggaussian}, using the reprojected \hi\ flux distributions, the 2-pixels Gaussian smoothed \ha\ velocity field and the single 2D Gaussian PSF to describe the new \hi\ angular resolution.

\begin{figure*}[th] 
\centering
    \includegraphics[height=5.8cm,trim={6.5cm 0.1cm 6cm 0.5cm},clip]{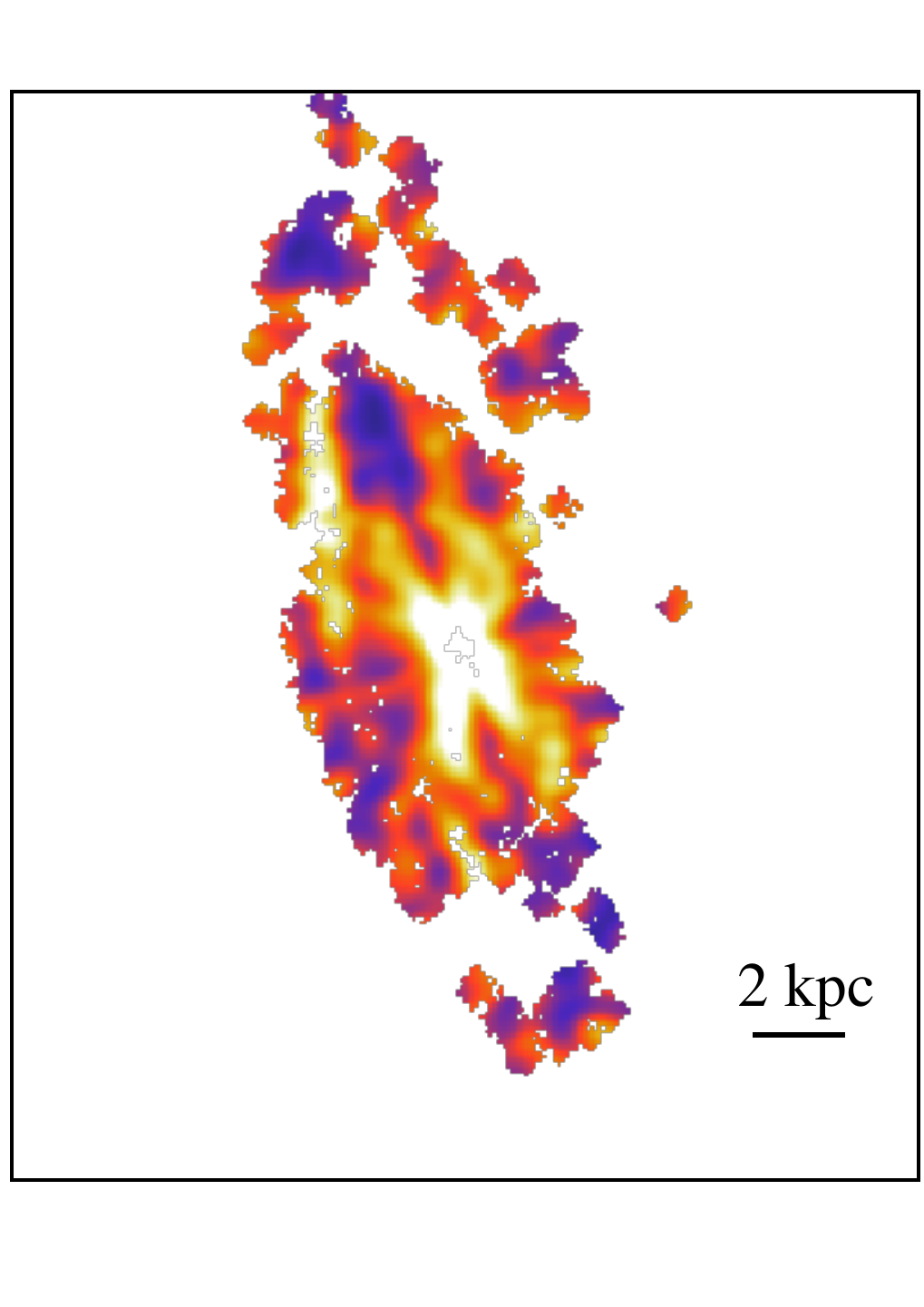}
    \includegraphics[height=5.8cm,trim={6.5cm 0.1cm 2cm 0.5cm},clip]{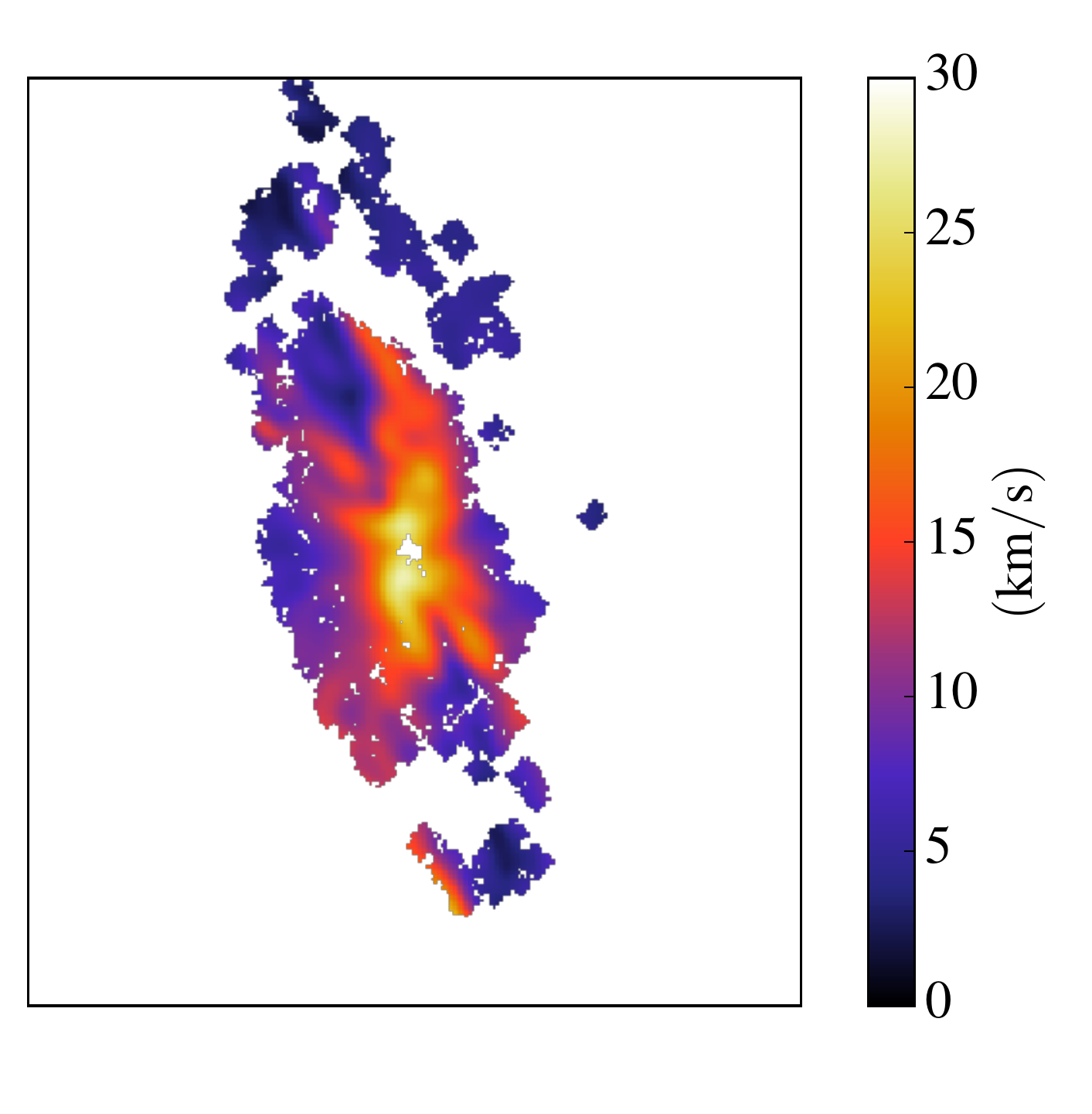}
    \includegraphics[height=5.8cm,trim={6.5cm 0.1cm 2cm 0.5cm},clip]{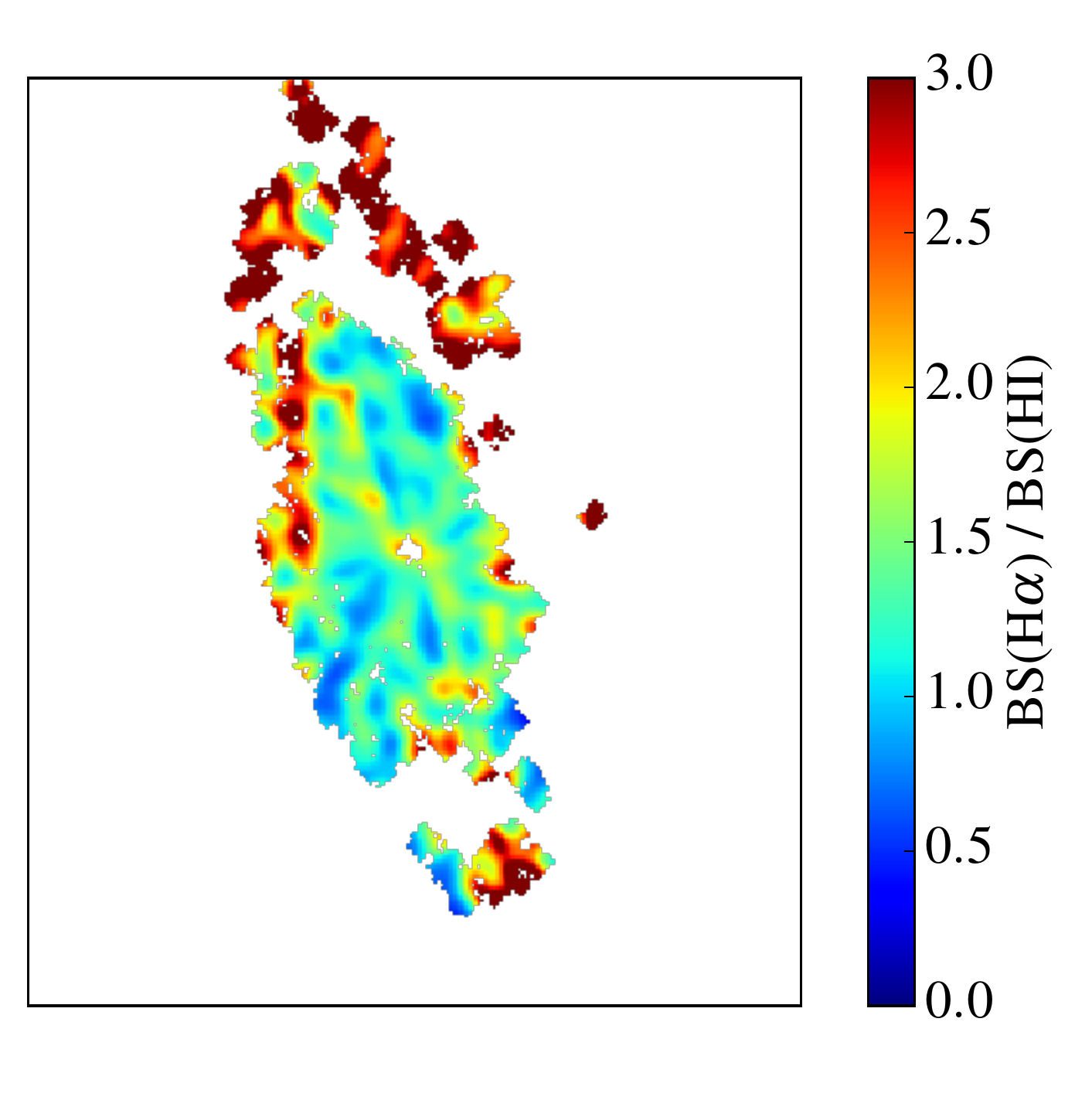} 
    \caption{Comparison of the BS models using optical and cm data for the galaxy NGC3521. \textit{Left panel:} H$\alpha$ data based BS model. \textit{Middle panel:} \hi\ data based BS model, trimmed at the same extent as the optical model. \textit{Right panel:} Ratio of the BS models, defined as BS(H$\alpha$) / BS(\hi).}
    \label{fig:bsha}
\end{figure*} 

\begin{figure}[th]
\centering
    \includegraphics[height=6cm,trim={3cm 0.1cm 2.5cm 0.5cm},clip]{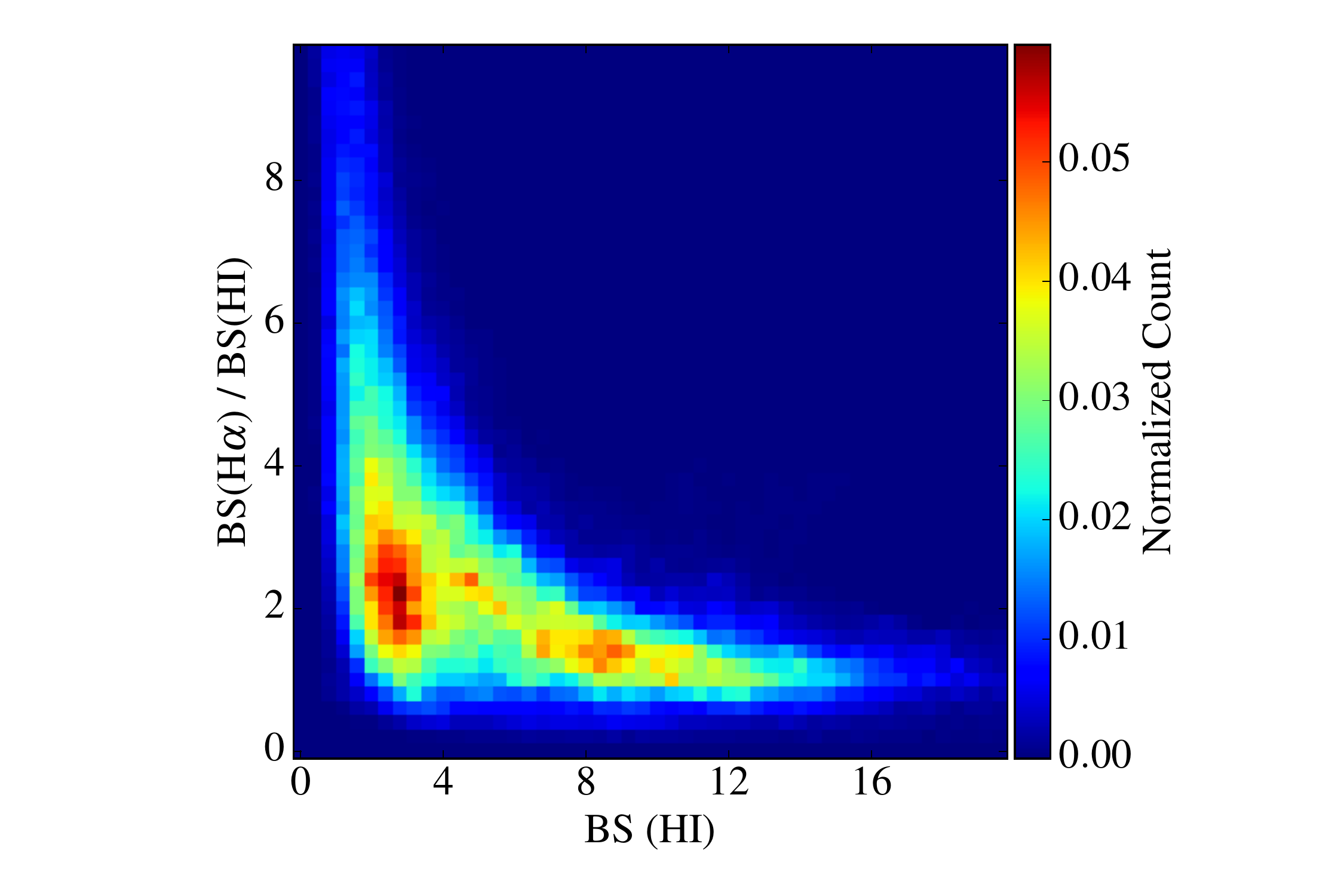}
    \caption{Pixel-by-pixel comparison of the BS contribution using H$\alpha$ SINGS and \hi\ THINGS data. The y-axis shows the ratio of the BS
    amplitudes H$\alpha/\hi$ and in the x-axis is the amplitude of the \hi\ model. }
    \label{fig:bs-ha-hi}
\end{figure}

In Fig.~\ref{fig:bsha}, we show the resulting \ha\ $\sigma_{\rm bs}$ map (left panel), compared to the  \hi\ $\sigma_{\rm bs}$ map (middle panel), as well as the ratio of the two maps, for an example galaxy (NGC3521). 
The BS pattern is similar in both cases, with the typical cross-shape in the centre.
While the BS correction inferred from \hi\ data is less peaked in the centre than for \ha, the latter also decreases more slowly with radius. This explains the large values in the normalised ratio found far from the centre. Beam smearing amplitudes are higher when modelling with the ionised gas. As the amplitude may be an important concern in our modelling, we study galaxy by galaxy the amplitude ratio for the two different cases.

In Fig.~\ref{fig:bs-ha-hi},  a comparison of the BS contributions is made for the THINGS sample. 
For each galaxy, a 2D histogram with the amplitude of the \hi\ BS correction on the x-axis and the ratio of the BS corrections in \ha\ to \hi\ on the y-axis, is made. Then, the histograms were normalised by the number of pixels in the maps, to avoid being biased by objects with a larger number of pixels. Fig.~\ref{fig:bs-ha-hi} is then obtained by summing all the 2D histograms.
It shows that the ratio is $>1$, which is not surprising given that the \ha\ BS  benefits from a higher angular resolution. On average, the \hi\ BS is $2.3$ times smaller than the \ha\ BS. 
We also observe a trend at low \hi\ BS  with high ratio induced either by noise or by discontinuities between regions in the \ha\ data. Indeed, a continuously increasing ratio with decreasing \hi\ BS  can be modelled by assuming that the \ha\ BS is equal to the quadratic sum of the \hi\ BS  and  a constant term of $\sim$ 3 to 10 \kms. This constant term can be related to pixel-to-pixel variations either induced by noise in the \ha\ data, or by true local velocity variations or discontinuities unseen in \hi. Amplitudes vary within the sample, due to various intrinsic velocity gradients in each galaxy, as well as its distance and geometry. 

Despite these findings, we decided to continue working with the \hi\ $\sigma_{\rm bs}$ for several reasons. First, the spatial extent of \hi\ gas is larger, making it possible to evaluate the local BS contribution of spiral arms or other features that are not located in the inner disc regions. A second reason is that the \ha\ gas does not trace the same interstellar component, therefore the geometry of the discs might differ, and the flux distribution are noticeably different.
Furthermore, it has been shown  that the imprint of the systematic phase along the minor axis of the galaxy M33 did not depend on the angular resolution \citep{2020chemin}, while a range of 7 in the size of the synthesised beam was probed.  
Using \hi\ maps to derive $\sigma_{\rm bs}$ should thus not be an issue in the analysis (see also Sect.~\ref{sec:BS_residuals}).

To assess the impact of BS under more conservative conditions, we then subtracted respectively twice and thrice the \hi\ BS from the THINGS velocity dispersions. 
The result of this process is that a large number of rings are removed in this case.  Indeed, in order to avoid sampling issues in the FFT, we removed rings having more than 5\% of pixels with undefined value after the BS correction, due to this quadratic correction being larger than the actual measured dispersion. The median fraction of such pixels per ring is 0.7\%, 4.0\% and 16.6\%\ when subtracting once, twice and thrice the BS contribution, respectively. Only 621 and 258 tilted rings remain in the two latter cases, respectively, with less than 5\% of pixels with an undefined value, decreasing drastically the statistics. With these particularly conservative BS subtractions, some galaxies totally disappear from the remaining rings, which indicates that the velocity dispersion would be totally dominated by BS, which should not be the case owing to the (high) \hi\ resolution. 

\begin{figure}[t]
    \centering
    \includegraphics[height=4.31cm,trim={0.3cm 0cm 0cm 0cm},clip]{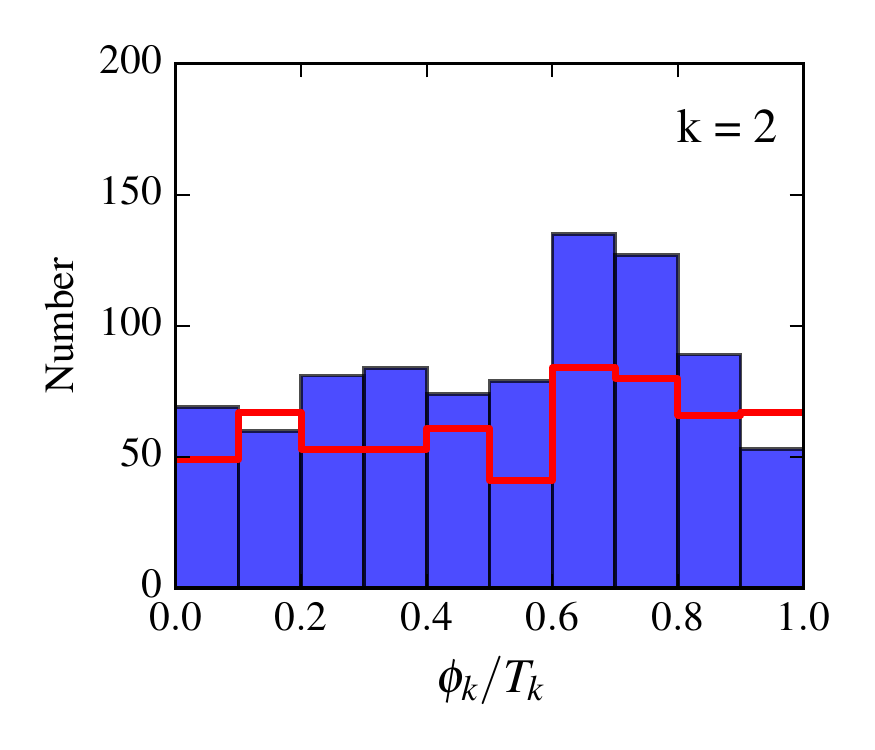}\includegraphics[width=4.44cm,trim={1.cm 0cm 0cm 0cm},clip]{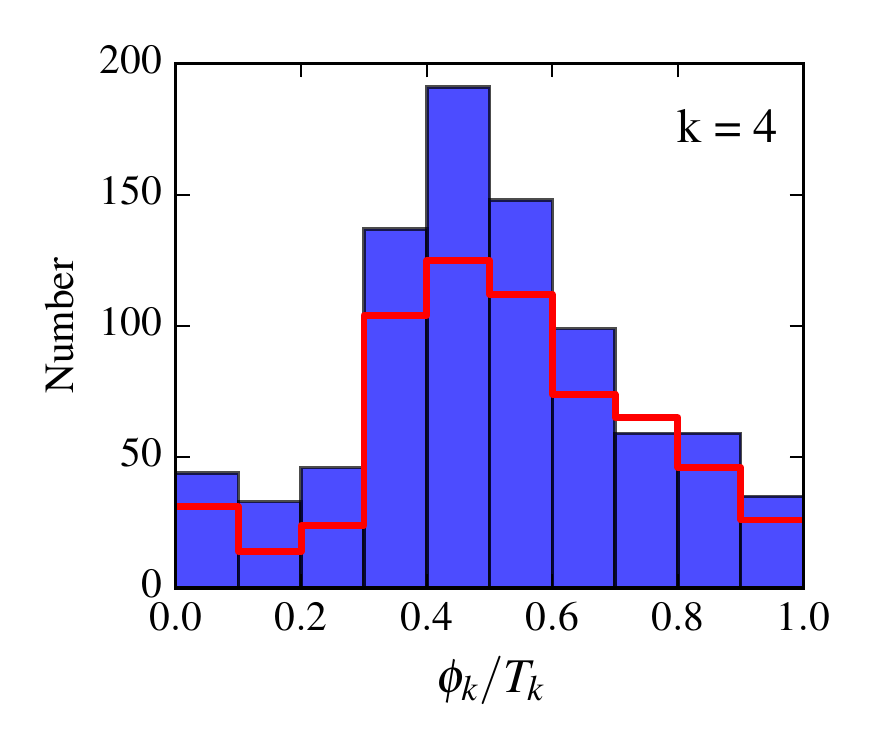}
    \caption{Comparison of the normalised phase angles histograms of orders $k = 2$  and $k = 4$ (left and right panels, respectively) in the velocity dispersion by subtracting once and twice the contribution from the BS effect (blue and red colours, respectively).}
    \label{fig:distrib-bs}
\end{figure}

In Fig.~\ref{fig:distrib-bs}, we show the phase angles for $k = 2$ and $k = 4$ resulting from the subtraction of once or twice the contribution $\sigma_{\rm bs}$ (red and blue histograms, respectively). We note that the amplitude distributions are not strongly affected. 
As mentioned before,  more than 200 rings were removed, which explains why the average number is lower for the red histograms. For  $k = 2$, the   peak seen in the blue distribution is no more present in the red one. This behaviour is also observed for $k = 1$ (not shown here). For $k = 4$, in both cases we observe the strong peak at  $\phi_4 / T_4 \sim 0.5$. The interpretation is not straightforward, as the observed phase angle can be explained either by a residual BS signature or by the correlation of the asymmetry with the BS. Nevertheless, the fact that a strong peak  remains in $\phi_4$  tends to indicate that BS alone cannot explain the observed trends.

We also smoothed the H$\alpha$ data at the average resolution of the \hi,  11\arcsec\ for all the sample with the goal to verify that the same level of smoothing of the gradients implies a similar BS effect. The results are as expected, being that the normalised residuals for each galaxy show a distribution peaked on 0, indicating that both modelling are consistent with each other.

Similarly, to verify the consistency of the findings described above, we also performed the Fourier analysis of an example galaxy (NGC2841) using the $6\arcsec$ angular resolution THINGS velocity and density maps, as obtained from robust weightings. The amplitude of the BS is larger than for natural weightings, which is  due to the combined effects from the higher angular resolution with the noisier velocity field for robust weightings than for natural weightings. This finding matches perfectly the result found with the \ha\ kinematics.  Nevertheless, no significant difference in the strength and phase angles of the Fourier modes is observed with respect to the natural weighting for this galaxy. 

\subsubsection{Amplitude of beam smearing residuals}
\label{sec:BS_residuals}

The correlation between BS and observed signatures found in Sect. \ref{sec:bs_properties} may indicate that BS residuals remain after BS correction, due to the use of observed flux maps and velocity fields rather than high resolution data to infer the correction (see Sect. \ref{sec:beamsmearinggaussian}). 
To quantify  residuals, we use the toy models described in Sect. \ref{sec:toy_models} and App. \ref{app:toymodels}. We focus on the isotropic case with no perturbation and a steep inner gradient of the rotation curve, for which the impact of BS is the largest, and added realistic noise to dispersion maps (cf. App. \ref{app:biaises}).
Correcting for BS as described in Sect. \ref{sec:beamsmearinggaussian} reduces the strengths of orders $k=2$ and $k=4$ by a factor between 2 to 4 and is more efficient for less inclined galaxies and for order $k=2$.
We also show that the maximum strength of orders $k=2$ and $k=4$ of BS residuals is reached at a lower radius than without BS correction, and that it decreases for decreasing inclinations.
For inclinations below 60\degr, residual BS cannot account for the $k=2$ signature at radii greater than a few beam FWHM and at radii $> 10$ beam FWHM for $k=4$. 
At 75\degr, however, the BS signature remains well above the uncertainties because (i) the projected velocity gets higher than at lower inclination, and (ii) the spatial resolution along the minor axis gets poor, inducing more hidden velocity gradients within the beam and pixel size.
We also performed an FFT analysis separately for the 50\% largest and smallest radii and found, as expected, that the amplitude of orders $k=2$ and $k=4$ is reduced (by a factor larger than two) and that the significance of the peak in phase histograms of the fourth order is much lower (by a factor around 4) at large radii. For the model with a weaker velocity gradient, when noise is added, all the systematic phase angles disappear and the distributions become flat.
The impact of BS induced by large-scale rotation depends on the physical resolution, the rotation curve shape, the projected maximum velocity, galaxy inclination and high resolution flux distribution. A complete study of BS on velocity dispersion maps depending on all these parameters is beyond the scope of the present study.

\begin{figure}[t]
\centering
    \includegraphics[width=0.48\columnwidth]{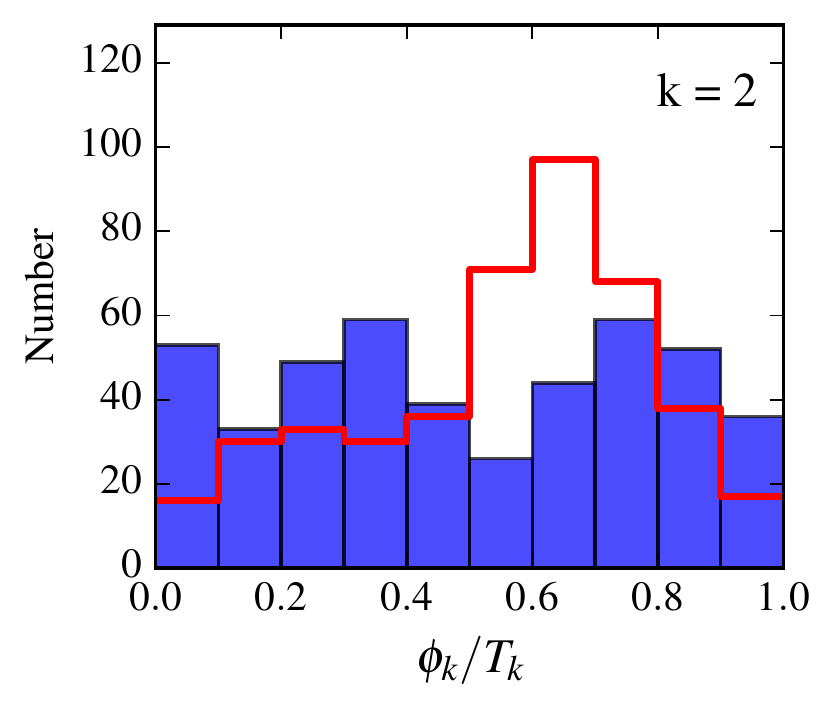}
    \includegraphics[width=0.48\columnwidth]{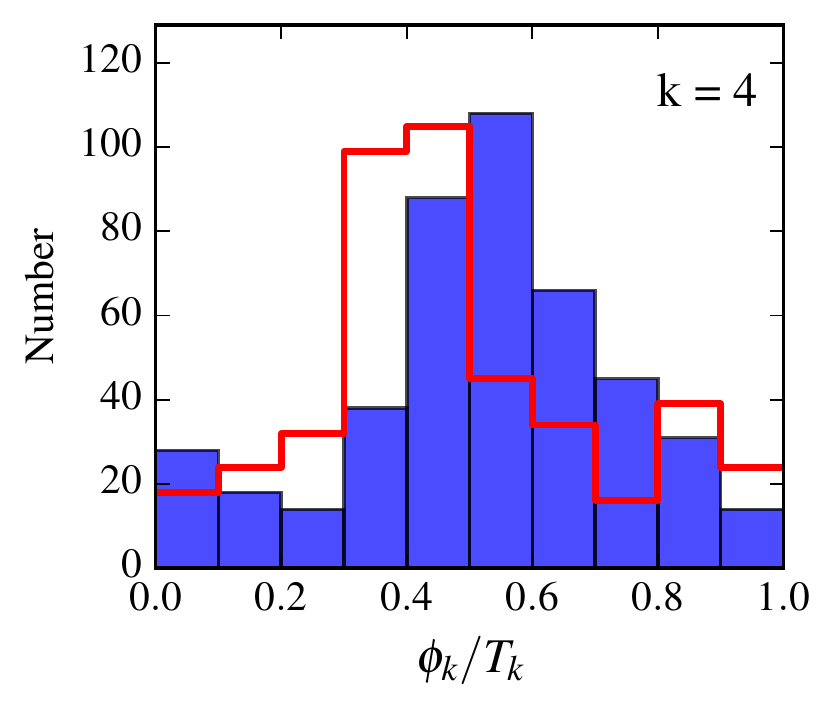}
    \caption{Histograms of normalised phase angles resulting from the FFT analysis for the THINGS sample splitting the rings as a function of radius (inner part in blue, outer part in red).}
    \label{fig:fftvsrad}
\end{figure}

About half galaxies in our sample have projected maximum rotation velocities above 150~\kms, as inferred from the profile width at 20\% of the peak intensity given in \cite{2008AJ....136.2563W}, and about 60\% have an inclination larger than 60\degr, which may therefore present BS residuals signatures in the fourth order.
We first split the sample between inner and outer rings. The FFT amplitudes of the innermost rings are 27\% to 43\% larger than for the outermost rings, depending on the order. The phase angle distributions are shown in Fig.~\ref{fig:fftvsrad}, with the innermost and outermost rings shown in blue and red, respectively. These distributions are different. 
For $k = 2$, the distribution for the outermost rings is similar to that of the  global distribution whereas the distribution does not peak on 0.5 $\phi_2 / T_2$ for the inner rings, as would be expect if the dispersion maps were dominated by BS residuals. 
The inner and outer regions both show a prominent peak $k = 4$ close to the half period, the peak being slightly offset towards lower phase angles in the outer parts.
We also split the sample in low versus high projected maximum velocities, leading to the same subsamples as for the absolute magnitude (see Sect.~\ref{sec:phasevsproperties} and  Fig.~\ref{fig:fftvsmabs}). As already discussed, the peak observed in the phase angle histograms of the fourth order is much more pronounced when projected velocities are high. This may be a proof that BS residuals are affecting our results at least for the fourth order. Nevertheless, given the large values of order four strengths $\sigma_4$ compared to what is expected from residuals in toy models, it may also be that such a signature is intrinsic to massive galaxies with high gas density contrasts. On the other hand, the peak in the second order around half period of galaxies with low projected velocities cannot be attributed to beam smearing.

\subsubsection{A possible link between velocity gradients and elevated velocity dispersion}
 
We have shown that BS may explain the signatures we observe if not properly taken into account, as there are strong correlations between the modelled BS pattern and the observed velocity dispersion. Nevertheless, we have also shown that the amplitude of BS induced by large-scale motions is expected to be too low to have a significant impact on the data.
The fact that BS seems to be coupled to observed signatures in the phase angles of asymmetries may indicate that the regions with the strongest velocity gradients or discontinuities, especially in the outer parts of galaxies, are also regions with an intrinsic large velocity dispersion at a higher level than what would be induced by BS from the observed gradients. These regions with large velocity gradients are often related to perturbations like bars, spiral arms, inter-arms and warps, especially at large radius, which means that there might be locally induced turbulence responsible for perturbations in both the velocity field and the velocity dispersion. 
This may also be that BS occurs on much smaller scale than that reached within our observations and that what we observe is not related to large-scale motions. 
A complementary way to investigate the effects of unresolved gradients of asymmetric velocities on the velocity dispersion is addressed in Sect.~\ref{sec:orderedvelocityeffect}.  
 
\subsection{Projection effects in random motions} 

Finding systematic phase angles of asymmetries in  velocity dispersion maps is a surprising result. 
 The projection of asymmetric patterns  along the line-of-sight should be randomly distributed. Flat  distributions of phase angles were then expected in Fig.~\ref{fig:distrib} from a theoretical perspective (see discussion in App.~\ref{app:anisotropy}). Privileged phases thus imply a dependency of the patterns on the  orientation of the disc principal axes with respect to the observer, because the major axis of discs has been chosen at the origin of the azimuthal angles ($\theta = 0$ along the major axis). In other words, this  suggests the presence of projection effects in the velocity dispersion maps.  In this section, we thus want to study the possible origin(s) of projection effects in the velocity dispersion. This is achieved by investigating the effect  of  asymmetries in the ordered motions on the random motions (Sect.~\ref{sec:orderedvelocityeffect}). We also address the effect of correlated velocity components (Sect.~\ref{sec:tiltedvelo}). 
 
\subsubsection{Effects from the asymmetric ordered motions}
\label{sec:orderedvelocityeffect}
 
It is usual to assume that the  velocity ellipsoid of gas is isotropic in galactic discs. Under this assumption, no deprojection of data is required, so that the observed velocity dispersion is a direct proxy of the gaseous random motions. However, this assumption is very idealised, and can only apply to axisymmetric kinematics, which rarely occurs in reality.  
Indeed, the velocity fields  (\vlos) always exhibit disturbances, as caused by the projection of asymmetric ordered motions ($V_R$, $V_\theta$, $V_z$) due to, e.g., bars, rings, spiral arms, or warps. 
The effect of any of such disturbances in \vlos\ should also propagate to \slos\ because of the implied gradients in one or several directions in the plane.
We test this hypothesis by studying the toy models with asymmetric kinematics perturbations described in Sect. \ref{sec:toy_models} and App. \ref{app:toymodels} in both isotropic and anisotropic cases.
We mainly focus hereafter on cases with the weakest velocity gradient (shown in Fig. \ref{fig:toy_models}) to reduce possible residual BS signatures related to unresolved rotation (see Sect. \ref{sec:BS_residuals}) and emphasize on local perturbations.

In the axisymmetric and isotropic model (used as reference), the strength of the Fourier coefficients of the various orders are comparable ($\sim 0.5$ \kms\ on average). In all other models, $\sigma_1$ and $\sigma_3$  barely vary from this value, while the $k=2$ and $k=4$ components are significantly larger, being mostly $> 2$ \kms, except in the axisymmetric case with uniform local anisotropy, where $\sigma_4$ is very comparable to the odd amplitudes. 
The distributions of the phase angles for the negligible modes $k = 1$ and $k = 3$ do not show systematic peaks.
The even orders thus always dominate the dispersion asymmetries, and the bisymmetry is the strongest perturbation, on average.

Any model with a velocity ellipsoid showing a radial bias $\sigma_\theta < \sigma_R$ creates a bisymmetry theoretically aligned with the minor axis.
The strength of $k=2$ asymmetries in \slos\ within mock data obtained from models combining effects from the uniform radial bias and $V_R$ and/or $V_\theta$ streamings is between 20 to 40$\%$ larger than those without the radial bias, on average, while the amplitude of the mode $k = 4$ is not affected. The distribution of $\phi_2$ is sharply peaked at $\pi/2$ for the axisymmetric case of anisotropy with the radial bias, and the peak is slightly enlarged when there are asymmetries in $V_R$ and/or $V_\theta$, regardless of their strength.  Moreover, in this latter case with the velocity perturbation exclusively set along the azimuthal direction ($\Delta V_R = 0$, $\Delta V_\theta \neq 0$), the incidence of $\phi_2=0$ or $1$  remains extremely low.  These findings can be explained with the help of Eq.~3 from \citet[][]{2020chemin}, where having $k=2$ asymmetries in \slos\ aligned with the minor axis is a genuine imprint of a radially biased velocity ellipsoid (see also Eq.~\ref{eq:slos_disc2} in App.~\ref{app:anisotropy}, but with null cross terms).   
Thus,  a local and uniform anisotropy of the velocity ellipsoid with a radial bias drives the phases of asymmetries in \slos\ more strongly than the effect of the  velocity anistotropy induced at larger scale by the projection of $V_R$ and $V_\theta$ gradients. This is because the radial bias is applied to all velocities in the mock data, while the $V_R$ and $V_\theta$ streamings are only restricted to the locations of the perturbations. 
If a tangential bias $\sigma_\theta > \sigma_R$ had been assumed, the bisymmetry would have been exclusively aligned along the disc major axis, to the detriment of the minor axis. Because THINGS and WHISP galaxies do not show a $\phi_2$ distribution  strongly  concentrated around $\pi/2$, anisotropic gaseous velocity ellipsoids consistent with a radial bias uniformly distributed everywhere in discs, like the one presented here, seem very unlikely. 
A solution to  avoid the $\phi_2/T_2$ concentration at $0.5$ is to have either an axis ratio $\sigma_\theta/\sigma_R$ which varies with radius, or tilted velocity ellipsoids. Section~\ref{sec:tiltedvelo} provides a hint of how this goal could be achieved.

The impact of velocity anisotropy arising from asymmetries in $V_R$ (or $V_\theta$, respectively) is to greatly decrease the likelihoods of finding $\phi_2$ close to zero ($\pi/2$), in the case the velocity perturbation is along the radial axis, i.e. $\Delta V_\theta = 0$ and $\Delta V_R \neq 0$, (azimuthal axis, $\Delta V_R=0$, $\Delta V_\theta \neq 0$). This is because there is little projection of $V_\theta$ ($V_R$, respectively) onto the minor (major) axis. These results remain valid regardless of the uniform isotropy/radial bias assumptions.  If there were an equal  number of tilted rings dominated by radial and azimuthal velocity streamings, this would lead to  distributions of $\phi_2$ with minima around 0 and $\pi/2$, and conversely with maxima $\phi_2=0.25\pi$ and $0.75\pi$, which values are not far from those observed in the sample within the quoted uncertainties. 
 
Then, still for models with asymmetries in $V_R$ and/or $V_\theta$, the distributions of $\sigma_4$ and $\phi_4$  are very similar in both cases of uniform isotropy or radial bias. 
Thus, the uniform velocity anisotropy has negligible effect on the fourth order perturbation of \slos, and this latter is an harmonics of the second order perturbation in our toy models. The information about the orientation of the bisymmetry (i.e. along major or minor axis) is nevertheless lost modulo $\pi/2$ in the fourth order\footnote{A 4th-order mode could be generated as an harmonic of a second order mode with a phase of either $\phi_2=\phi_4$ or $\phi_2=\phi_4 + \pi/2$.}.
Moreover, it is obvious that asymmetries have destroyed the systematic peak at $\phi_4 = 0.5T_4$ seen in the reference axisymmetric models. This is because this peak could be caused by small residuals from BS correction, as mentioned previously. Unless our velocity asymmetries in the toy models are significantly overestimated with respect to the reality, the occurrence of a systematic peak at $\phi_4 = 0.5T_4$ in the observations indicates that the fourth order asymmetry is very likely an harmonic of a second order which would have preferential phase angles $\phi_2$ close to either $0.25\pi$ and $0.75\pi$, which values remains close to the observed $\phi_2 \sim 0.35\pi$ and $0.65\pi$ within the quoted uncertainty of $0.1 T_2$.

We note that combining a steep velocity gradient top the perturbations (see Fig. \ref{fig:toy_models2}) does not change qualitatively the previous results for the order $k=2$ though the contrast between high and low probability is a bit reduced, especially in the isotropic case with the velocity perturbation along $V_R$, due to residual BS inducing a larger probability for null phase angles whatever the direction of the perturbation. Anisotropy always dominates order $k=2$ when present. For order $k=4$, a peak related to residual BS clearly remains around $\phi_4/T_4 = 0.5$, which increases the trend already observed with weak rotation-induced BS.

Finally, although not shown here, we observe  that when the BS correction is made with flux density and velocity maps at the highest angular resolution of the mock data (native resolution of $\sim 50$ pc, before spatial smoothing), no significant residuals due to beam smearing remain in the distribution of phase angles. Note here that this value of 50 pc is only applicable to our toy models, and not to real observations, because we do not know precisely the angular scale of velocity streamings in galaxies.  
This indicates that the velocity anisotropy from the streamings  can only affect the phase angles  when the streamings are not properly resolved.  Therefore, our method consisting in searching for systematic effects in the phase angles of \slos\ from observations of `limited' resolutions like THINGS is a powerful tool to predict the directions in the plane towards which the perturbations and anisotropy are more significant. In particular, it implies that the lopsidedness perturbation of \hi\ kinematics in THINGS is more likely to be stronger along the tangential dimension than along the radial direction, hence a preferentially large-scale tangential bias in the velocity anisotropy, in order to explain the higher incidence of $\phi_1$ along the major axis of the galaxies. Similarly, it is less probable to find a bisymmetry perturbation   stronger in the tangential direction, to explain the lower incidence of  $\phi_2$ towards the major axis of the galaxies. The observations of $\phi_2$ aligned near the disc minor axis in lower mass galaxies (Sect.~\ref{sec:phasevsproperties}), similarly to the outer part of M33, would thus likely be explained by dominant radial streamings, hence a stronger large-scale radial bias in the velocity anisotropy  \citep{2020chemin}.
Another consequence of the analysis is that velocity asymmetries  in the first moment maps of galaxies must be observed on a scale of about the PSF width, though of reduced amplitude. We thus expect that systematic effects  are present in the phase angles of $V_{\rm los}$ asymmetries, by analogy with those evidenced in \slos, tracing the direction of dominant bias of the velocity anisotropy from the perturbations. Comparing the properties of asymmetries of $V_{\rm los}$ and  \slos\ is beyond the scope of this article, however.
 
\subsubsection{Effects from anisotropic velocity ellipsoids with tilted axes}
\label{sec:tiltedvelo}

Until now, the tests on the shape of  the  velocity ellipsoid  has only considered components of the velocity tensor which are independent from each other. However, the asymmetries in $V_R$, $V_\theta$ and  $V_z$ all result from the same perturbed gravitational potential. By consequence, the components must show a certain degree of correlation, at least at the vicinity of perturbations. Correlated velocities are evidenced in the kinematics of stellar populations of the Milky Way, through tilted axes of the stellar velocity ellipsoid. The observed tilt angle of the axes of the radial and vertical velocity distribution of Galactic halo stars can inform on the 3D shape of the mass distribution and the dark matter halo \citep[]{2008siebert, 2009smith,2019wegg}, while that of the radial and tangential velocity distribution (the vertex deviation) is caused by a bisymmetric perturbation in the disc, such as a bar, a spiral structure, a warp, or more simply an elliptical disc \citep{1991kui,2019debattista}. 

We can test the effect of correlations between \hi\ velocities by means of the uniform velocity anisotropy presented above, by considering the possibility of tilted ellipsoids, as we do for stars. This has a consequence on the projection of the ellipsoid on the line-of-sight, and following the complete mathematical description from App.~\ref{app:anisotropy}, the squared velocity dispersion is given by:
\begin{equation}
\begin{split} 
\sigma_{\mathrm{los}}^2 = \left(\frac{\sigma_{\theta}^2-\sigma_{R}^2}{2}\cos{2\theta} +\sigma^2_{R\theta}\sin{2\theta}\right)\sin^2{i}\\
+\left(\sigma^2_{R z}\cos{\theta}+\sigma^2_{\theta z}\sin{\theta}\right)\sin{2i}\\
+\sigma_{z}^2\cos^2{i} + \frac{\sigma_{\theta}^2+\sigma_{R}^2}{2}\sin^2{i} \text{~,}
\end{split} 
\label{eq:slossqcros}
\end{equation}
where $\sigma_{R\theta}^2$, $\sigma_{\theta z}^2$ and $\sigma_{R z}^2$ are the covariance terms marking the correlation between the components of the velocity tensor in the cylindrical frame of a disc. This equation can then be expressed as a sum of three terms of orders 0, 1 and 2, with an analogue formalism as the one we used to study asymmetries:
\begin{equation}
\label{eq:sigloscorrel}
\sigma_{\mathrm{los}}^2 =  k_0 + k_1 \cos{(\theta - \alpha_1)} + k_2 \cos{(2(\theta - \alpha_2))} \text{~,}
\end{equation}
where $k_i$ are the amplitudes of these terms and $\alpha_i$ are their phases, that all depend on the various velocity dispersion terms. More specifically, we have
\begin{equation}
\label{eq:vertexdev}
 \tan{2\alpha_2} = 2 \sigma^2_{R\theta}/(\sigma_{\theta}^2-\sigma_{R}^2) \, .
\end{equation} 

Assuming no tilted axes for the uniform velocity anisotropy in the previous section left no choice than having a dispersion bisymmetry   systematically aligned with the minor or major axis ($\alpha_2 = \pi/2$ or $0$, respectively), depending on the sign of  $\sigma_\theta^2 - \sigma_R^2$. This is at odds with the   observation that   the second order phase angles $\phi_2= 0.35 T_2$ and $\phi_2=0.65 T_2$ are particularly more likely. 
Equations~\ref{eq:sigloscorrel} and~\ref{eq:vertexdev} thus imply that it is  possible to get various $\phi_2$ orientations if the velocity anisotropy is not as uniform through the disc, and more importantly if the covariance varies.
In the Milky Way disc, the measurement of the axis ratio of the tangential-to-radial velocity dispersions of young stars as measured with Gaia spectro-astrometric kinematics, is $\sigma_\theta/\sigma_R = 0.66$ \citep{2022gaiadrimmel}. Since young stars kinematics are expected to be comparable to that of the gas, we further assume for this exercise that the velocity anisotropy of the planar components of the interstellar gas for any galaxies in our sample is comparable to the value found for young stars in the MW disc. 
Reminding now that $\sigma_{R\theta}^2 = \rho_{R\theta} \sigma_R\sigma_\theta$, with $\rho_{R\theta}$ the correlation coefficient, an appropriate choice of covariance and correlation coefficient which explain the higher incidence of $\phi_2 \equiv \alpha_2 = 0.35\pi$ and $0.65\pi$ for gas is $\rho_{R\theta} = \pm 0.6$. Furthermore,  Eq.~\ref{eq:sigloscorrel} implies that  correlations between the vertical and the two planar components should also be considered to get systematic phases of first order phases. In order to explain those seen in Fig.~\ref{fig:distrib}, we would thus need to have $\alpha_1 \sim 0$ or $\pi$ to match the observations, that is,  
$\tan \alpha_1 = \sigma^2_{\theta z}/\sigma^2_{Rz} \sim 0$, following Eq.~\ref{eq:ki} in App.~\ref{app:anisotropy}. This is possible if $\rho_{\theta z}\sigma_\theta << \rho_{Rz}\sigma_R$. Here again, studies of the kinematics of young stars in the Galaxy should help to test these predictions, which is relevant only if the shape of gaseous velocity ellipsoids are comparable to those of young stars in nearby discs. It would then be interesting to investigate how these predictions on $\rho_{R\theta}$  compare with the observed planar correlations of young stars of  the Galactic disc, and how  $\rho_{\theta z}\sigma_\theta$ compare with $\rho_{Rz}\sigma_R$ for the same stellar populations. To our knowledge, none of such  distributions have ever been measured yet for the Milky Way. 

\section{Summary and concluding remarks}
\label{sec:conclusion}
 
This article has performed the first systematic search and characterisation of asymmetries in velocity dispersion maps of \hi\ gas in nearby galaxies. 
We used the best \hi\ data available in the archives, namely 32 galaxies from THINGS and WHISP samples with a spatial resolution better than 1.5 kpc and a neutral gas distribution extended by at least seven times the resolution, to carry out this pioneering work.  
We performed a Fourier analysis of the observed second moment maps previously corrected from both the line spread function and the two-dimensional beam smearing contribution tracing the  velocity gradients occurring at 
large angular scale. This allowed us to measure the strength and phase angle for each of the first four Fourier harmonics, which are the most important perturbations in the random motions of \hi\ gas.

We find a wide range of strength of asymmetries, up to $\sim 35$ \kms. The shape of the likelihood distributions of the  amplitude of the Fourier modes are similar, with a peak around $5-10$\kms, and an extended tail towards larger values. 
Overall, the strongest Fourier mode is the bisymmetry ($k=2$), with a mean amplitude of $\sim 11$ \kms, followed by the lopsidedness $k=1$ asymmetry.
On the other hand, the shape of the likelihood distributions of phase angles strongly depends on the Fourier mode, and the  distributions present significant variations.
For the $k=1$ asymmetry, the probability to find a phase angle near the major axis of the discs is larger, while finding an asymmetry orientated near 0.15 and $0.75\times 2\pi$ is less probable. For $k=2$, the likelihood is larger at phase angles of $\sim 0.35$ and $0.65 \times \pi$, and lower near the disc major axis, whereas that of the $k = 4$ mode  is   concentrated at a value of $0.5 \times \pi/2$, implying an orientation mainly  lying at an equivalent separation from the disc major and minor axes. 
These systematic phase angles are robust against the number of titled rings and against the significance of the detected signal. Using a uniform number of rings for all galaxies lowers the significance of the result, but the trends on the shape of the  probability distributions are preserved.
More uniform likelihood distributions were expected for randomly distributed perturbations in the galaxies and with isotropic velocity ellipsoids for the \hi\ gas. This is evidence that strong projection effects imprint on the velocity dispersion maps of \hi\ gas in nearby galaxies. 

Finding the origin(s)  of the   asymmetries and its systematics  is a difficult task because of the nature of the velocity dispersion itself, which stems from instrumental effects and various processes inherent to the galactic random motions (dynamical, hydrodynamical, and thermal).
In that aspect, we showed that large-scale induced BS might produce non negligible residuals, especially in the fourth order for highly inclined galaxies with high projected rotation velocities. Typical signatures of BS on phase angles are orientated near the major axis ($k=1$), the minor axis ($k=2$) and at $0.5 \times \pi/2$ ($k=4$).
Nevertheless, these orientations are not fully consistent with the observations, especially for the second order, and the strength of Fourier modes in the BS contribution is much smaller than the observed asymmetries. This suggests that unresolved velocity streamings could be responsible for some variations in the likelihood distributions of the phase angles.
The correlations found between Fourier decomposition of \hi\ gas density and velocity dispersion maps suggest that the observed bisymmetry, and thus streaming motions, are induced by spiral arms, bars or warps.

We thus further addressed the impact  on the random motions from such velocity streamings  present in the ordered motions $V_R$ and $V_\theta$. To do so,  toy models mocking discs with a barely resolved spiral perturbation in both the density and kinematics were made, from which  mock moment maps  of discs seen under various projection angles, strength and orientation of the perturbations were derived, and Fourier analyses performed, by analogy to the observations. When streamings in the ordered motions are stronger in the radial dimension, hence in presence of a radially biased velocity anisotropy on large scales, they propagate to the random motions through the beam smearing effect, but on small scales, and it is more likely to find phase angles of the $k=2$ mode orientated along the minor axis. The same effect applies to the major axis in presence of a tangential large-scale anisotropy as caused by stronger streamings in the tangential dimension. When  both $V_R$ and $V_\theta$ streaming motions are present, the phase angles of the bisymmetry of velocity dispersion reflect the competition between the two directions. Therefore, producing distributions of $k=2$ phase angles similar to the observations would probably require specific mixing between the radial and tangential perturbations in discs, like having different regions of dominance of streamings in one direction over the other. Also, no $k=4$ phase angles as concentrated around $\pi/4$ as  in the observations could be produced within the toy models. Our results suggest that the fourth order asymmetry is likely a combination between BS residuals and harmonics of the second order asymmetry in the velocity dispersion.

This analysis thus shows that it is possible to constrain which asymmetry between $V_R$ or $V_\phi$ dominates the streamings by studying the systematic orientations of $k=1$ and $k=2$ phase angles of \slos.
In particular, to explain the  systematic orientation of $k=1$ phase angles  along the major axis observed in the samples, this work suggests that lopsidedness in $V_{\rm los}$ must be dominated by tangential perturbations, while the systematic orientations of $k=2$ phase angles outside the major axis require elliptical streamings in $V_{\rm los}$ affected by a stronger contribution from radial perturbations.
By consequence, the observation of velocity asymmetries  more significant in one direction in the plane than the other implies that \hi\ velocities are highly anisotropic, at least at the angular scales probed by the Fourier analysis and the resolution of the observations.  Yet it is possible that the velocity ellipsoid of gas remains isotropic  at  angular scales much smaller than those probed by the data, thus very locally,   perhaps at the scale of individual clouds. Moreover, we predict that the signature of velocity anisotropy on large scales is also present in velocity fields of galaxies, again in the shape of systematic orientations of asymmetries near the principal axes of the discs. This prediction should greatly benefit from analyses of residual velocity fields of galaxies. 

We further show that an anisotropic velocity ellipsoid with correlations between velocity components can generate asymmetries in the velocity dispersion with any orientation. Within this formalism, we predict that the probability to find coefficients of a correlation between $V_R$ and $V_\theta$ must be larger towards $|\rho_{R\theta}| \sim 0.6$, in order to explain the larger occurrence of gaseous $\phi_2$ at $\sim 0.35\pi$ and $\sim 0.65\pi$, assuming $\sigma_\theta/\sigma_R \sim 0.7$ like for the youngest populations of stars in the Milky Way.
It would mean that the correlations between velocities in nearby gaseous discs are not random values, thus that perturbations of the gravitational potential  share common properties among the galaxies. This prediction could be compared with the velocity correlations measured for the youngest stellar populations of the Milky Way,  to which the kinematics of interstellar gas should be comparable in many aspects.  

This pioneering work has highlighted the importance of velocity anisotropy in shaping the asymmetries of \hi\ velocity dispersions,  through velocity streamings in the \hi\ gas which are dominant in particular directions in the plane, and/or tilted velocity ellipsoids.
The present study is intended to serve as a stepping stone towards the design and the analysis of future large-scale surveys of galaxies, observed at higher sensitivity and resolution than those samples studied here, like those upcoming with the Square Kilometre Array and its precursors for the \hi\ gas, or those already observed in the molecular interstellar medium of galaxies by ALMA, as well as in ionised gas and stellar velocity fields. In that aspect, comparisons of the shapes of asymmetries in \slos\ from both the stellar and gaseous components will be crucial to support the proposal that gas velocities are anisotropic. Harmonic decompositions of $V_{\rm los}$ will also be helpful. This will be the subject of future papers from this work.  
 
\begin{acknowledgements}
We are grateful to F. Walter for sharing the maps of the dirty beams of THINGS observations. P. Adamczyk and L. Chemin warmly acknowledge the Comit\'e Mixto European Southern Observatory-Gobierno de Chile. The research of L. Chemin is funded by the Chilean Agencia Nacional de Investigaci\'{o}n y Desarrollo (ANID) through the Fondo Nacional de Desarrollo Cient\'{\i}fico y Tecnol\'{o}gico (FONDECYT) Regular Project 1210992.
\end{acknowledgements}

%
\bibliographystyle{aa} 
\bibliography{biblio} 
%

\begin{appendix}
\section{Significance of FFT results with noisy data}
\label{app:biaises}

In inner regions of galaxies, rings are smaller and contain less pixels than in the outermost parts. Moreover, the average velocity dispersion tends to decrease with radius, and can thus vary significantly from one ring to another, so that the properties of rings are not necessarily uniform as a function of radius.
The results of a FFT, and in particular the precision that can be obtained on Fourier amplitudes, is impacted these parameters.
It is essential to understand how the noise in the data affects the results of the FFT and to estimate to which limits we can recover the real signal, or in other words the reliability of the recovered signal.In our study, we apply the FFT implemented in the Python package FFTPACK from the Scipy library on the squared velocity dispersion $\sigma^2$. To have a realistic estimate of the noise $\sigma^2$, we express the uncertainties on $\sigma^2$ using the differential of the squared velocity dispersion as:
\begin{equation}
d(\sigma^2) = 2 \sigma_0 d\sigma \text{~,}
\label{eq:d_sig2}
\end{equation}
where $\sigma_0^2 = \langle\sigma^2\rangle$ is the value of the zeroth order of the FFT, because all higher orders cancel out on average over a ring, and where $d\sigma$ is the uncertainty on the observed velocity dispersion and that roughly corresponds to the uncertainty on the instrumental resolution at a given signal-to-noise ratio. 
Since we do not have uncertainty maps due to the processing techniques (moments method) on THINGS data, $d\sigma$ is not known precisely. 
Since the spectral resolution of both THINGS and WHISP data corresponds to about $\sigma_{\rm LSF}~\sim 2-5$~km~s$^{-1}$, and because the actual uncertainty on the line width depends on the SNR, we assume that the uncertainty on the velocity dispersion is of the order $d\sigma \sim 1$~km~s$^{-1}$.

The way the uncertainties propagate from the data to the FFT coefficients depends on how the FFT is implemented. A discrete Fourier transform is used in FFTPACK to compute Fourier coefficients:
\begin{equation}
 C_k = \sum_{m=0}^{N-1} c_m \cos{\left( - \frac{2\pi k m}{N} \right)} + j \sum_{m=0}^{N-1} c_m \sin{\left(-\frac{2\pi k m}{N}\right)} \text{~,}
 \label{DFT_direct_def}
\end{equation}
where $j$ is the unit imaginary number, $C_k$ is the Fourier coefficient of order $k$ as computed by FFTPACK, and $c_m$ are the $N$ elements in the discrete array. In our study, $c_m$ are the squared velocity dispersion values along rings, $\sigma^2(m)$, and we deduce the amplitude of each order $k\ne 0$ as $\sigma^2_k = 2 |C_k|/N$, and $\sigma^2_0 =  |C_0|/N$ as the average of the squared velocity dispersion values along the ring. From this formalism, it can be shown that for $k\ne 0$:
\begin{equation}
\begin{split}
&\textrm{Var}[\sigma^2_k] = \frac{4}{N^2} \sum_{m=0}^{N-1} \textrm{Var}[\sigma(m)^2] \\
& +\frac{8}{N^2} \sum_{\Delta m=1}^{N-1}\sum_{m = 0}^{N-1-\Delta m} \textrm{Cov}[\sigma(m)^2, \sigma(m+\Delta m)^2] \cos{\left( \frac{2\pi k \Delta m}{N} \right)} \text{~,}
\end{split}
\label{variance_general}
\end{equation}
with $\textrm{Var}$ and $\textrm{Cov}$ being respectively the variance and covariance.
Assuming that the variance is constant across the ring ($\textrm{Var}[\sigma(m)^2]=\textrm{Var}[\sigma^2]$) and that covariance is null between different pixels, we can get at first approximation that:
\begin{equation}
\textrm{Var}[\sigma^2_k] = \frac{4}{N} \textrm{Var}[\sigma^2] \text{~,}
\label{variance_approx}
\end{equation}
leading to the uncertainty on coefficients:
\begin{equation}
d(\sigma^2_k) = \frac{4 \sigma_0 d\sigma}{\sqrt{N}}  \text{~.}
\label{uncertainty_approx}
\end{equation}
Interestingly, this equation shows that the uncertainty is the same for all orders, that it is proportional to the zeroth order, and that it decreases with radius since the number of pixels increases linearly with radius and because $\sigma_0$ rarely tends to decrease with radius.

\begin{figure}[t]
    \centering
        \includegraphics[width=.49\textwidth]{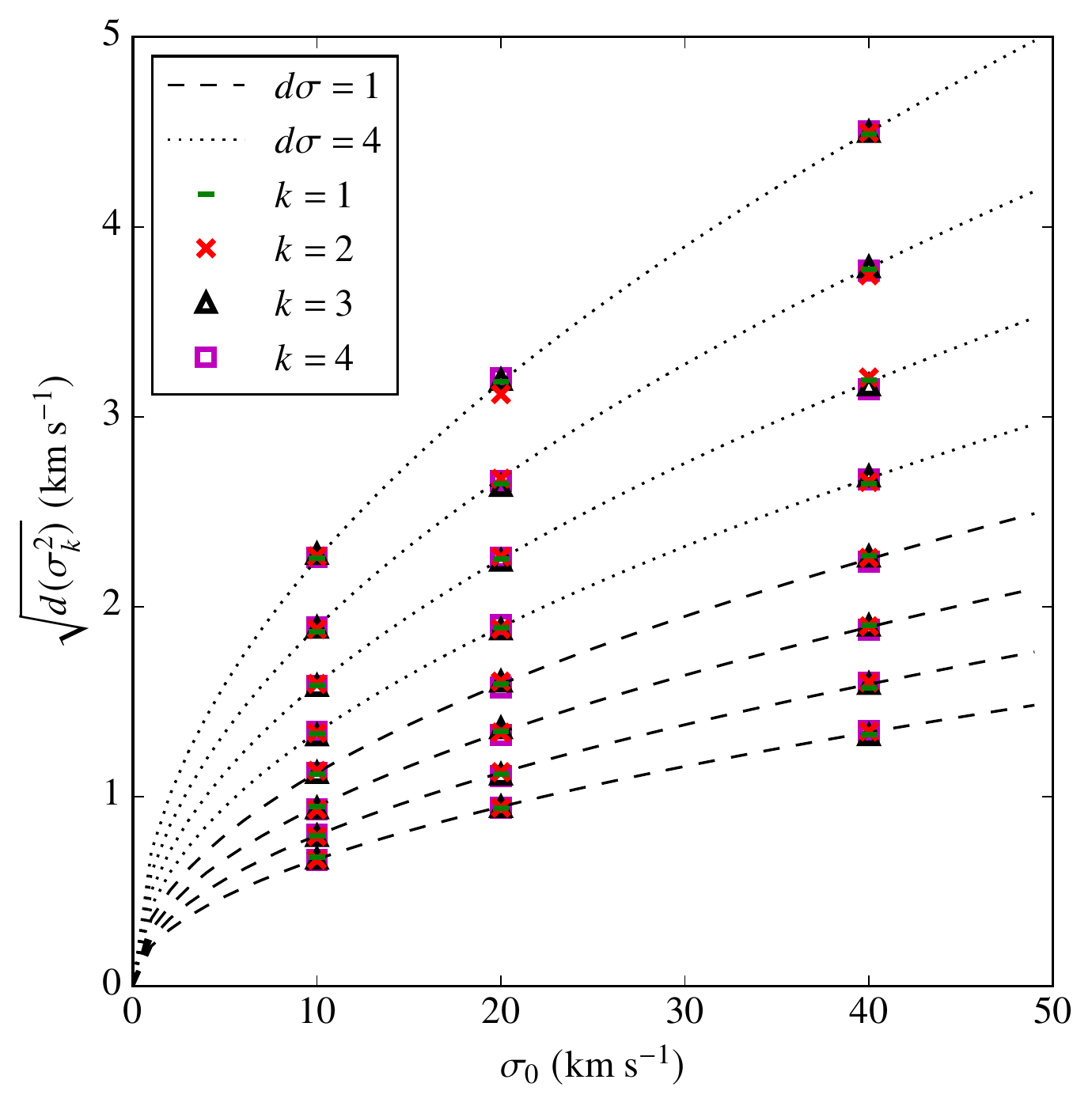}
        \caption{Square root of the noise on the squared velocity dispersion amplitude for orders $k=$~1 to 4 of the FFT as a function of $\sigma_0$ for two distinct values of $d\sigma$ and various number of points $N$ in the FFT. The curves correspond to Eq.~\ref{uncertainty_approx} with $d\sigma = 1$~km~s$^{-1}$ (dashed) and $d\sigma = 4$~km~s$^{-1}$ (dotted), each corresponding to a different value of $N$ (from top to bottom : 1000, 2000, 4000 and 8000).
        Coloured symbols correspond to different orders measured on toy models. The similar behaviour of orders leads to superposed points. }
        \label{fig:toymod2}
\end{figure}

We have checked numerically the consistency between this expression and the actual results from FFT using a toy model.
We define the noise $\delta$ on $\sigma_{obs}^2$ as a Gaussian distribution centred on 0 and with a standard deviation of $\sigma_{Gauss} = d(\sigma^2)$.
We create a toy model of the observed velocity dispersion with a constant velocity dispersion to which we added noise:
\begin{equation}
\sigma_{obs}^2 = \sigma_0^2 + \delta \text{~,}
\label{eq:toymodel_significance}
\end{equation}
with $\delta$ a random value in the noise distribution. Doing so, the distribution of each observed point is taken independently. We used several values for $d\sigma$, $\sigma_0$, and $N$ the number of points in the FFT, since $N$ varies from one ring to another and from one galaxy to another. We do 1000 iterations of the FFT and analyse the amplitudes for orders 1 to 4, which are null in our model, as shown in Eq. \ref{eq:toymodel_significance}. Fig.~\ref{fig:toymod2} shows the measured uncertainties of the toy models and confirms that Eq. \ref{uncertainty_approx} is a good description of the uncertainties.
Equation~\ref{uncertainty_approx} can therefore be used to infer at first order the significance of Fourier coefficients, $s_k = \sigma_k^2/d(\sigma^2_k)$, depending on the number of rings and on the mean velocity dispersion value.

\section{Toy model description to study projection and resolution effects on random motions}
\label{app:toymodels}

\begin{figure}
\centering
\includegraphics[width = 8.5cm]{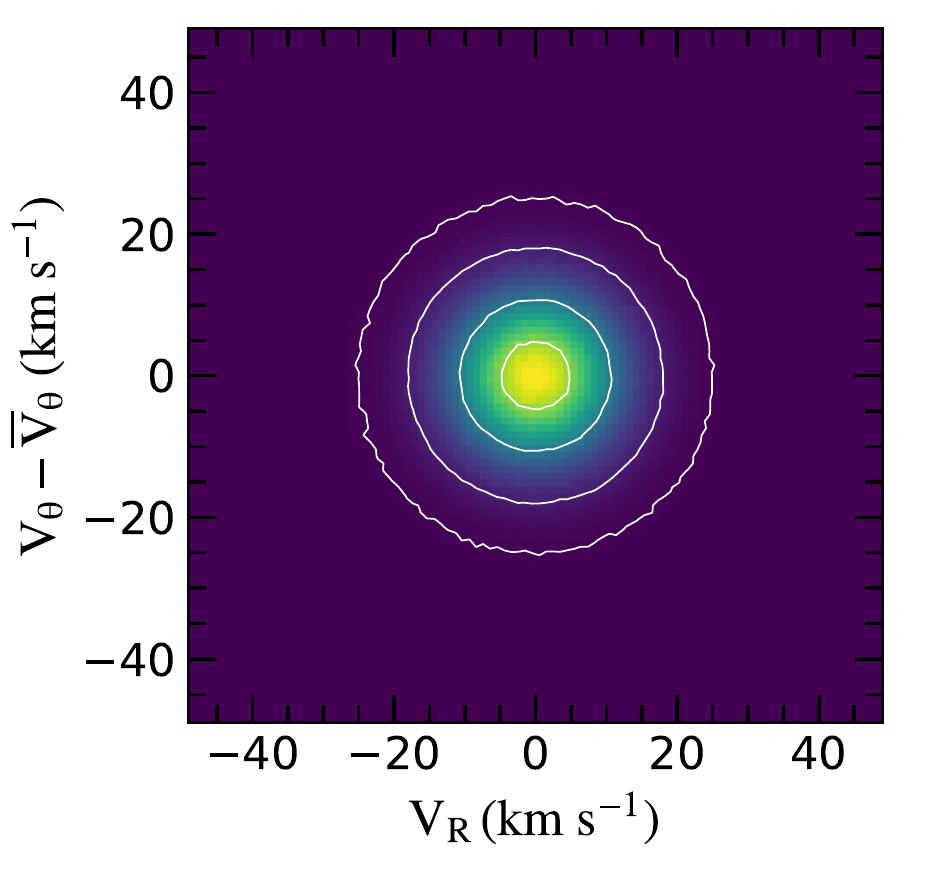}\\
\includegraphics[width = 8.5cm]{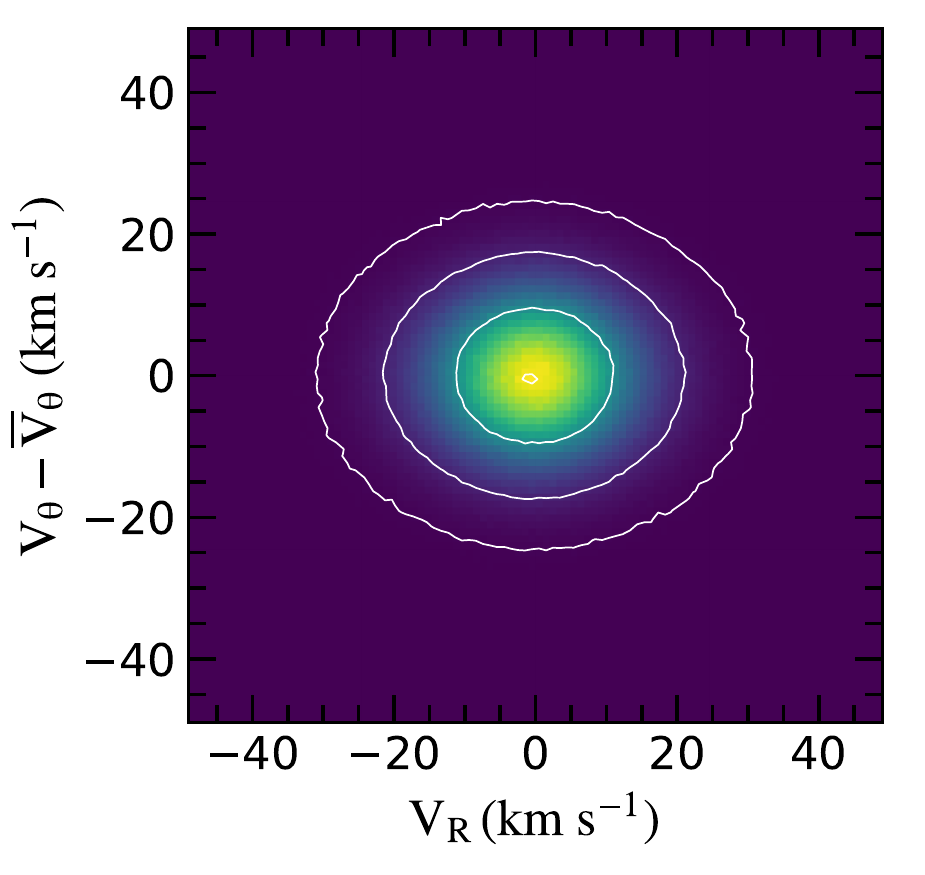}
\caption{Velocity ellipsoid of the mock disc with local velocity isotropy for the cases without large-scale anisotropy (upper panel), and with a large-scale anisotropy induced by the spiral $V_R$ streamings (bottom panel), assuming radial motions are null on average. The white contours represent densities of 100, 1000, 5000, and 10000 toy-model particles.}
\label{fig:ellipsoidtoy}
\end{figure}

Here, we describe the detailed generation of the toy models presented in Sect. \ref{sec:toy_models} and used in Sections \ref{sec:bs_properties} and \ref{sec:orderedvelocityeffect} to investigate the impact of projection effects on the study of random motions in both axisymmetric and asymmetric cases as well as with and without local anisotropy.
We first built disc-like kinematics with $5 \times 10^6$ points representing individual particles or gas clouds uniformly distributed over a disc of 200\arcsec\ radius.
The adopted axisymmetric contribution of azimuthal velocities $V_\theta$  follows $V_{\theta}(R) = v_0/(1+(r_s/R)^\gamma)^{1/\gamma}$ for two different sets of parameters to describe either (i) a rotation curve slowly rising to a moderate maximum rotation velocity with $v_0=200$ \kms, $r_s=100\arcsec$ and $\gamma = 1$, or (ii) a rotation curve with a steep inner gradient and a sharp transition to a high velocity plateau reached at small radius with $v_0=250$ \kms, $r_s=20\arcsec$ and $\gamma = 2$. We considered those two rotation curves to have either a low or a high impact of large-scale rotation on BS. The disc was not assumed contracting or expanding, that is, $V_{R}(R) = 0$, and  a negligible average vertical component, $V_z(R) = 0$. 
A few solutions exist to describe the departure of \vaz\ and \vrad\ from axisymmetry at a given $R, \theta$ position in the plane. For example, planar asymmetries can follow Fourier harmonics of second order to mimic elliptical orbits in a bar or spiral potential \citep[e.g.][]{2007spekkens,2022gaiadrimmel}. This is useful to describe large-scale perturbations of \vaz\ and \vrad, but not necessarily appropriate to mock sharp velocity gradients across and along spiral arms \citep{2016chemin}. Therefore, we adopted a different prescription than the simple cylindrical harmonics, and defined a spiral pattern by:
\begin{equation}
    \theta = \phi_{sp} + 2\pi R / r_0 \text{~,}
    \label{eq:spiral}
\end{equation}
where $R$ and $\theta$ are radial and azimuthal coordinates in the plane of the galaxy, $\phi_{sp}$ is the phase of the pattern at the centre, and $r_0$ is the radius interval necessary for the spiral to complete $2\pi$. 
We used $r_0/2\pi = 50$\arcsec\ in all models, and since the interesting parameter to probe here is the difference between the orientation of the velocity perturbation and a reference axis in the disc, we vary the initial angle of the spiral perturbation with a sampling of 15\degr, from 0\degr\ to 165\degr\ with respect to the major axis of the galaxy, leading to 12 possible orientations for the pattern. 
For each $R,\theta$ position, the closest point along the spiral in the galaxy frame was found to infer the distance $D_{sp}$ to the spiral and the unitary vector $\vec{u}$ between those points.
A density perturbation following the spiral pattern was introduced as a constant over-density for all points having their distance $D_{sp}$ lower than a maximum distance $\Delta D$ where the spiral has an impact on the axisymmetric model. We note that the distribution of points is not important in this exercise, as it does not affect the kinematics. It can only impact the derivation of the BS model, though with negligible overall effect.
We did not model the case of asymmetric vertical motions, and defined the amplitude of the velocity perturbation by $\Delta V = \Delta V_0 \times (1 - D_{sp} / \Delta D)$ for $D_{sp} \le \Delta D$, $\Delta V_0$ being the maximum amplitude of the velocity perturbation. We used $\Delta V_0=25$~km~s$^{-1}$ and $\Delta D=25$\arcsec. The value of 25 \kms\ was chosen to match the maximum variation of velocity through the spiral arms of the grand-design spiral Messier 99 with respect to the axisymmetric circular velocity \citep[Fig.~8 of][]{2016chemin}, or the maximum strength of the elliptical motions  of RGB stars in the Galactic bar \citep[Fig.~19 of][]{2022gaiadrimmel}.
We assumed the 6 following configurations:
\begin{itemize}
\item (1) Axisymmetric models with no perturbation ($\Delta V = \Delta V_R = \Delta V_\theta = 0$). These models are used as references from which comparisons can be made, with respect to previous asymmetric models. 
\item Asymmetric velocity models with a spiral velocity perturbation along either (2) the azimuthal direction  ($\Delta V_\theta = \Delta V \times \vec{u}.\vec{u_\theta} / |\vec{u}.\vec{u_\theta}|$, $\Delta V_R = 0$), $\vec{u_\theta}$ being the unitary azimuthal vector, or (3) the radial direction  ($\Delta V_R = \Delta V \times \vec{u}.\vec{u_R} / |\vec{u}.\vec{u_R}|$, $\Delta V_\theta = 0$), $\vec{u_R}$ being the unitary radial vector. These cases are necessary to identify the impact of velocity gradients along one planar direction, independently from the other direction. This thus represents a departure from $V_\theta(R)$  of at least $\sim 13\%$, and significantly more from $V_R(R)$.
\item Asymmetric velocity models with a spiral velocity perturbation along both azimuthal and radial directions for the cases (4) $\Delta V_\theta = 2 \Delta V_R = 2 \Delta V \times \vec{u}.\vec{u_\theta} / |\vec{u}.\vec{u_\theta}|$, (5) $\Delta V_R = 2 \Delta V_\theta = 2 \Delta V \times \vec{u}.\vec{u_R} / |\vec{u}.\vec{u_R}|$, and (6) with the spiral velocity perturbation $\Delta V$ along $\vec{u}$, i.e. $\Delta V_R = \Delta V \times \vec{u}.\vec{u_R}$ and $\Delta V_\theta= \Delta V \times \vec{u}.\vec{u_\theta}$. In the latter case, the orientation of the velocity perturbation changes with radius, being more azimuthal in the inner parts and more radial in the outer parts. These three models are useful to identified the combined effect of planar asymmetries with different strengths.
\end{itemize}
The azimuthal velocity at the coordinate $R,\theta$ was thus obtained by $V_\theta(R,\theta) =  V_\theta(R) \pm |\Delta V_\theta(R,\theta)|$, and similarly for the radial component. This creates a sharp kinematic discontinuity through the spiral feature, and the asymmetries in  $V_\theta$ or $V_R$ make the velocity ellipsoid intrinsically anisotropic on scales larger than that of single gas clouds.
To illustrate this, Fig.~\ref{fig:ellipsoidtoy} shows the velocity ellipsoid in the case of  local velocity isotropy without large-scale anisotropy due to a spiral streaming (top panel), and in the case of local velocity isotropy but with anisotropy induced at large scale by the $V_R$ streaming described above (bottom panel). The signature of the radial bias of the anisotropy is clearly seen as contours of density elongated in the radial dimension of the velocity tensor.

Each particle was then assigned $V_\theta, V_R$ and $V_z$ components, which were drawn randomly following a Gaussian probability distribution centred on the local values described above, $V_\theta(R,\theta), V_R(R,\theta)$ and $V_z(R)$, and with standard deviation  $\sigma_\theta, \sigma_R$ and $\sigma_z$, respectively. The two following cases have been explored:  $\sigma_\theta = \sigma_R = \sigma_z = 8$ \kms, corresponding to traditional isotropic velocity ellipsoids in the absence of asymmetries, and  $\sigma_R = 8$ \kms, $\sigma_\theta = 0.7 \sigma_R = 5.6$ \kms, and $\sigma_z = 5$ \kms, which is a direct way to also generate a velocity anisotropy, uniformly, and in addition to the velocity streamings. To keep the modelling simple, we thus did not allow direct variations of $\sigma_\theta$ and $\sigma_R$   within the disc. Combined to the six previously enumerated configurations, this leads to a total of 12 cases for each rotation curve. We did not explore the possibility $\sigma_\theta > \sigma_R$, as this situation only occasionally occurs in the stellar populations of the Milky Way or the Large Magellanic Cloud \citep[][]{2021luri,2022gaiadrimmel}. 

Then, the mock cubes of data were built, giving  each particle a \vlos\ following Eq.~\ref{eq:vlos_disc}. Each data cube contains 200 spectral channels of 3\kms\ width, and $400 \times\ 400$ squared pixels, with a pixel scale of 1\arcsec. The assigned channel map of a particle corresponds to the one closest to \vlos, and the line profile in a spatial pixel was obtained by summing the counts of all individual particles from this pixel.  Constant position angle of the disc major axis and disc inclination were assumed as a function radius. The position angle was fixed at 45\degr\ in all models, and inclinations  of 45\degr, 60\degr, and 75\degr\ were used. 
For each inclination, 12 mock data cubes were generated, corresponding to the 12 orientations of the spiral perturbation described above. 
To mimic an observational BS effect, the data cubes were then smoothed by a Gaussian function with a FWHM of 8 pixels (corresponding to $\approx 465$ pc, assuming a distance of 12 Mpc). These are idealised data cubes to which no further smoothing by synthesised light and point spread functions were applied, and no further random noise was added. The first and second moments of the data cubes were then derived, the BS correction applied, using the low-resolution flux and velocity maps, to remain consistent with the methodology applied to the observed data.

\begin{figure*}[t]
\noindent
\centering
    \includegraphics[height = 3.5cm]{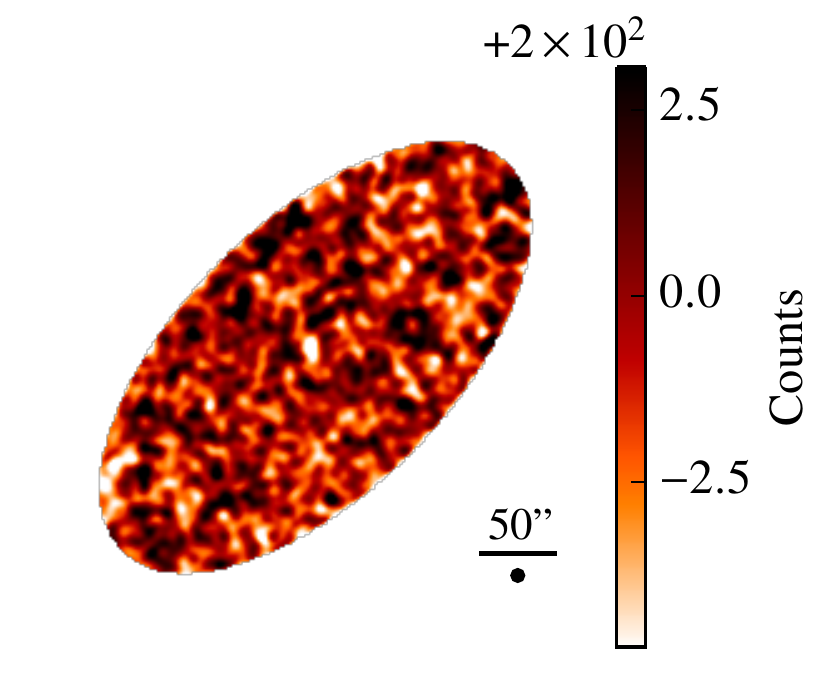}
    \includegraphics[height = 3.5cm]{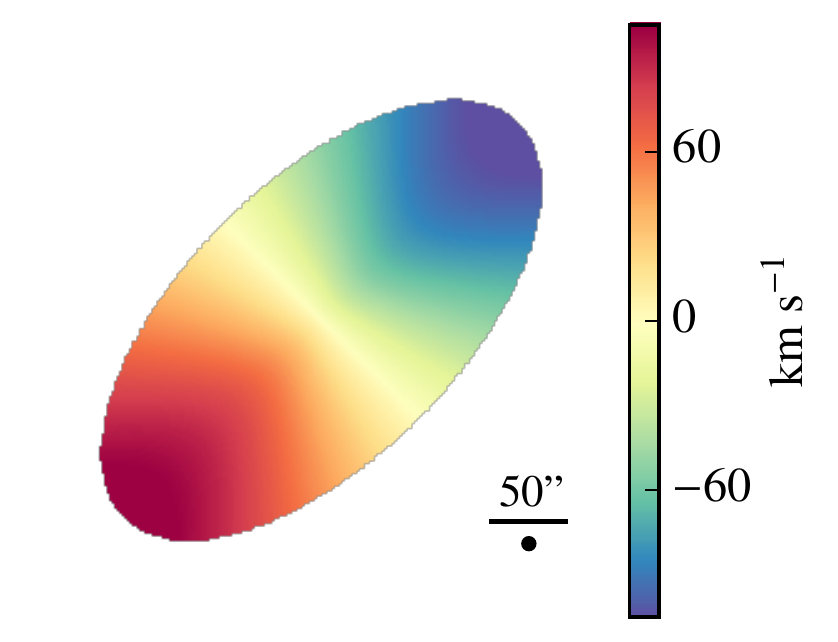}
    \includegraphics[height = 3.5cm]{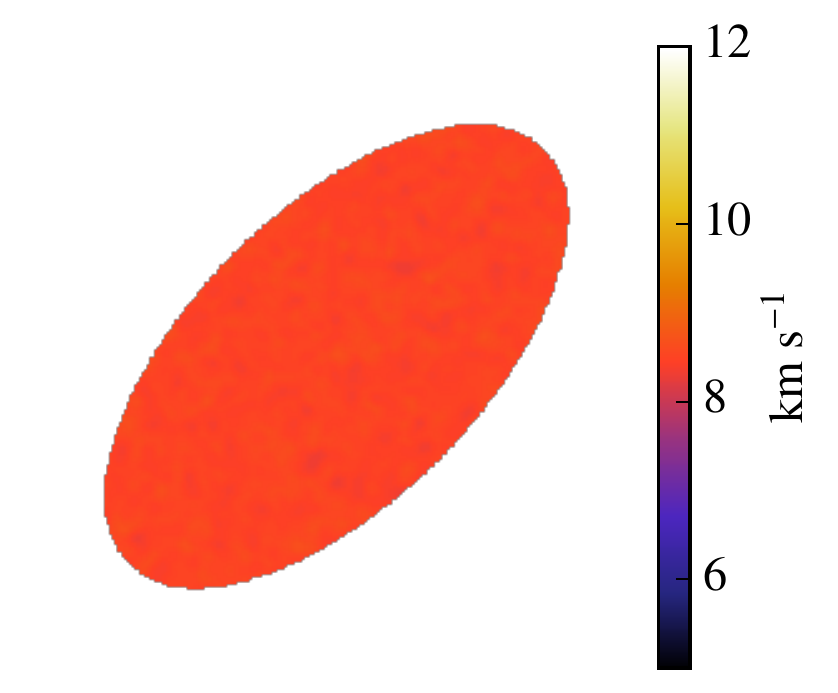}
    \includegraphics[height = 3.5cm]{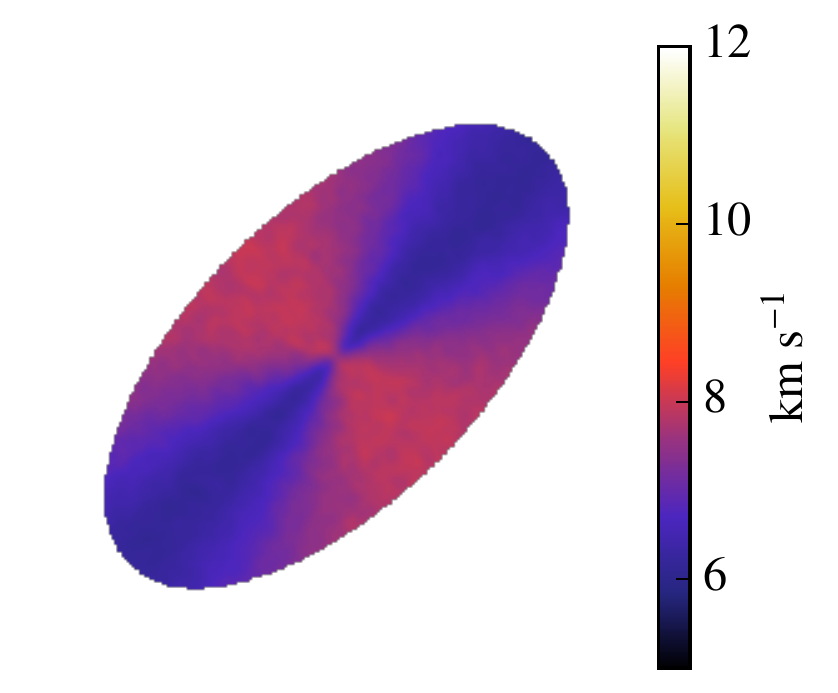} \\
    \includegraphics[height = 4cm]{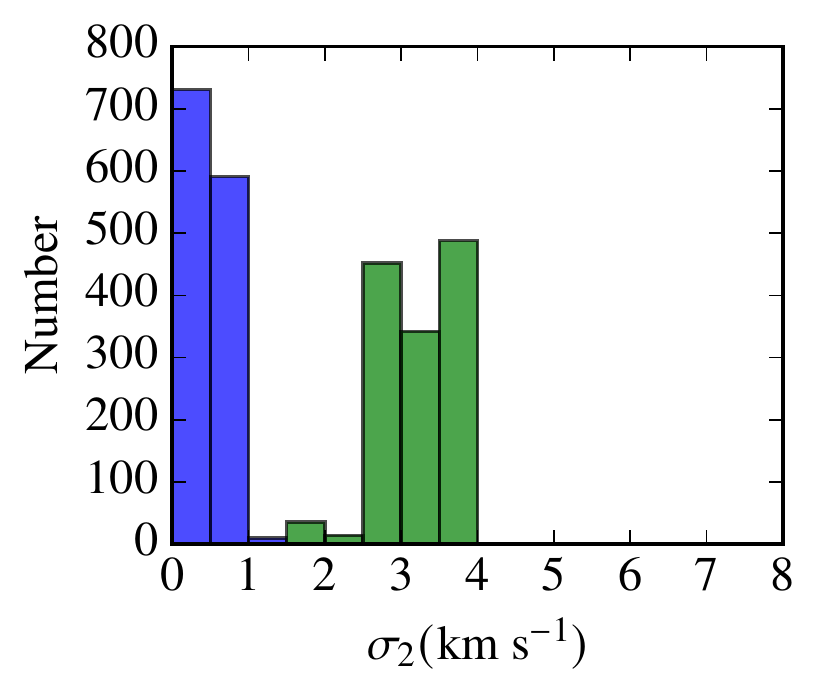}
    \includegraphics[height = 4cm, trim = 0.6cm 0 0 0 , clip]{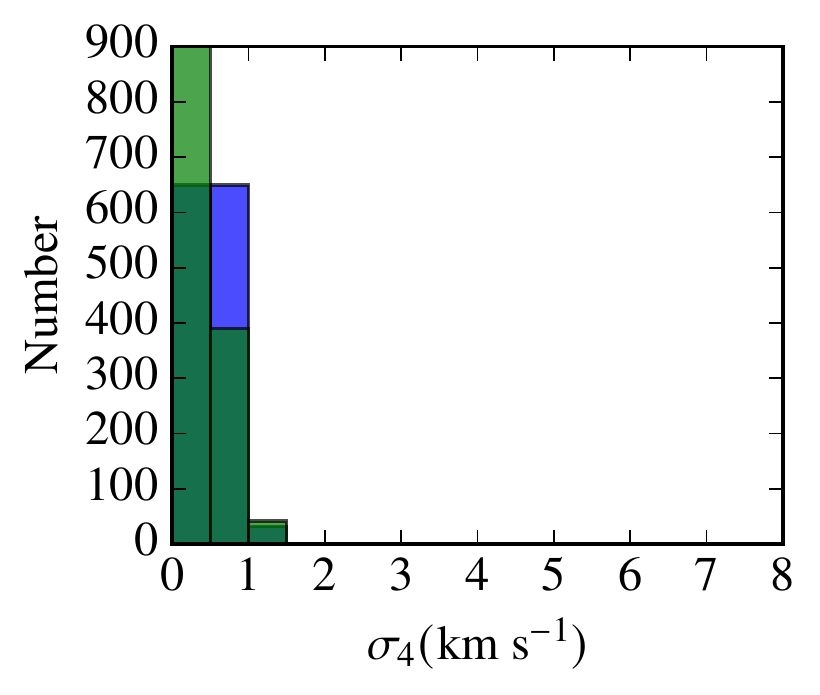}
    \includegraphics[height = 4cm, trim = 0.6cm 0 0 0 , clip]{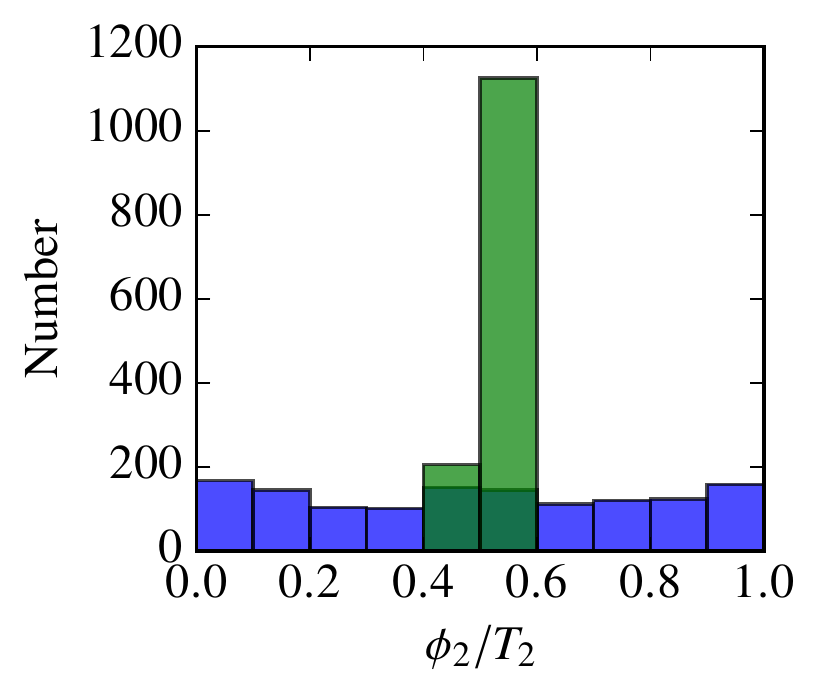}
    \includegraphics[height = 4cm, trim = 0.6cm 0 0 0 , clip]{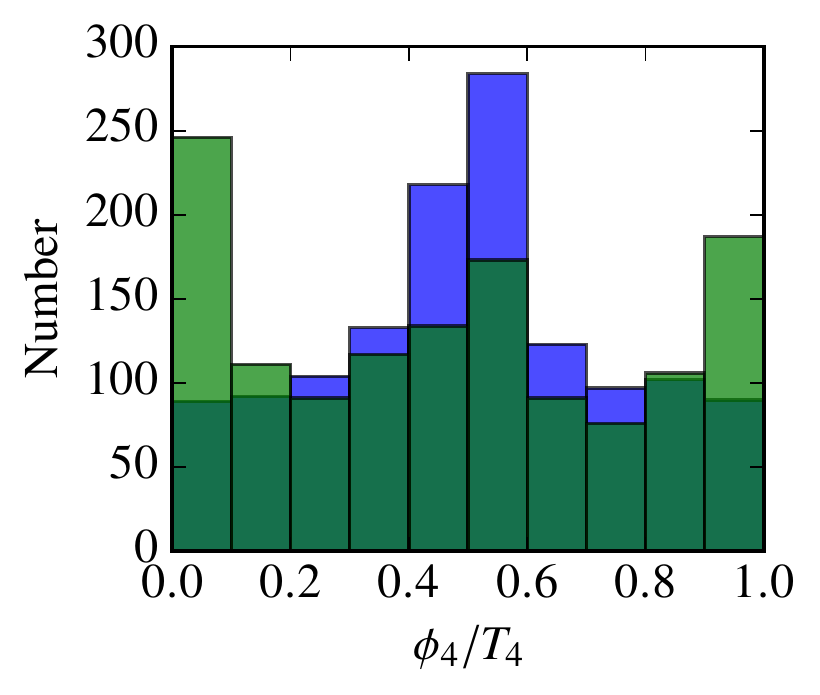}
    
    \includegraphics[height = 3.5cm]{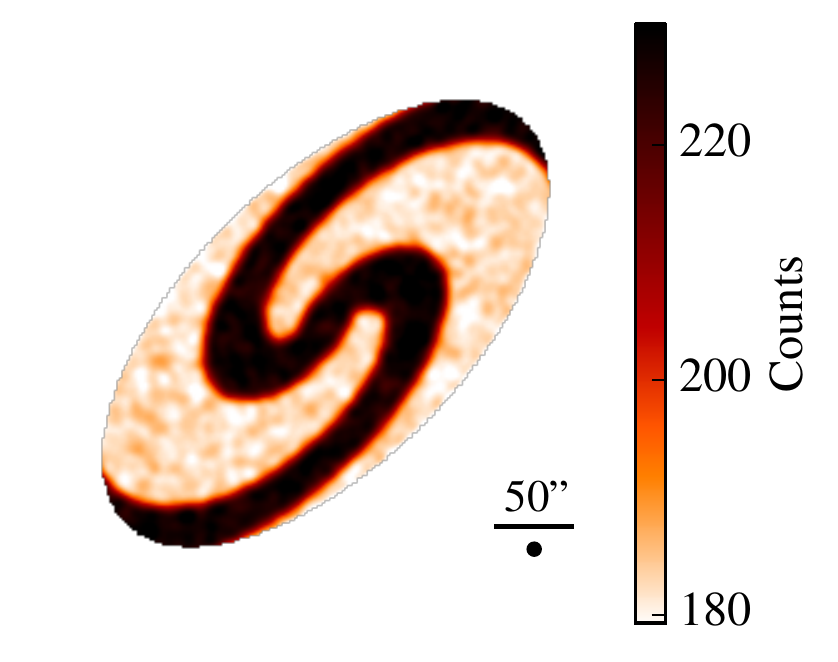}
    \includegraphics[height = 3.5cm]{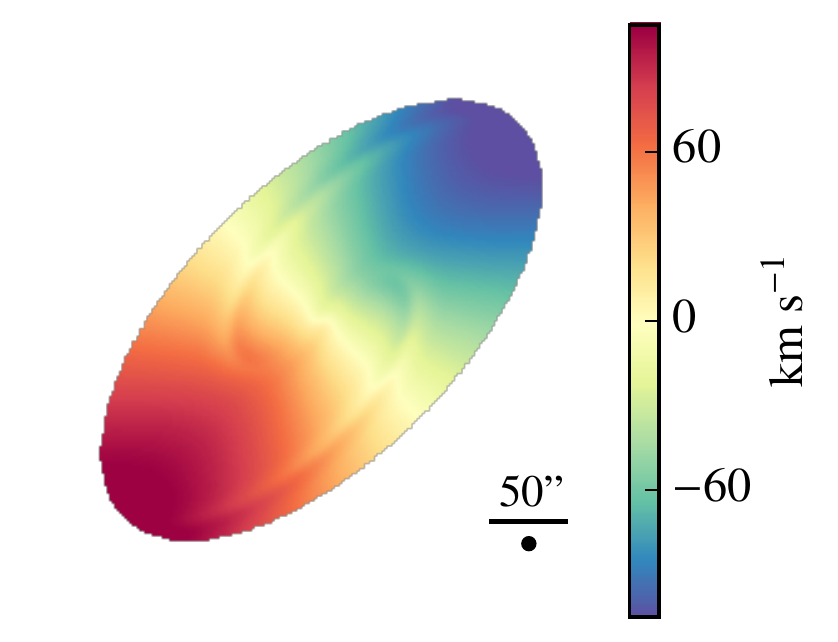}
    \includegraphics[height = 3.5cm]{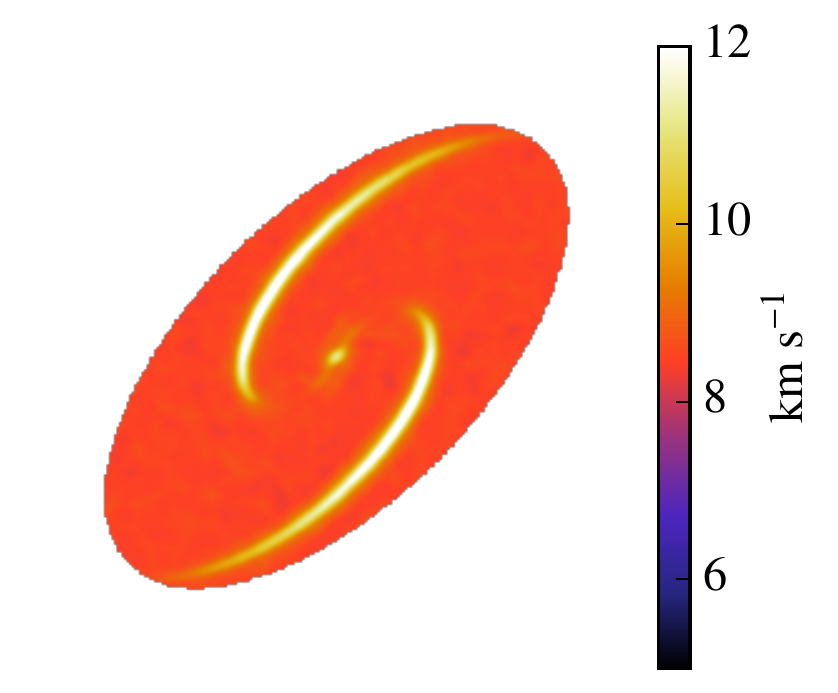}
    \includegraphics[height = 3.5cm]{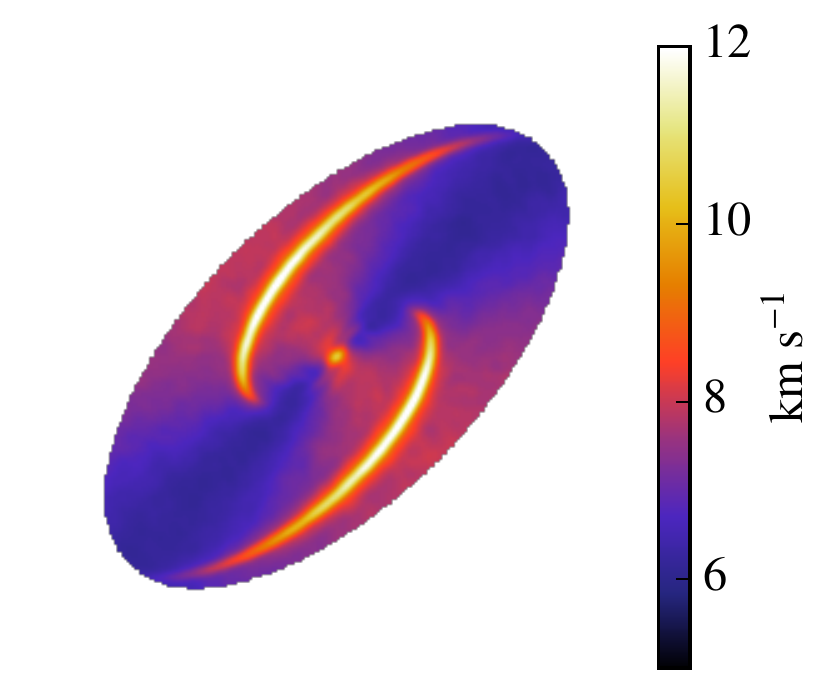} \\
    \includegraphics[height = 4cm]{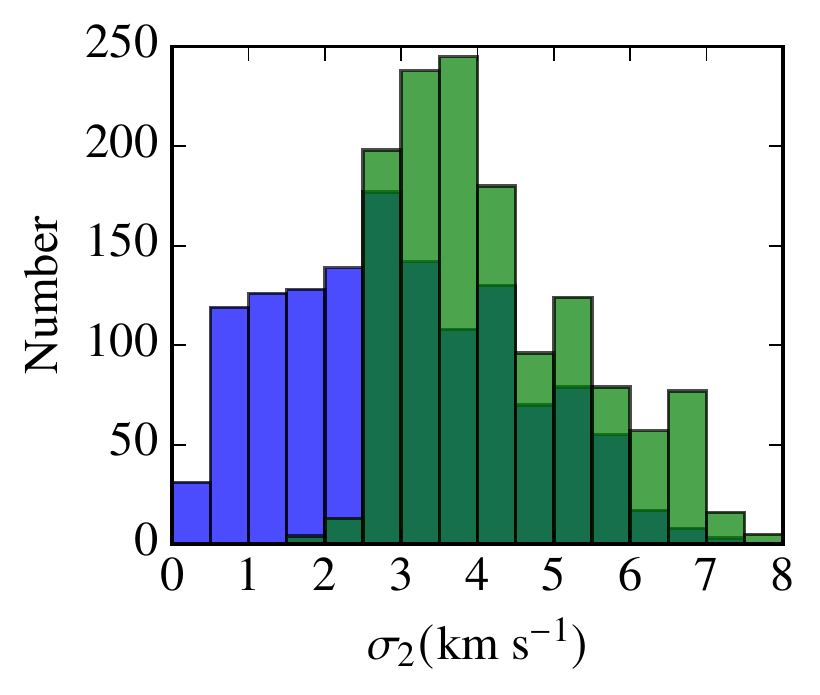}
    \includegraphics[height = 4cm, trim = 0.6cm 0 0 0 , clip]{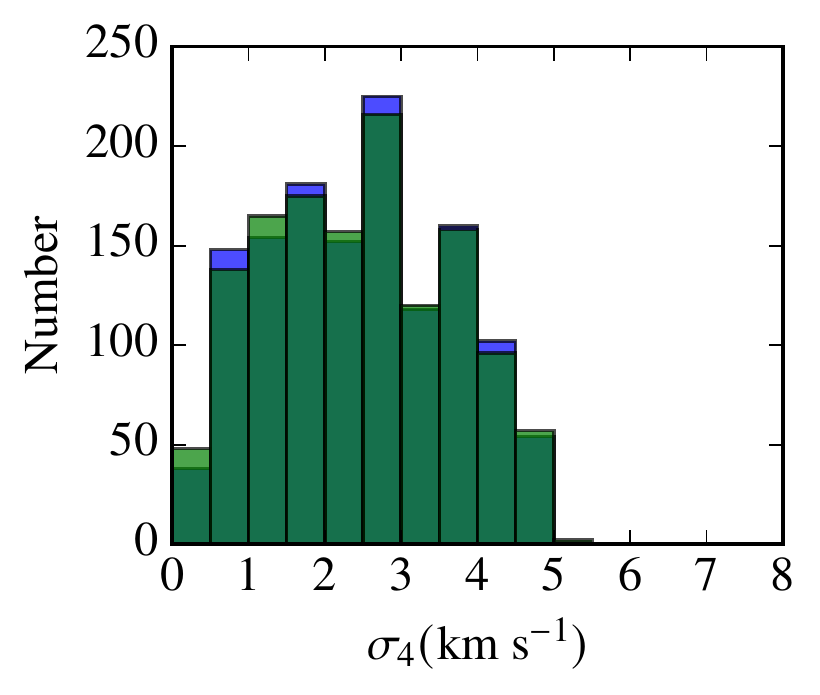}
    \includegraphics[height = 4cm, trim = 0.6cm 0 0 0 , clip]{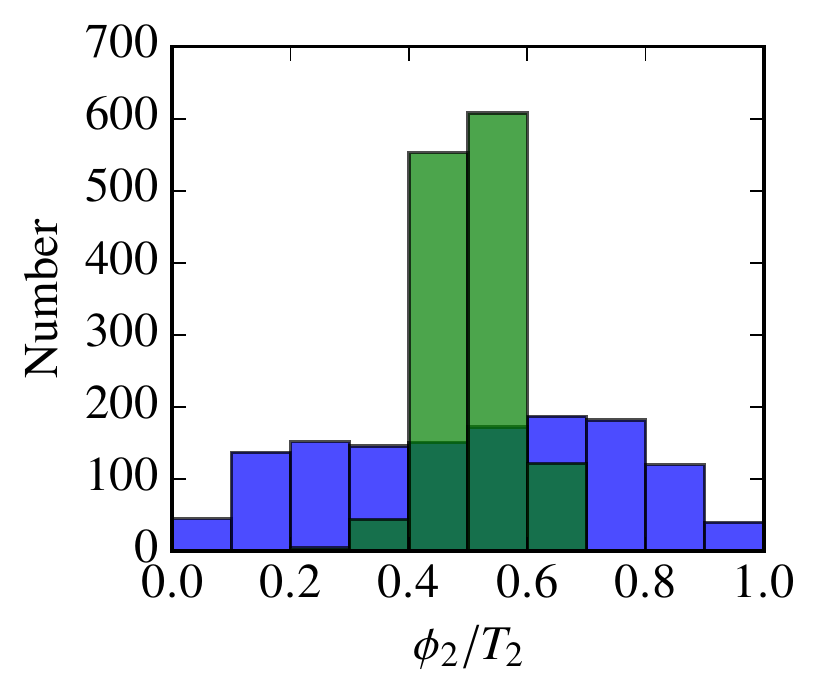}
    \includegraphics[height = 4cm, trim = 0.6cm 0 0 0 , clip]{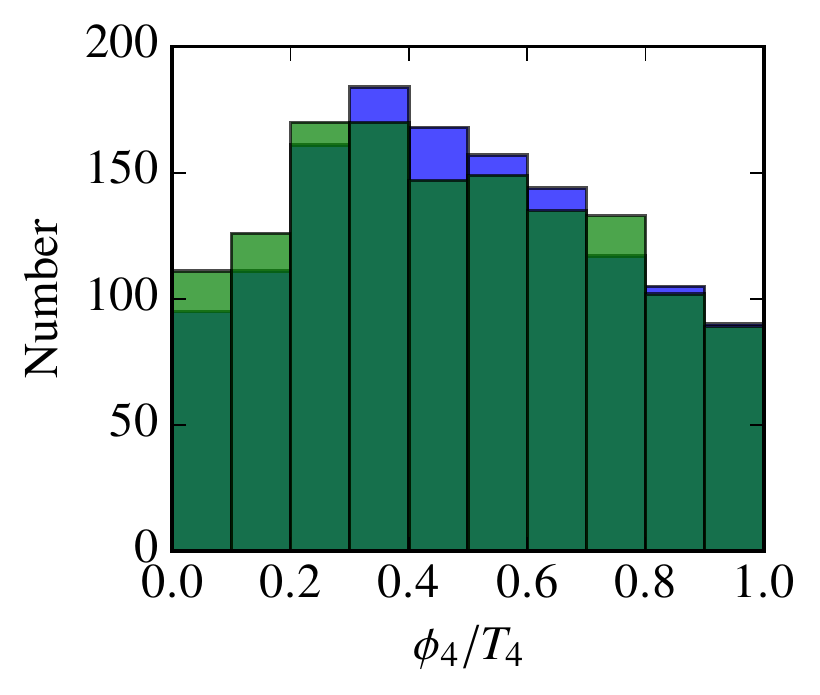}
    
    \includegraphics[height = 3.5cm]{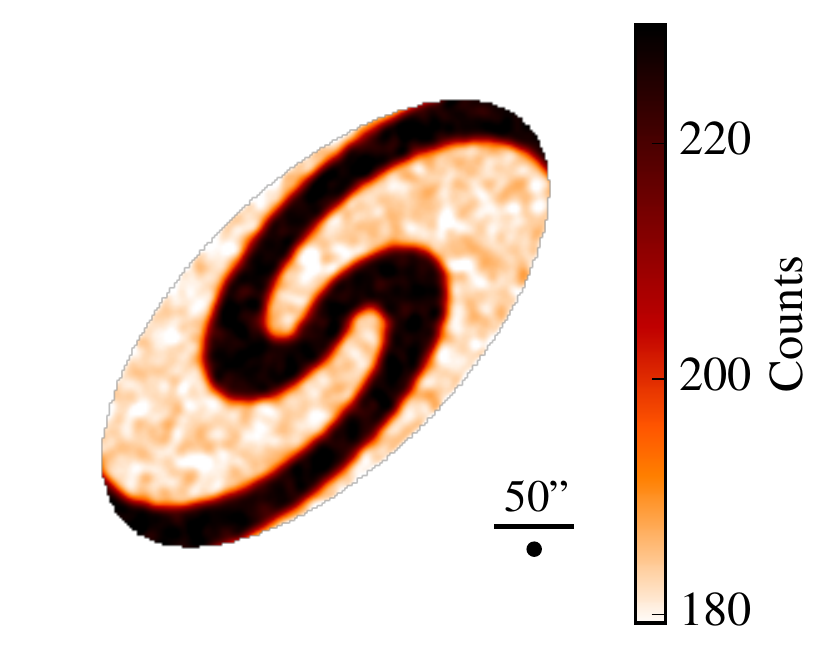}
    \includegraphics[height = 3.5cm]{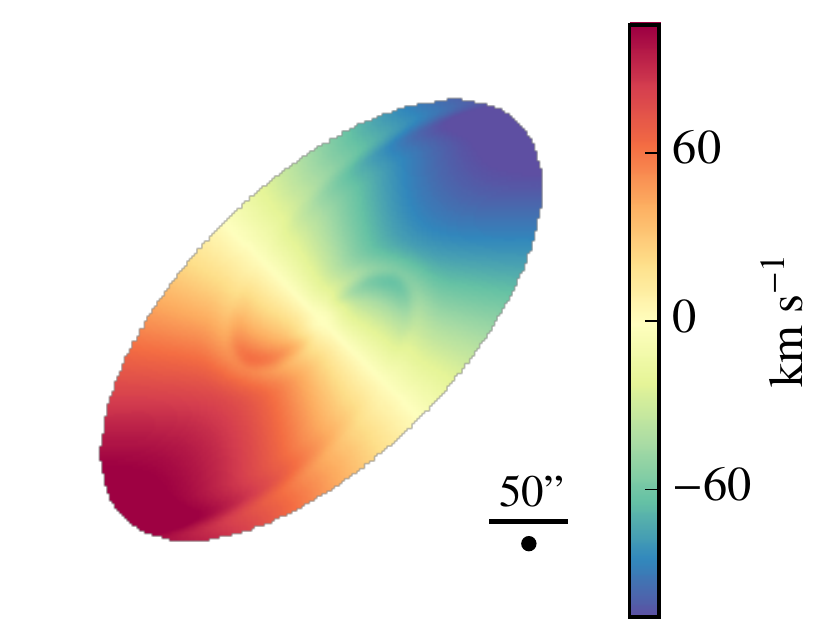}
    \includegraphics[height = 3.5cm]{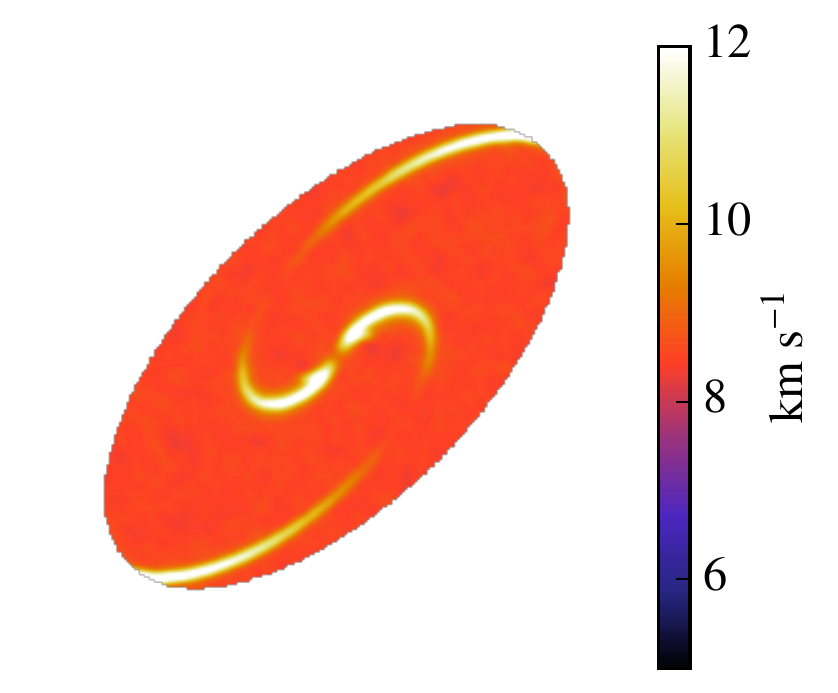}
    \includegraphics[height = 3.5cm]{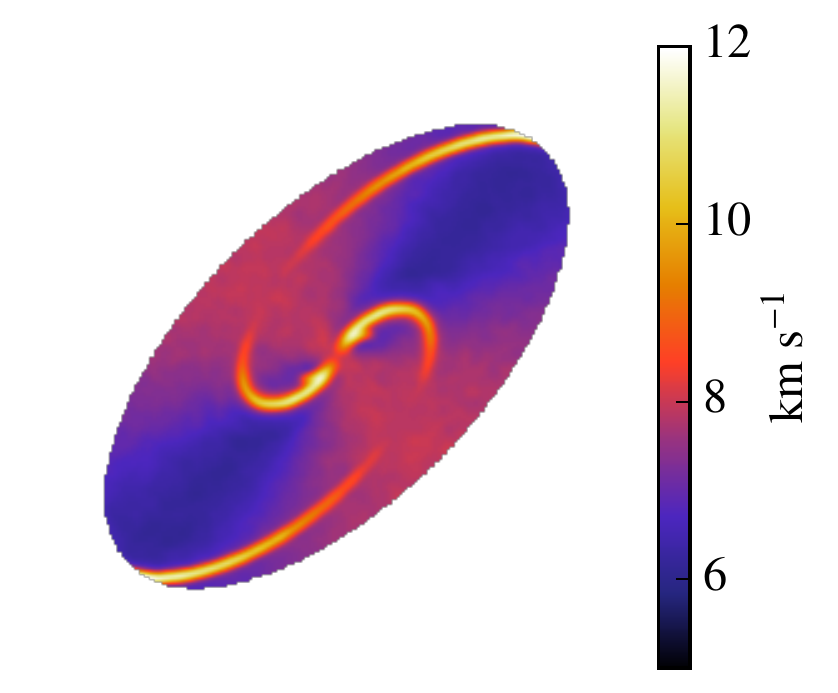} \\
    \includegraphics[height = 4cm]{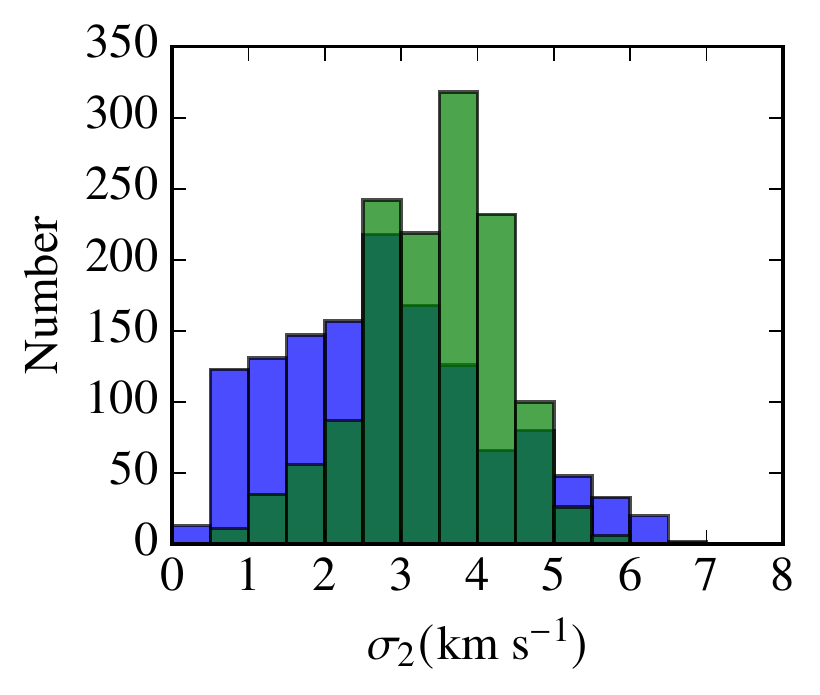}
    \includegraphics[height = 4cm, trim = 0.6cm 0 0 0 , clip]{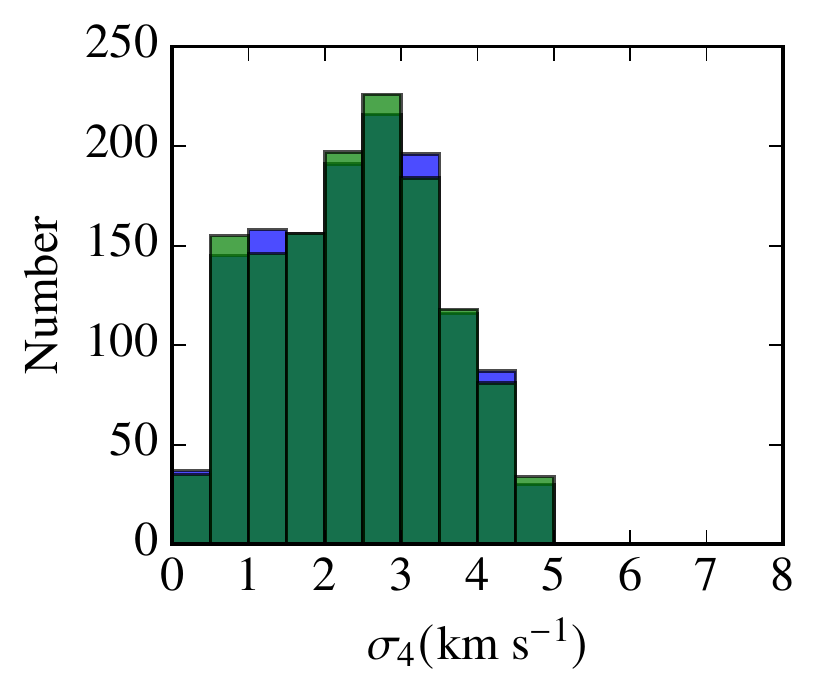}
    \includegraphics[height = 4cm, trim = 0.6cm 0 0 0 , clip]{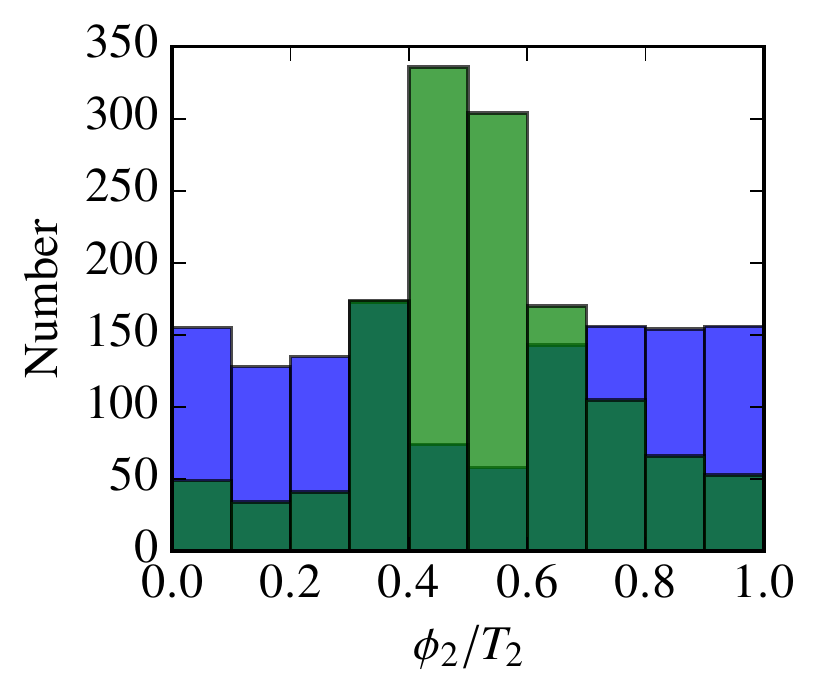}
    \includegraphics[height = 4cm, trim = 0.6cm 0 0 0 , clip]{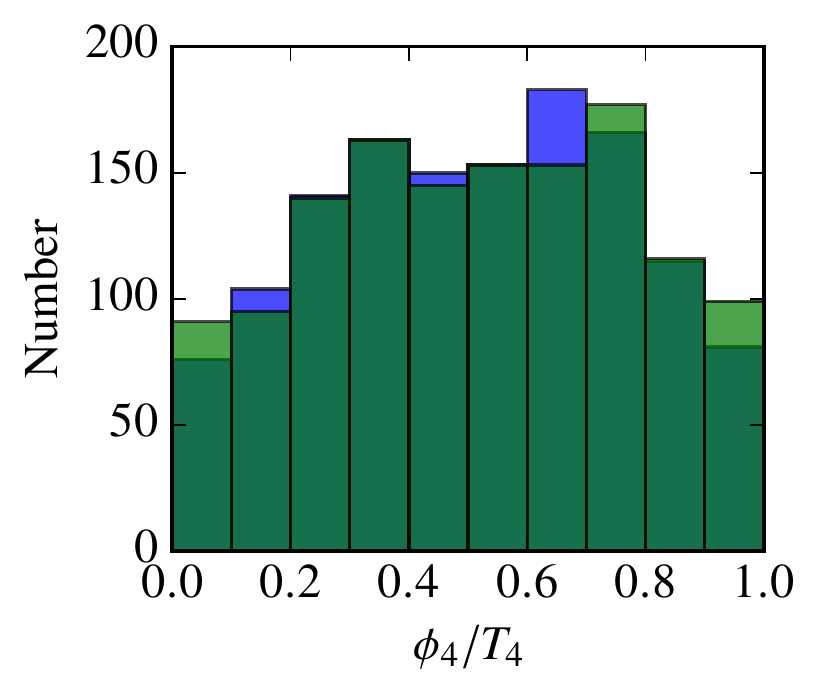}

    \caption{Kinematics maps and velocity dispersion Fourier analysis histograms for the toy models with a weak velocity gradient rotation curve ($v_0=200$ \kms, $r_s=100\arcsec$ and $\gamma = 1$) without any perturbation (top), and with a spiral velocity perturbation towards either radial (middle) or azimuthal (bottom) components.
    We show, from left to right the density map, the velocity field and the BS corrected dispersion maps in the isotropic and anisotropic cases, obtained with an inclination of 60\degr\ and with $\phi_{sp}=135$\degr.
    We show from left to right amplitude histograms of orders $k = 2$ and $k = 4$ and normalised phase angle histograms of these orders, obtained from the FFT analysis of toy models dispersion maps.
    Blue and green histograms correspond to isotropic and anisotropic cases respectively.}
    \label{fig:toy_models}
\end{figure*}  

\begin{figure*}[t]
\noindent
\centering
    \includegraphics[height = 3.5cm]{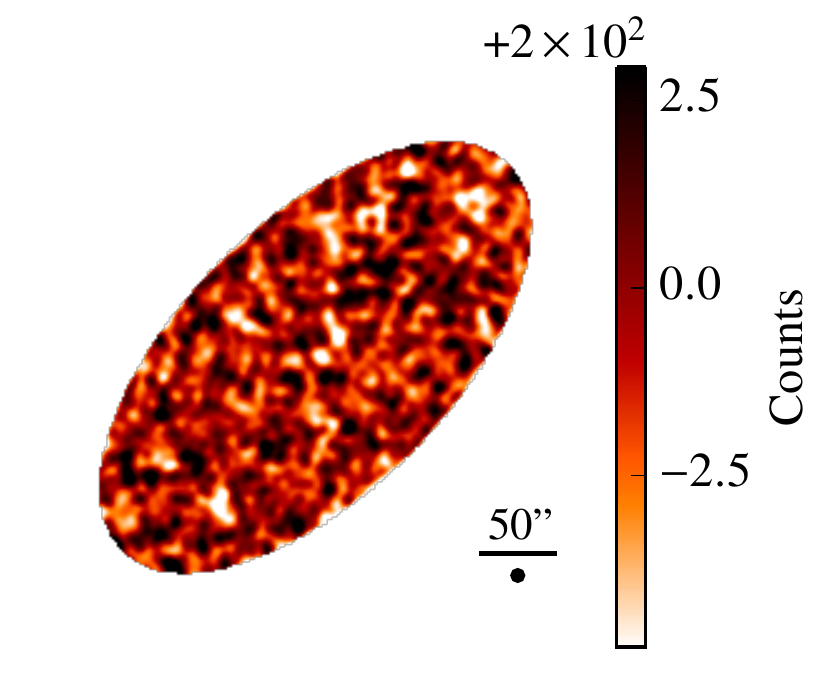}
    \includegraphics[height = 3.5cm]{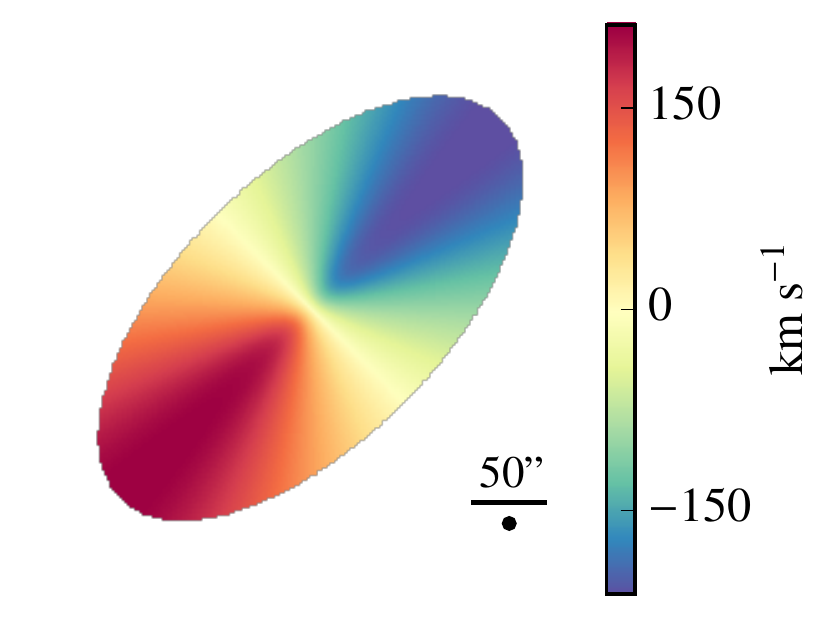}
    \includegraphics[height = 3.5cm]{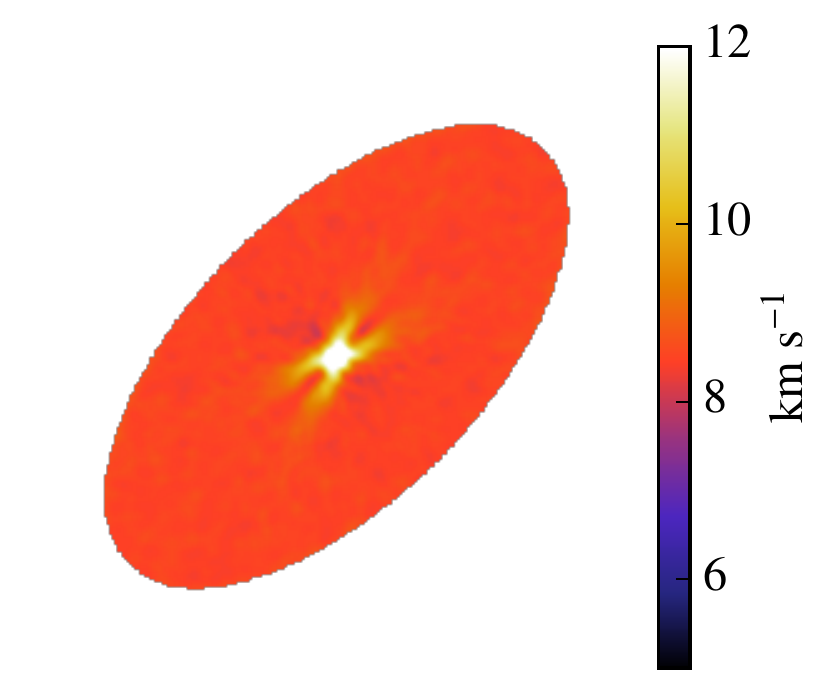}
    \includegraphics[height = 3.5cm]{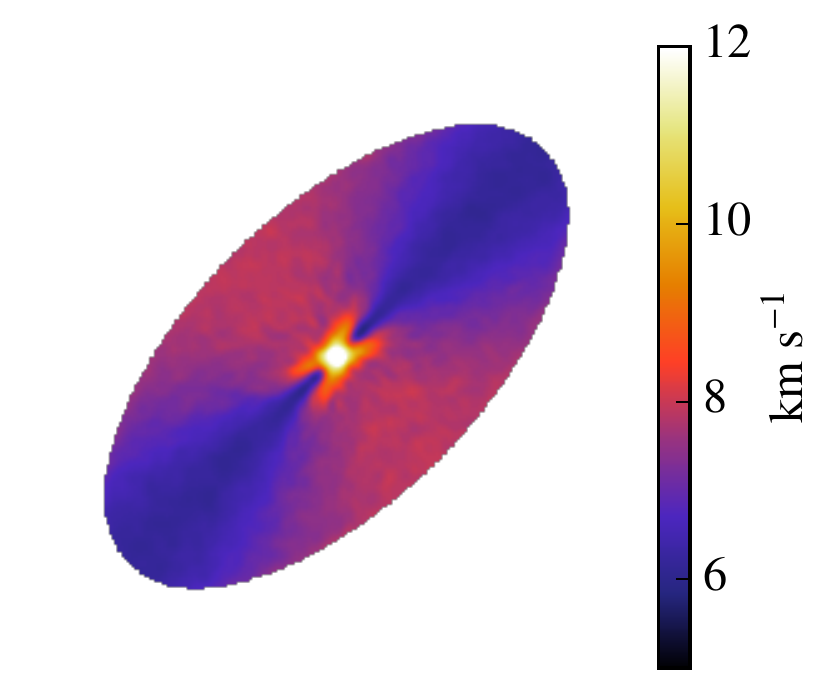} \\
    \includegraphics[height = 4cm]{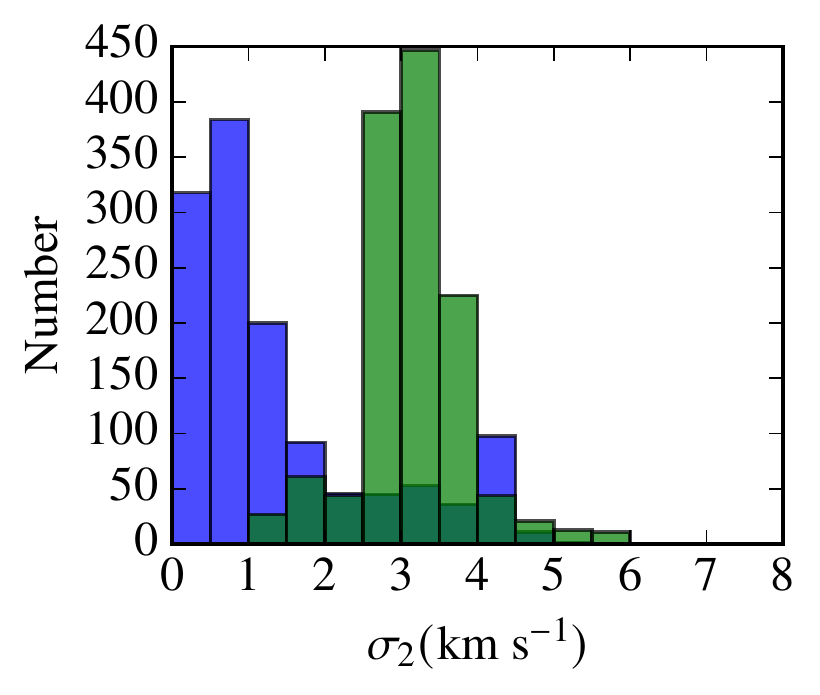}
    \includegraphics[height = 4cm, trim = 0.6cm 0 0 0 , clip]{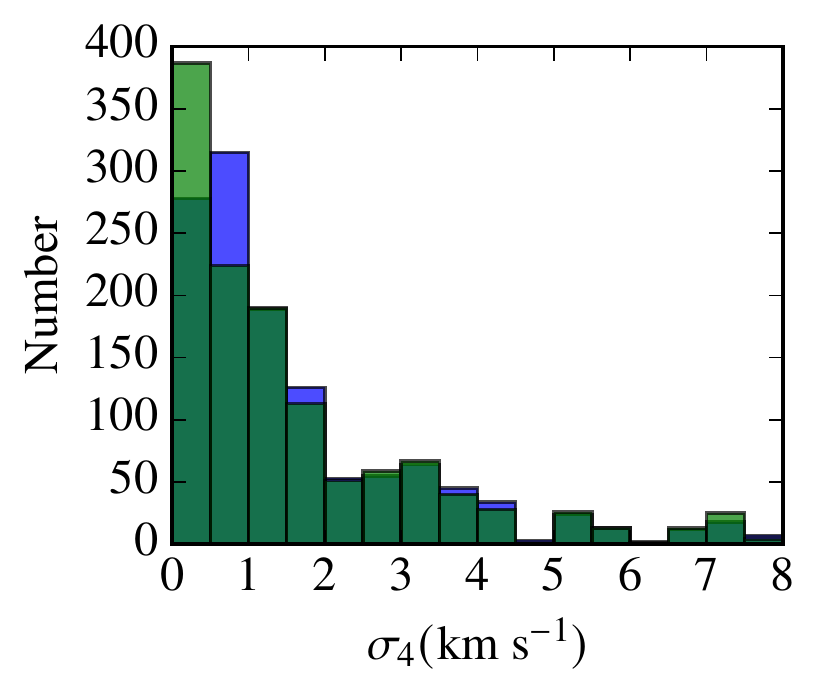}
    \includegraphics[height = 4cm, trim = 0.6cm 0 0 0 , clip]{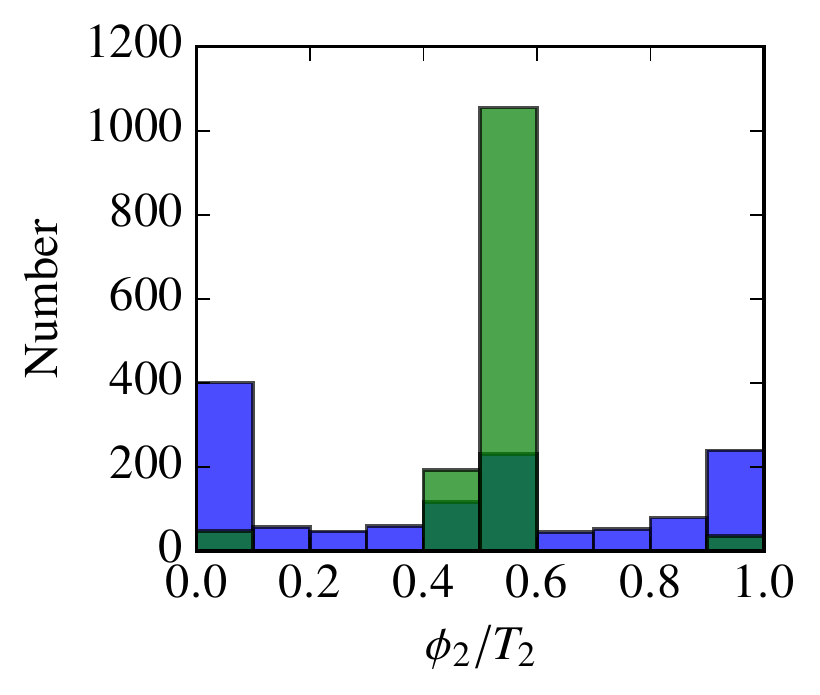}
    \includegraphics[height = 4cm, trim = 0.6cm 0 0 0 , clip]{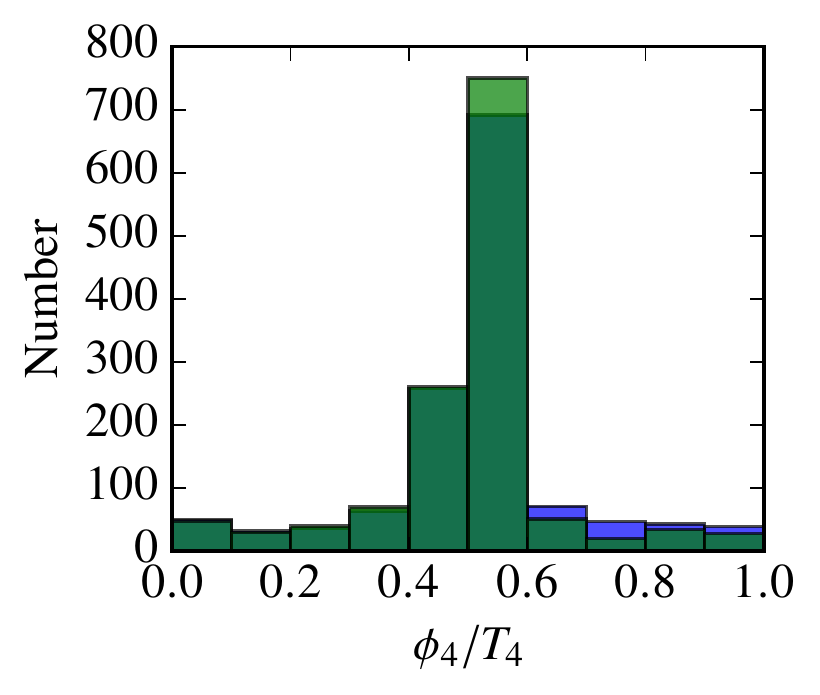}
    
    \includegraphics[height = 3.5cm]{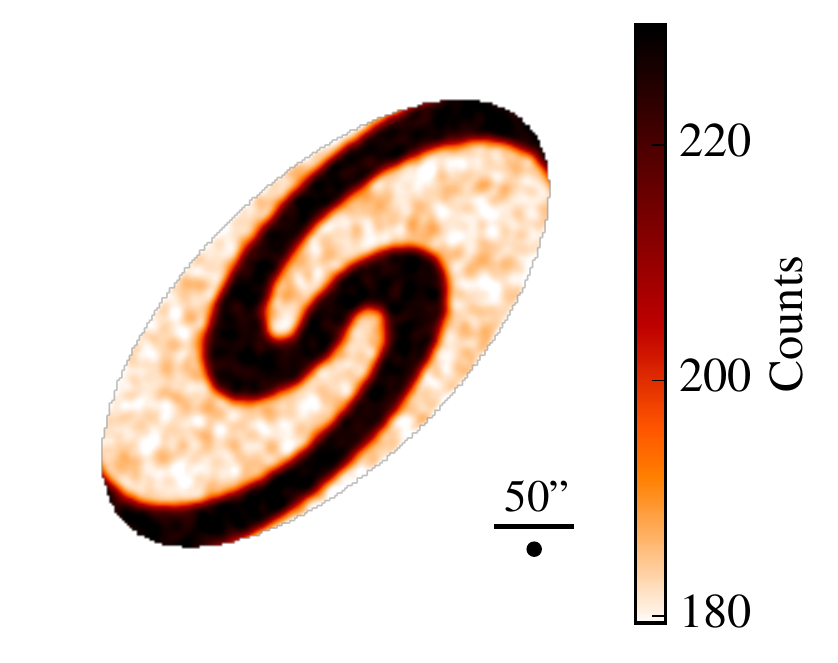}
    \includegraphics[height = 3.5cm]{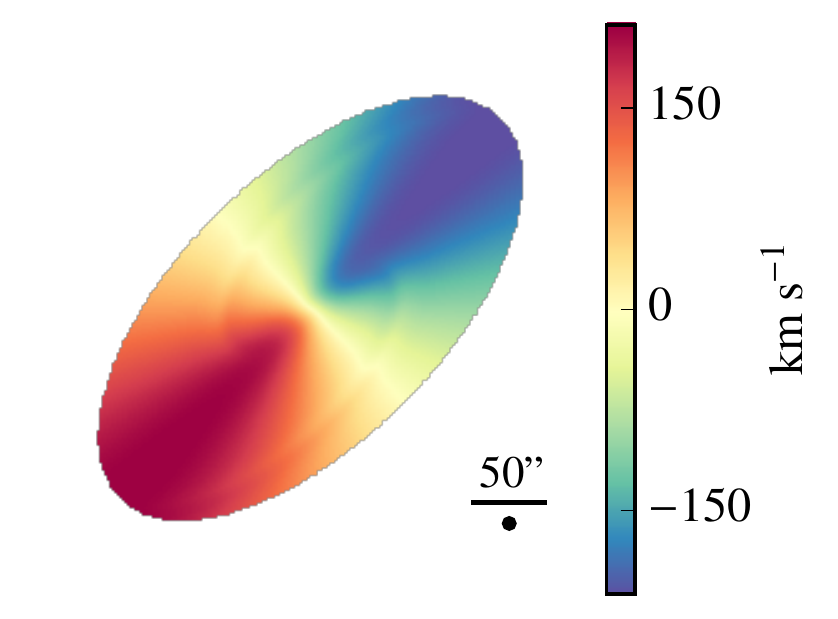}
    \includegraphics[height = 3.5cm]{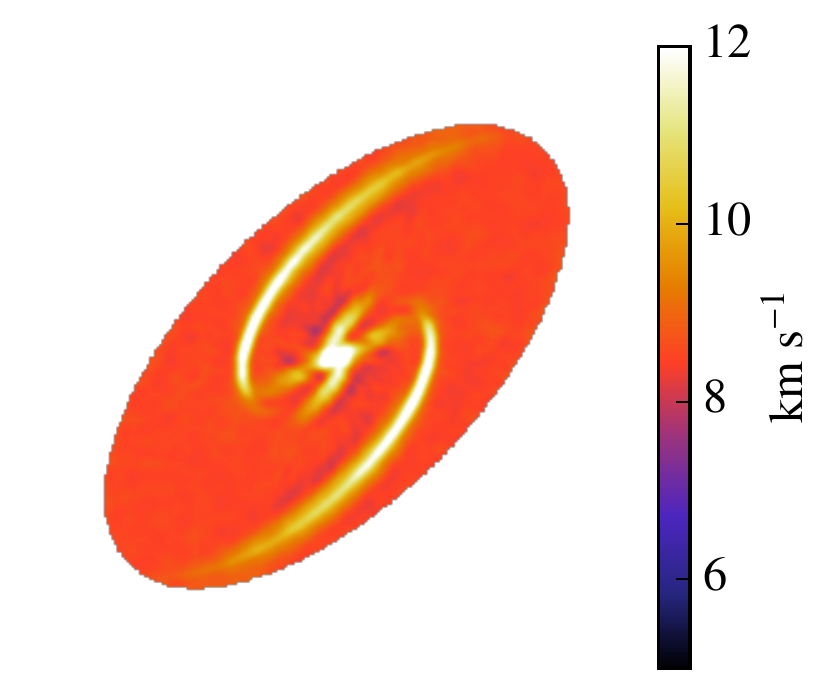}
    \includegraphics[height = 3.5cm]{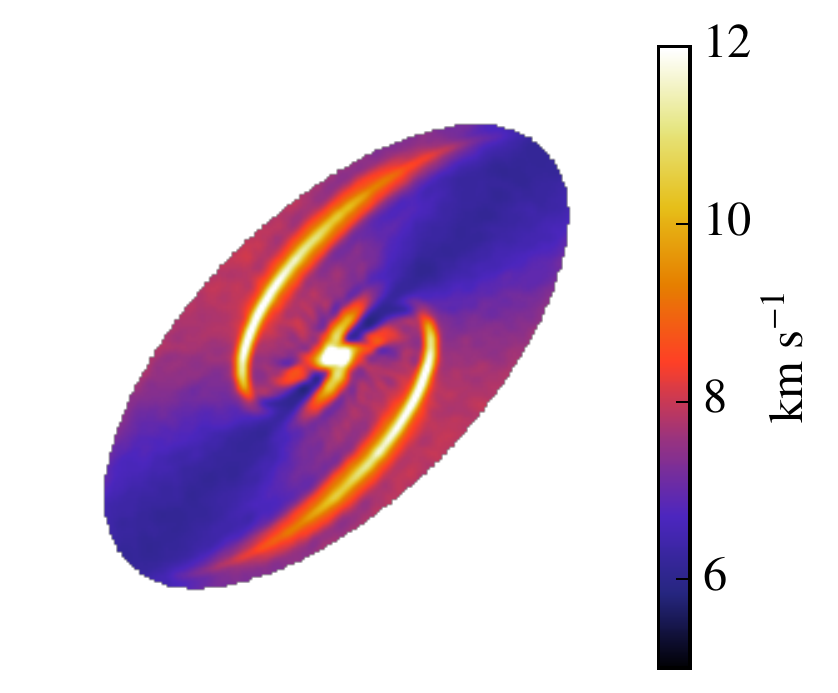} \\
    \includegraphics[height = 4cm]{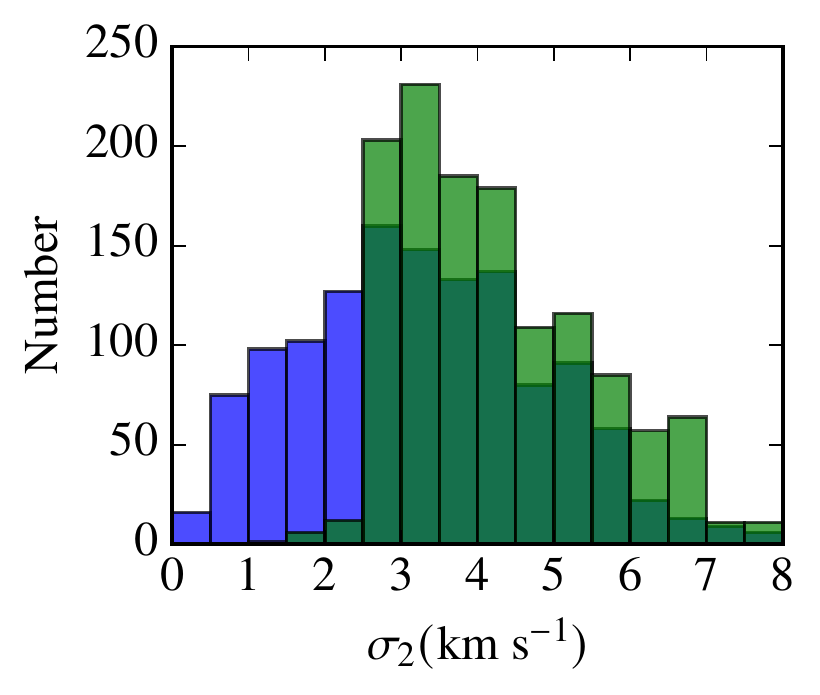}
    \includegraphics[height = 4cm, trim = 0.6cm 0 0 0 , clip]{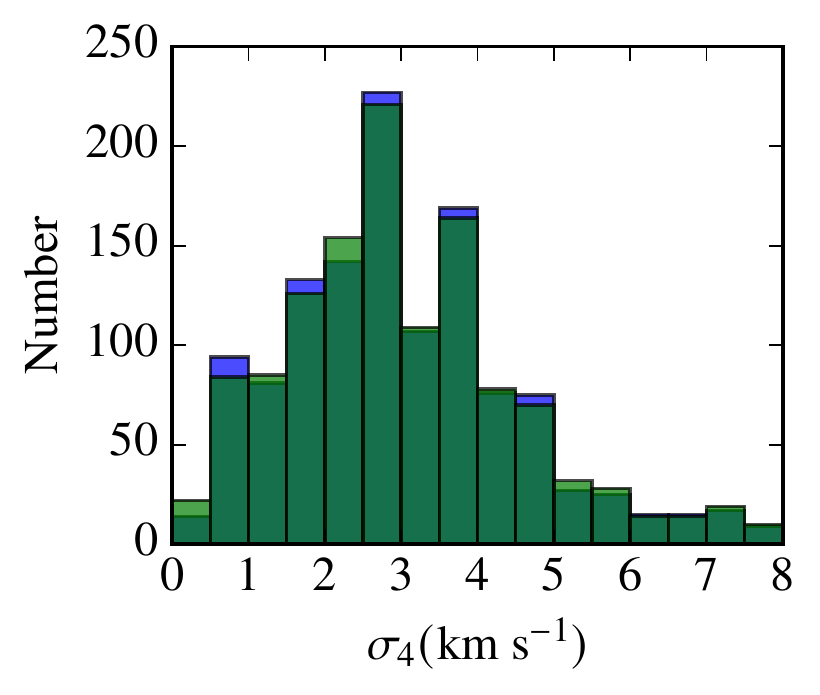}
    \includegraphics[height = 4cm, trim = 0.6cm 0 0 0 , clip]{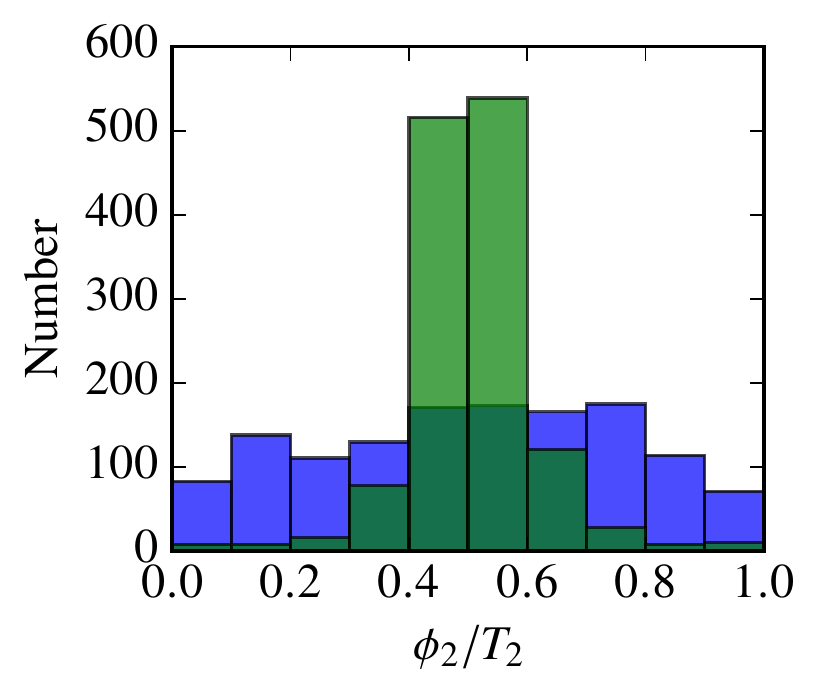}
    \includegraphics[height = 4cm, trim = 0.6cm 0 0 0 , clip]{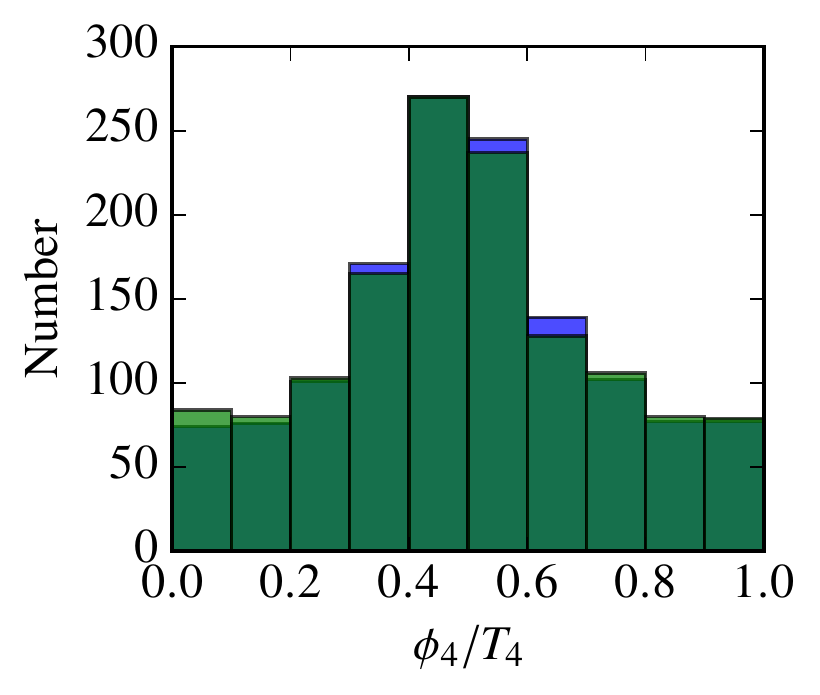}
    
    \includegraphics[height = 3.5cm]{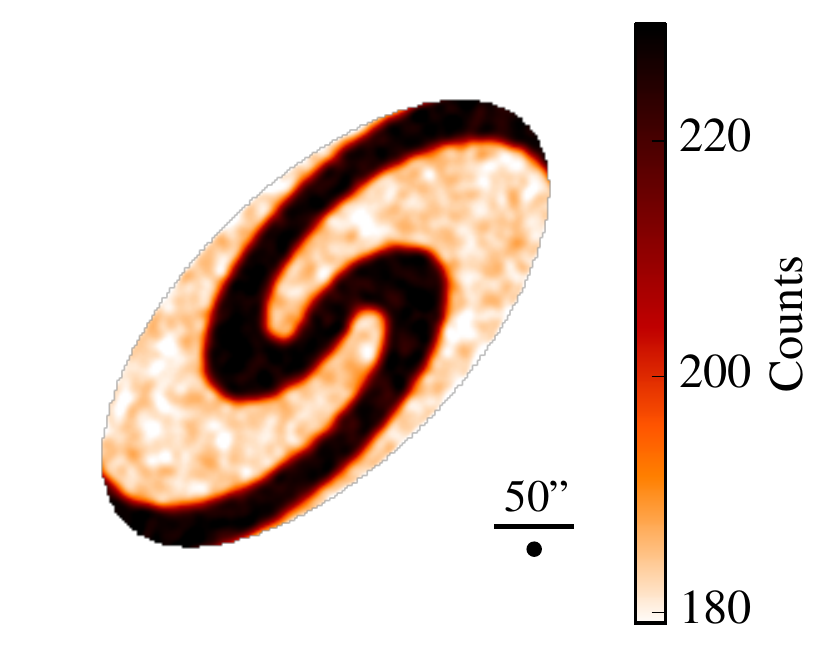}
    \includegraphics[height = 3.5cm]{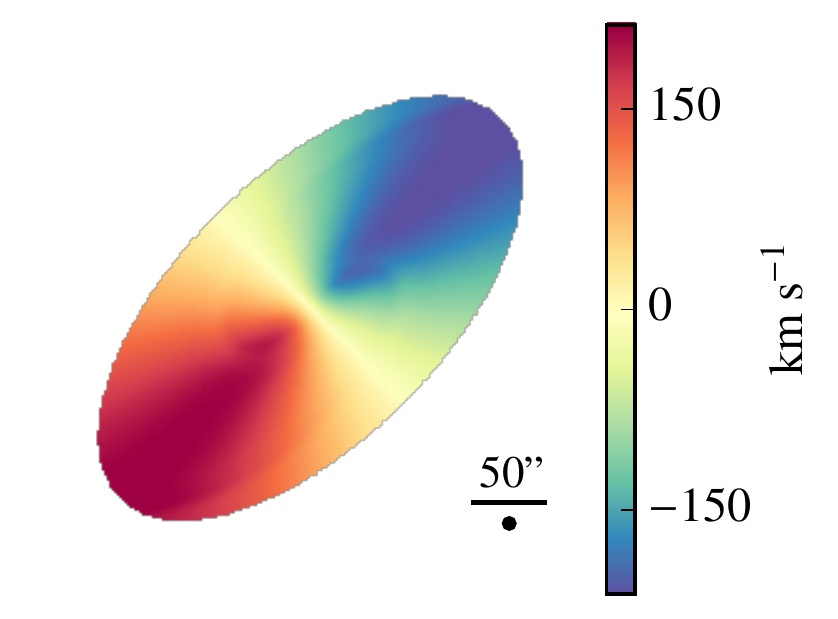}
    \includegraphics[height = 3.5cm]{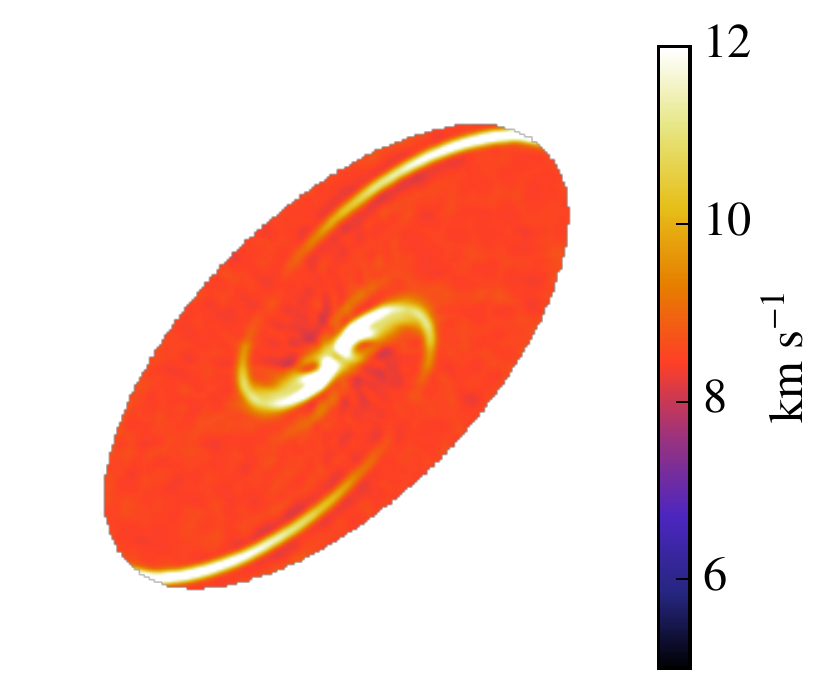}
    \includegraphics[height = 3.5cm]{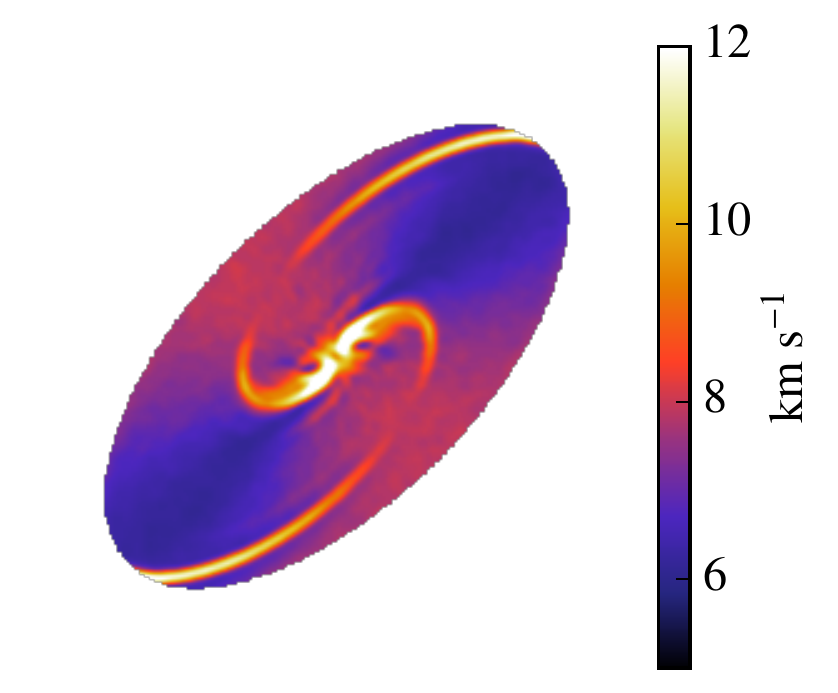} \\
    \includegraphics[height = 4cm]{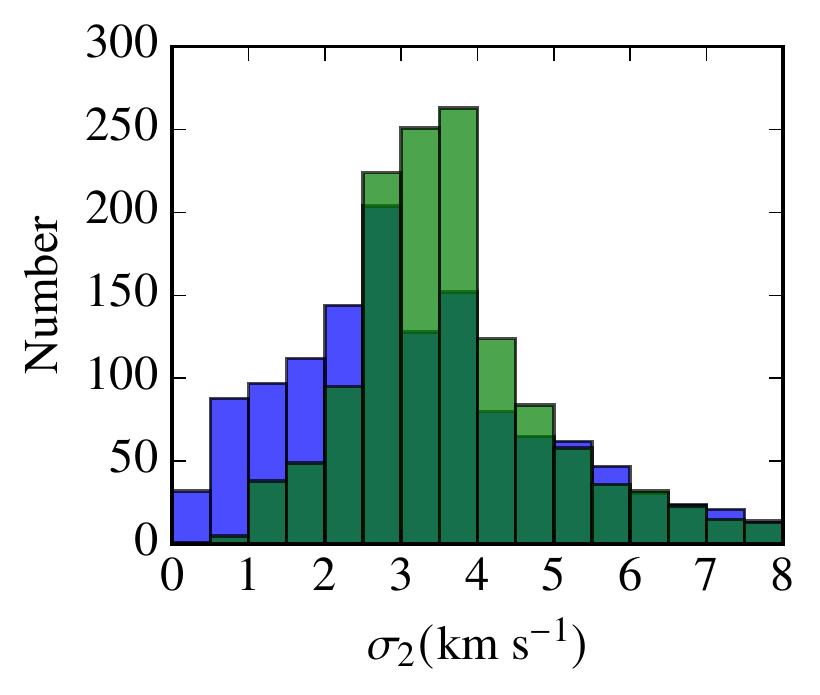}
    \includegraphics[height = 4cm, trim = 0.6cm 0 0 0 , clip]{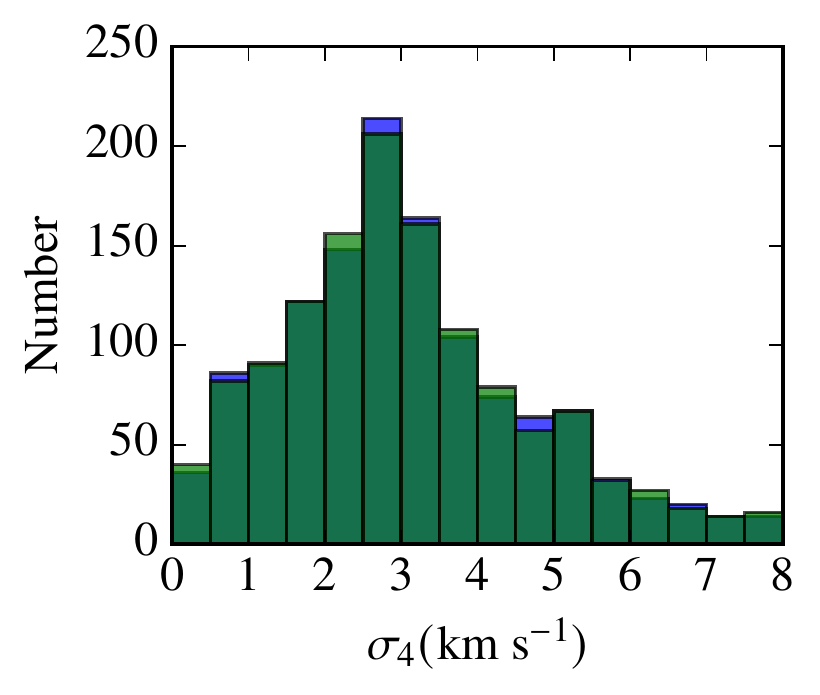}
    \includegraphics[height = 4cm, trim = 0.6cm 0 0 0 , clip]{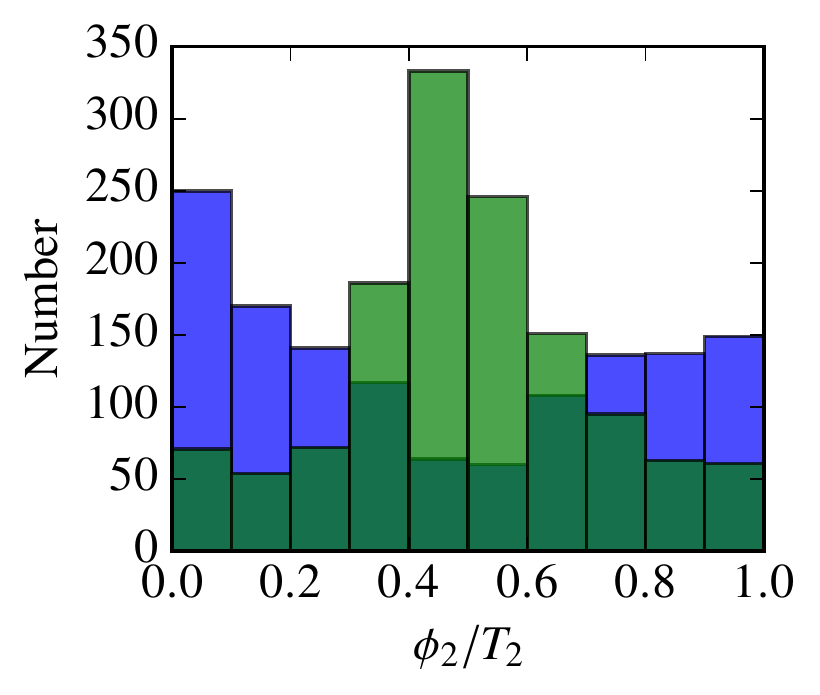}
    \includegraphics[height = 4cm, trim = 0.6cm 0 0 0 , clip]{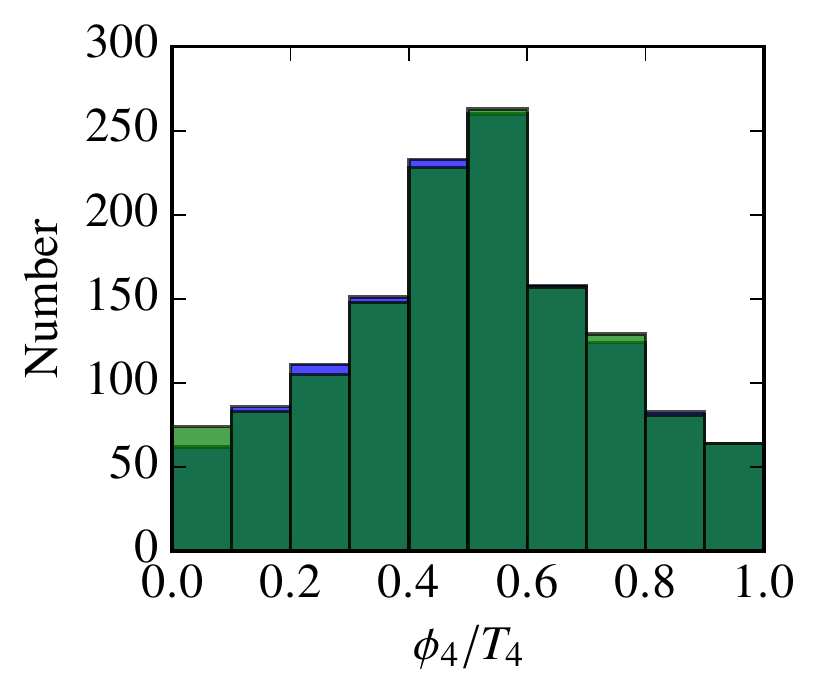}
    \caption{Same as Fig. \ref{fig:toy_models} for the toy models with a steep velocity gradient rotation curve ($v_0=250$ \kms, $r_s=20\arcsec$ and $\gamma = 2$).}
    \label{fig:toy_models2}
\end{figure*}  

In Figures~\ref{fig:toy_models} (weak velocity gradient, moderate maximum rotation velocity) and \ref{fig:toy_models2} (steep inner gradient, high maximum rotation velocity reached at small radius), we show both the effects of models of uniform isotropy and anisotropy for (1) the axisymmetric configuration, and for the spiral  velocity perturbation along either (2) radial or (3) azimuthal directions. An example with the velocity perturbation along $\vec{u}$ is shown in Fig.~\ref{fig:modele_spivrvphi} for the rotation curve with the steep gradient. We show for each case the maps obtained with an inclination of 60\degr\ and with $\phi_{sp}=135$\degr.
Whereas no obvious BS residuals due to rotation are visible in Fig. \ref{fig:toy_models}, stronger residuals are observed in Fig. \ref{fig:toy_models2}, especially in the inner parts. This is due to both the shape and amplitude of the rotation curve. Nevertheless, we checked that using the moment maps of the high-resolution mock data cubes, i.e. those obtained before the spatial smoothing, leads, as expected, to an accurate correction of the BS effect (see Sect.~\ref{sec:beamsmearinggaussian}), except at the very centre for the steeper rotation curve. We also point out that when
the spiral perturbation is set along the radial (azimuthal) direction, the impact is minimum along the major (minor) axis in both the velocity field and the velocity dispersion map. Fig. \ref{fig:toy_models} also shows the histograms of amplitudes and phases for orders two and four resulting from the Fourier analysis of the corresponding models, described and discussed in Sections~\ref{sec:bs_properties} and~\ref{sec:orderedvelocityeffect}.

\FloatBarrier

\section{Line-of-sight velocity dispersion for a disc in cylindrical coordinates}
\label{app:anisotropy}

In this Appendix, we introduce and discuss the geometrical and mathematical framework used to study asymmetries and anisotropies in the neutral gas spatially resolved velocity dispersion fields. Velocity dispersion is a second order moment, and therefore is a more complex quantity to study than flux and radial velocity which are respectively the zero and first moment order. 

Due to its hydrodynamical properties, the gas likely lies in a plane. The velocity vector in the frame of the galactic plane is described by two components both lying in this plane. Thus, in the cylindrical frame, the observed velocity along the line-of-sight is expressed as:
\begin{equation}
V_{\mathrm{los}} = V_{\mathrm{sys}} + V_\theta\cos\theta\sin i + V_R\sin\theta\sin i + V_z\cos i \text{~,}
 \label{eq:vlos_disc}
\end{equation}
where, 
\begin{enumerate}
    \item $V_\theta$ is the azimuthal velocity, i.e. the rotation motions;
    \item $V_R$ is the radial velocity, that is the inward or outwards motions;
    \item $V_z$ is the vertical velocity, that is motions perpendicular to the galactic plane.
\end{enumerate}
The observed velocity $V_{\mathrm{los}}$ (los for line-of-sight) is linked to the projection of 
$V_\theta$, $V_R$ and $V_z$ along the line-of-sight through 5 additional parameters:
\begin{enumerate}
\setcounter{enumi}{3}
    \item $PA$, the position angle of the major axis of the galaxy (measured counterclockwise from the North to the direction of receding side of the galaxy);
    \item $i$, the inclination of the galactic disc with respect to the sky plane;
    \item $V_{\mathrm{sys}}$, the systemic velocity of the galaxy;
    \item $x_c$ and  $y_c$, the coordinates of the rotation centre in Cartesian coordinates (sky projection).
\end{enumerate}

Both the radial, azimuthal, and vertical components can vary with $R$ and $\theta$, which are the polar coordinates in the plane of the galaxy with respect to the centre, choosing the major axis as reference $\theta=0$ (receding side). The azimuth in the plane of the galaxy, $\theta$, is linked to the position angle $PA$, the inclination $i$, the position x (east-west),y (north-south) and centre $x_c,y_c$ in the sky by the set of equations~\ref{eq:links_s} to \ref{eq:links_e}:
 \begin{eqnarray}
 \label{eq:links_s}
 R\cos{\theta}= r \cos{\psi} \text{~,}\\
 R\sin{\theta}= r \frac{\sin{\psi}}{\cos{i}} \text{~,}\\
 \cos{\psi}=\frac{(y-y_c) \cos{PA} - (x-x_c) \sin{PA}}{r} \text{~,}\\
 \sin{\psi}=-\frac{(x-x_c) \cos{PA} + (y-y_c) \sin{PA}}{r} \text{~,}\\
 r=\sqrt{(x-x_c)^2+(y-y_c)^2} \text{~,}\\
 R=r\sqrt{\cos^2{\psi}+\frac{\sin^2{\psi}}{\cos^2{i}}} \text{~,}
 \label{eq:links_e}
 \end{eqnarray}
$\psi$ being the counterclockwise angle from the major axis, and $r$ being the distance to the centre, both in the plane of the sky.

It is possible to define velocity dispersion components:
\begin{equation}
\sigma^2_{ij}=\overline{V_i V_j}-\overline{V_i}~\overline{V_j} \text{~,}
\label{eq:var}
\end{equation}
where the subscripts $i$ and and $j$ denote the different coordinate directions\footnote{by construction, $\sigma_{ij}^2$ is not necessarily positive, this notation was nevertheless used for the sake of homogeneity and because it was used by other authors \citep[e.g.][]{2009smith}.}: radial ($R$), azimuthal ($\theta$), and vertical ($z$), whereas $V_i$ and $V_j$ give the corresponding velocities. When $i \ne j$, the term is the covariance between $V_i$ and $V_j$, whereas in the specific case of $i = j$, the term is the variance of $V_i$ and we note $\sigma_{ii}^2 = \sigma_i^2$.
By definition, the line-of-sight velocity dispersion is thus defined as
\begin{equation}
\sigma_{\mathrm{los}}^2 = \overline{V_{\mathrm{los}}^2}-\overline{V_{\mathrm{los}}}^2 \text{~,}
\end{equation}
which leads, by replacing the terms by their corresponding expression using Eq.~\ref{eq:vlos_disc}, by using Eq.~\ref{eq:var}, and because $V_{\mathrm{sys}}$ is constant, to: 
\begin{equation}
\begin{split} 
\sigma_{\mathrm{los}}^2 = \sigma_{\theta}^2\cos^2\theta\sin^2{i}+\sigma_{R}^2\sin^2{\theta}\sin^2{i}+\sigma_{z}^2\cos^2{i}\\
+2 (\sigma^2_{R\theta}\cos{\theta}\sin{\theta}\sin^2{i} +\sigma^2_{\theta z}\sin{\theta}\cos{i} \sin{i} \\
+  \sigma^2_{R z}\cos{\theta}\cos{i} \sin{i}) \text{~,}
\end{split} 
\label{eq:slos_disc}
\end{equation}
where $\sigma^2_{R\theta}, \sigma^2_{Rz}$ and $\sigma^2_{\theta z}$ are the squared cross terms of the velocity dispersion tensor. Those terms can be expressed as $\sigma^2_{R\theta} = \rho_{R\theta}\sigma_R\sigma_\theta$, $\sigma^2_{Rz} = \rho_{Rz}\sigma_R\sigma_z$, $\sigma^2_{\theta z} = \rho_{\theta z}\sigma_\theta\sigma_z$, owing to the definition of the correlation coefficients between the components $\rho_{R\theta}$, $\rho_{R z}$ and $\rho_{\theta z}$ that are bounded between -1 and 1. The term $\sigma^2_{R\theta}$ can thus be negative, depending on the sign of the correlation coefficient.

Equation~\ref{eq:slos_disc} can be recast in a sum of trigonometric polynomials of second degree:
\begin{equation}
\begin{split} 
\sigma_{\mathrm{los}}^2 = \left(\frac{\sigma_{\theta}^2-\sigma_{R}^2}{2}\cos{2\theta} +\sigma^2_{R\theta}\sin{2\theta}\right)\sin^2{i}\\
+\left(\sigma^2_{R z}\cos{\theta}+\sigma^2_{\theta z}\sin{\theta}\right)\sin{2i}\\
+\sigma_{z}^2\cos^2{i} + \frac{\sigma_{\theta}^2+\sigma_{R}^2}{2}\sin^2{i} \text{~.}
\end{split} 
\label{eq:slos_disc2}
\end{equation}

We can formulate differently this equation to match better the formalism introduced with Fourier Transforms (see Sect.~\ref{sec:modfft}):
\begin{equation}
\sigma_{\mathrm{los}}^2 =  k_0 + k_1 \cos{(\theta - \alpha_1)} + k_2 \cos{(2(\theta - \alpha_2))} \text{~,}
\label{eq:slos_disc_simpl}
\end{equation}
with:
\begin{eqnarray}
\label{eq:ki}
 k_0 =  \sigma_{z}^2\cos^2{i}+\frac{\sigma_{\theta}^2+\sigma_{R}^2}{2}\sin^2{i} \text{~,}\\
 k_1=\sqrt{\sigma_{R z}^4 +\sigma_{\theta z}^4}\times \sin{2i} \text{~,}\\
 \cos{\alpha_1}=\frac{\sigma_{R z}^2}{k_1} \sin{2i} \text{~,}\\
 \sin{\alpha_1}=\frac{\sigma_{\theta z}^2}{k_1} \sin{2i} \text{~,}\\
 k_2=\sqrt{\left(\frac{\sigma_{\theta}^2-\sigma_{R}^2}{2}\right)^2 + \sigma_{R\theta}^4} \times \sin^2{i} \text{~,}\\
 \cos{2\alpha_2}=\frac{(\sigma_{\theta}^2-\sigma_{R}^2)}{2k_2} \sin^2{i} \text{~,}\\
 \sin{2\alpha_2}= \frac{\sigma_{R\theta}^2}{k_2} \sin^2{i} \text{~.}
\end{eqnarray}
This formalism further motivated our choice of performing Fourier Transforms on squared velocity dispersion. With these definitions, we find
\begin{equation}
 \tan{2\alpha_2} = 2 \sigma^2_{R\theta}/(\sigma_{\theta}^2-\sigma_{R}^2),
\end{equation}
so that $\alpha_2$ is the tilt angle of the velocity ellipsoid in $(R,\theta)$ \citep[see e.g.][]{2009smith}, i.e, the vertex deviation \citep{1991kui,1994kui,1998bin}, analogue to the tilt angle of the velocity ellipsoid in $(R,z)$ plane used e.g. in \citet{2019hagen}.

In case there are asymmetries, those are embedded in the terms $k_i$ and $\alpha_i$. For the sake of simplicity, let's assume that cross terms are null, which leads to
$$k_1=0 \text{~,~} k_2=\frac{\sigma_{\theta}^2-\sigma_{R}^2}{2} \times\sin^2{i} \text{~,~and~} \alpha_2=0 \text{~.}$$
If we only have an asymmetry of order $k$ (sinusoidal) in both $\sigma_R^2$ and $\sigma_\theta^2$ with similar amplitudes $\sigma_{R,k} = \sigma_{\theta,k} = \sigma_k$ and with phases $\phi_{R,k}$ and $\phi_{\theta,k} - \phi_{R,k} = \Delta \phi_k$, we can therefore express $k_0$ and $k_2$ as:
\begin{equation}
 k_0 = \sigma_{z}^2\cos^2{i} + \sigma_k^2 \cos{(\Delta \phi_k/2)}\cos{(k\theta - \phi_{R,k}-\Delta \phi_k/2)}\sin^2{i} \text{~,}
\end{equation}
\begin{equation}
 k_2 = \sigma_k^2 \sin{(\Delta \phi_k/2)}\sin{(k\theta - \phi_{R,k}-\Delta \phi_k/2)}\sin^2{i} \text{~,}
 \label{eq:k2}
\end{equation}
and Eq.~\ref{eq:slos_disc_simpl} becomes:
\begin{equation}
\begin{split}
\sigma_{\mathrm{los}}^2 = \sigma_{z}^2\cos^2{i} + \sigma_k^2 \cos{(\Delta \phi_k/2)}\cos{(k\theta - \phi_{R,k}-\Delta \phi_k/2)}\sin^2{i} \\
+ \sigma_k^2/2 \sin{(\Delta \phi_k/2)} \sin{((k+2)\theta - \phi_{R,k}-\Delta \phi_k/2)} \sin^2{i} \\
+ \sigma_k^2/2 \sin{(\Delta \phi_k/2)} \sin{((k-2)\theta - \phi_{R,k}-\Delta \phi_k/2)} \sin^2{i} \text{~.}
\end{split}
\label{eq:slos_disc_asym}
\end{equation}
Orders $k$, $k-2$ and $k+2$ appear. Nevertheless, even if $\Delta\phi_k$ is constant, the phase $\phi_{R,k}$ is supposed to remain random, so no peculiar direction should be favored.

The general case without the assumption that $\sigma_{R,k} = \sigma_{\theta,k}$ leads to:
\begin{equation}
    k_2 = c_k \cos{(k\theta-\phi_{R,k})} + s_k \sin{(k\theta-\phi_{R,k})} \text{~,}
\end{equation}
with
\begin{equation}
c_k = \frac{\sigma_{\theta, k}^2 \cos{\Delta\phi_k} - \sigma_{R, k}^2}{2} \text{~, and}
\end{equation}
\begin{equation}
s_k = \frac{\sigma_{\theta, k}^2 \sin{\Delta\phi_k}}{2} \text{~,}
\end{equation}
which can also be written
\begin{equation}
 k_2 = x_k \cos{(k\theta - \phi_{R,k} - \phi_k)} \text{~,}
\end{equation}
with
$$x_k=\sqrt{c_k^2 + s_k^2} \text{~,~} \cos{\phi_k}=c_k/x_k \text{~,~and~} \sin{\phi_k}=s_k/x_k  \text{~.}$$
The general case would therefore also lead to orders $k-2$ and $k+2$.

If $\sigma_{R,k} = \sigma_{\theta,k} = \sigma_k$, we obtain that $c_k = -\sigma_k^2 \sin^2{(\Delta\phi_k/2)}$ and $s_k = \sigma_k^2 \sin{(\Delta\phi_k/2)}\cos{(\Delta\phi_k/2)}$, and therefore $x_k= \sigma_k^2 \sin{(\Delta\phi_k/2)}$ and $\phi_k = \Delta\phi_k/2 + \pi/2$, which enables us to recover Eq.~\ref{eq:k2}.

\section{Results of the FFTs for individual galaxies in THINGS}
\label{sec:mapsgalaxies}

In this section we show the various kinematic product for each THINGS galaxy as presented in Figure \ref{fig:n2841-ex} and Figure \ref{fig:n2841-exfft0}.
The upper row plots represents from left to right the observed flux density map, the observed velocity field, the observed velocity dispersion, the BS model, and the velocity dispersion map corrected from the BS effect.
The second row shows the observed squared velocity dispersion and its modelling through FFT up to $k=4$, and their residuals. 
The third row  shows the individual orders of the FFT in the galaxy. 
The bottom row presents the amplitudes (left) and phase angles (right) of the FFT coefficients as a function of radius.

\begin{figure*}[t]
    \includegraphics[width=\textwidth]{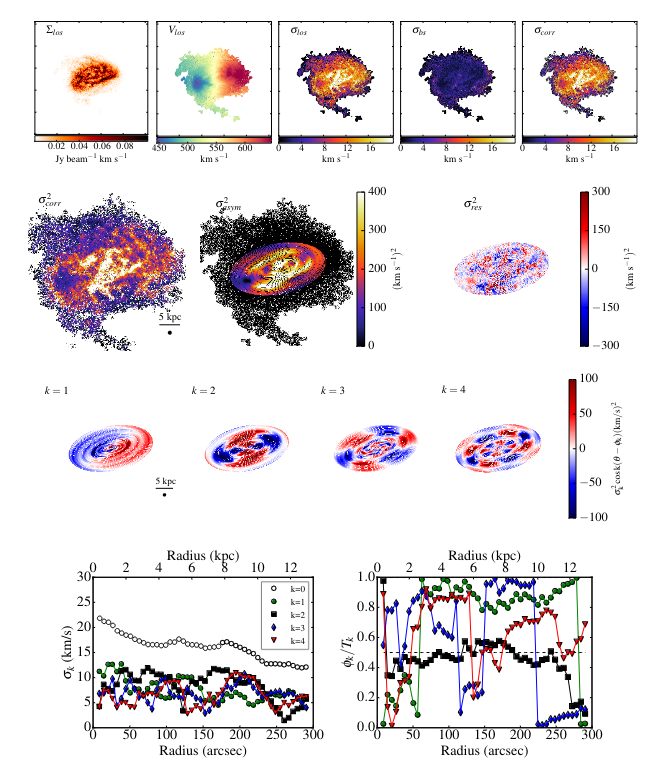}
    \caption{
    \hi\ density and velocity maps and FFT results of NGC925. (Top panel, from left to right) Observed flux density map, observed velocity field, observed velocity dispersion, beam smearing model, and velocity dispersion map corrected from the beam smearing effect. (Second panel, from left to right) Squared observed velocity dispersion and its corresponding modelling through FFT up to the order 4, and the residuals between those two maps. (Third panel) Individual squared orders of the FFT modelling projected in the plane of the galaxy. (Bottom panel) Radial variation of the FFT amplitudes (left) and phase angles (right) for orders $k = 0$ to $k = 4$.}
    \label{fig:n925-exfft}
\end{figure*}  

\begin{figure*}[t]
    \includegraphics[width=\textwidth]{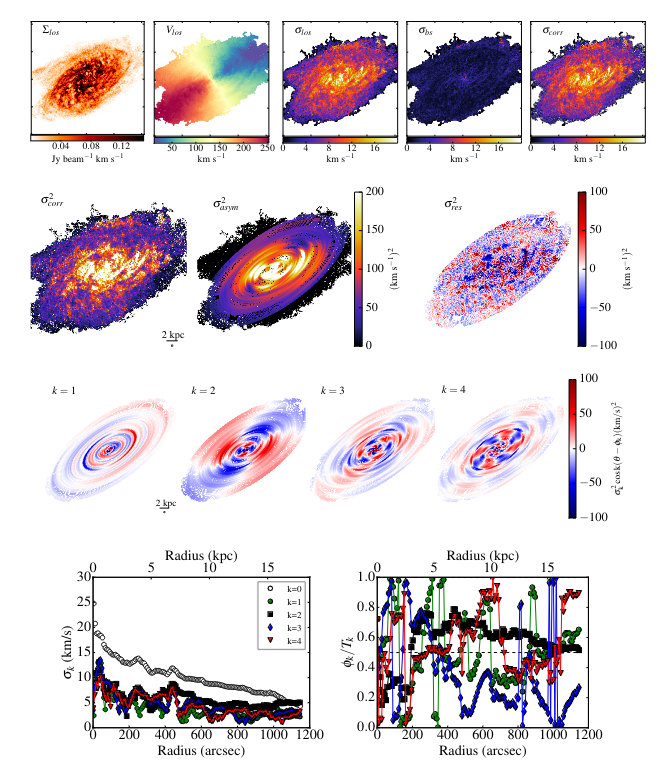}
    \caption{Same as Fig. \ref{fig:n925-exfft} for NGC2403.}
    \label{fig:n2403-exfft}
\end{figure*}  

\begin{figure*}[t]
    \includegraphics[width=\textwidth]{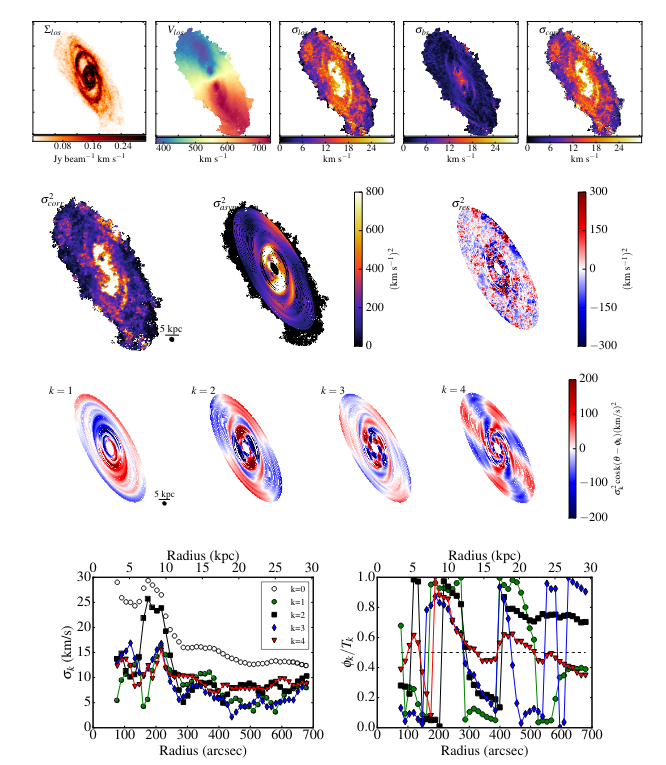}
    \caption{Same as Fig. \ref{fig:n925-exfft} for NGC2903.}
    \label{fig:n2903-exfft}
\end{figure*}  

\begin{figure*}[t]
    \includegraphics[width=\textwidth]{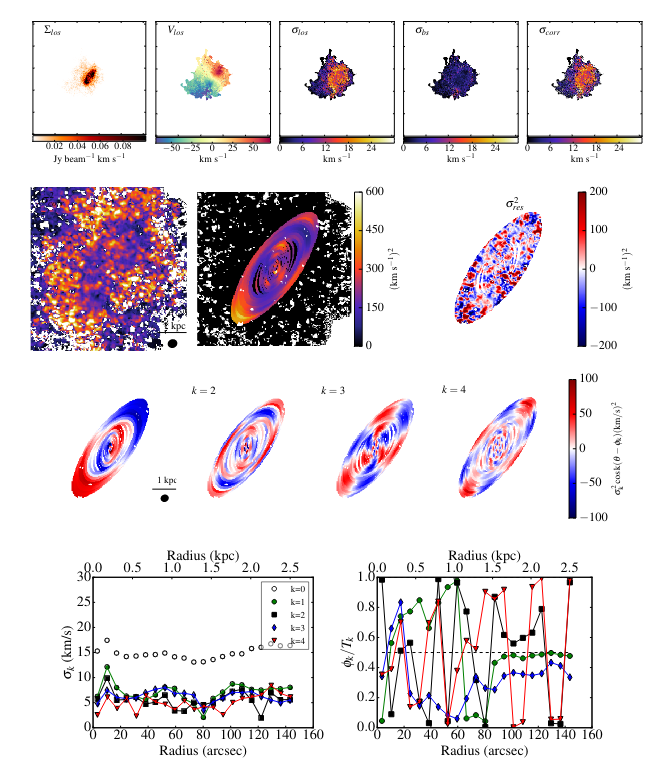}
    \caption{Same as Fig. \ref{fig:n925-exfft} for NGC2976.}
    \label{fig:n2976-exfft}
\end{figure*}  

\begin{figure*}[t]
    \includegraphics[width=\textwidth]{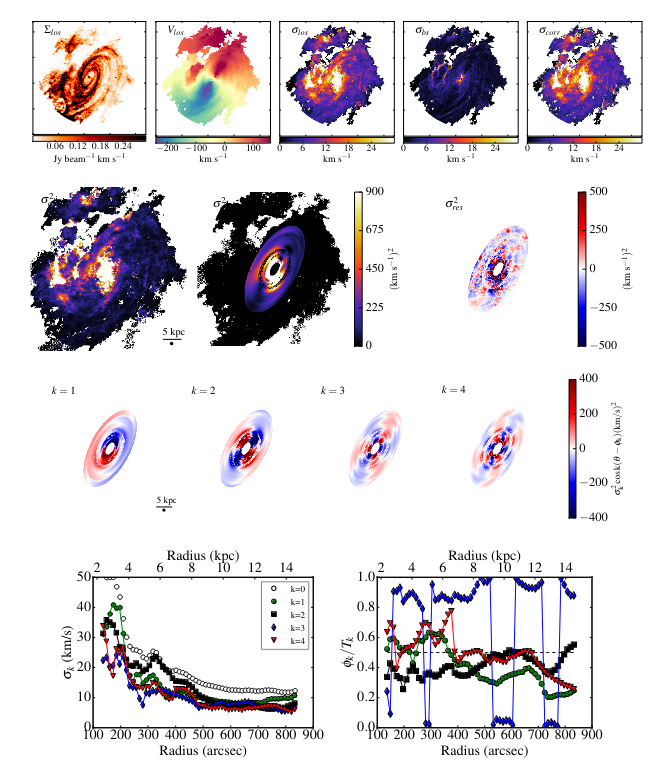}
    \caption{Same as Fig. \ref{fig:n925-exfft} for NGC3031.}
    \label{fig:n3031-exfft}
\end{figure*}  

\begin{figure*}[t]
    \includegraphics[width=\textwidth]{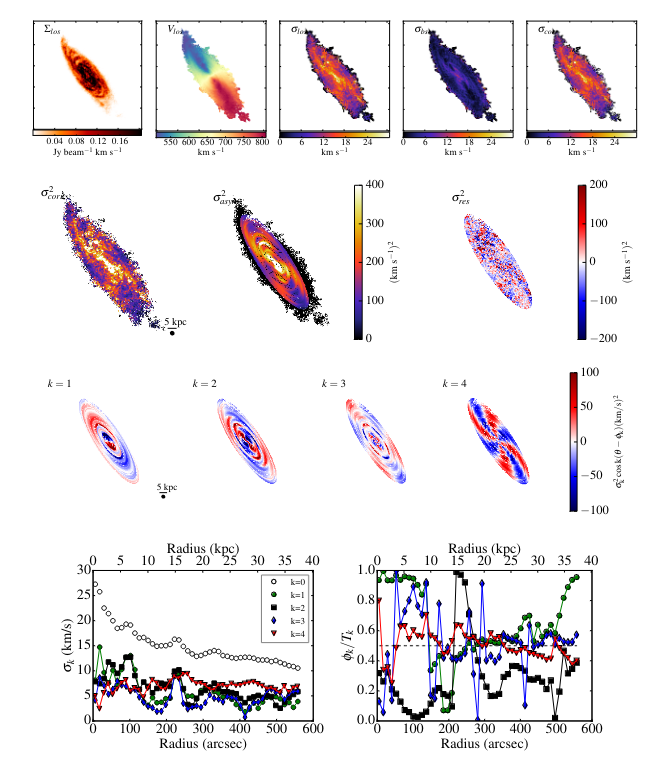}
    \caption{Same as Fig. \ref{fig:n925-exfft} for NGC3198.}
    \label{fig:n3198-exfft}
\end{figure*}  

\begin{figure*}[t]
    \includegraphics[width=\textwidth]{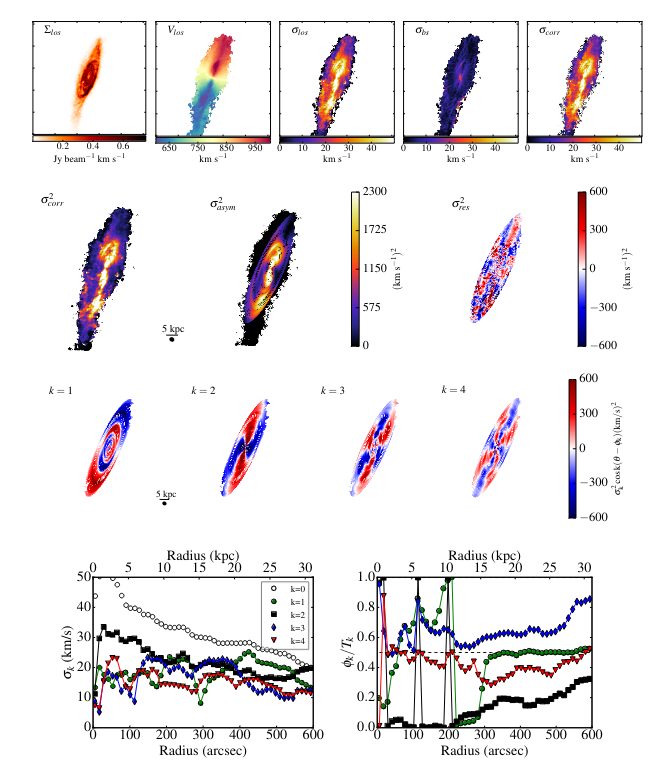}
    \caption{Same as Fig. \ref{fig:n925-exfft} for NGC3521.}
    \label{fig:n3521-exfft}
\end{figure*}  

\begin{figure*}[t]
    \includegraphics[width=\textwidth]{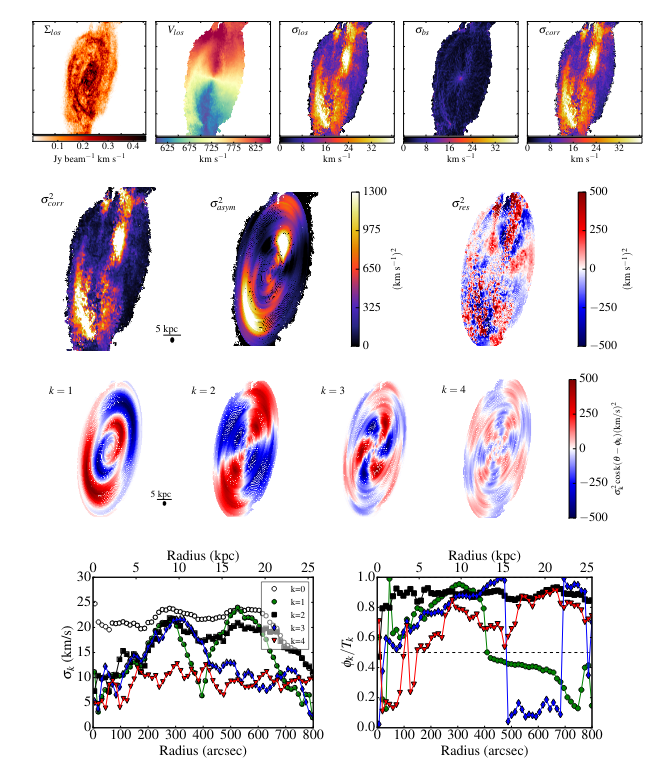}
    \caption{Same as Fig. \ref{fig:n925-exfft} for NGC3621.}
    \label{fig:n3621-exfft}
\end{figure*}  

\begin{figure*}[t]
    \includegraphics[width=\textwidth]{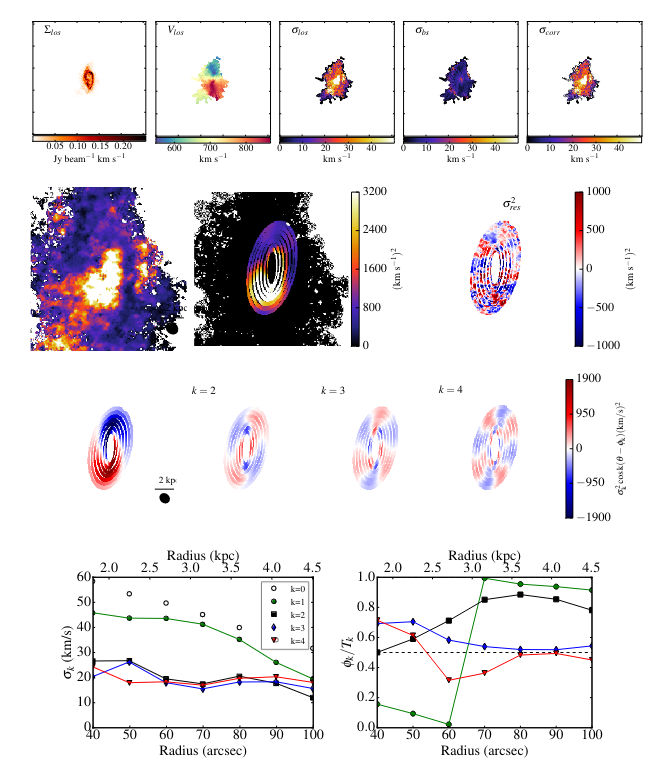}
    \caption{Same as Fig. \ref{fig:n925-exfft} for NGC3627.}
    \label{fig:n3627-exfft}
\end{figure*}  

\begin{figure*}[t]
    \includegraphics[width=\textwidth]{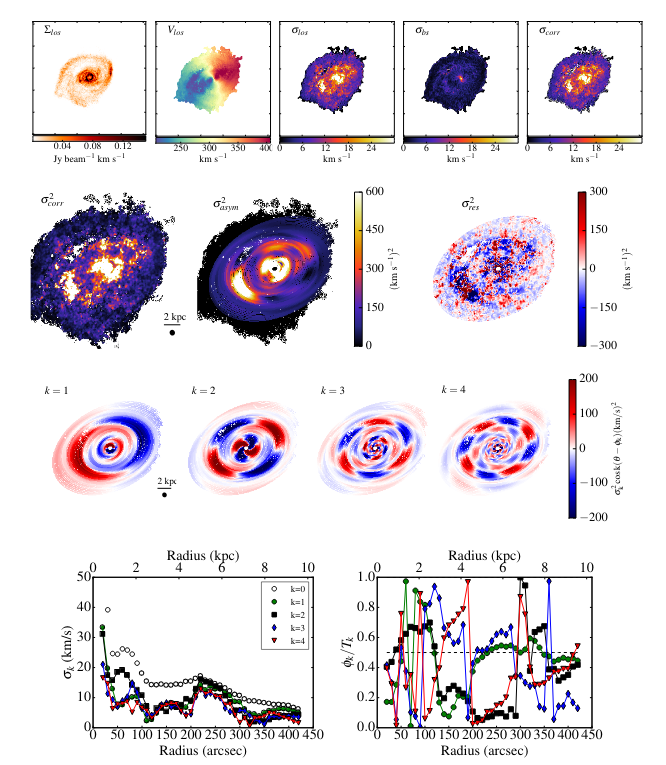}
    \caption{Same as Fig. \ref{fig:n925-exfft} for NGC4736.}
    \label{fig:n4736-exfft}
\end{figure*}  

\begin{figure*}[t]
    \includegraphics[width=\textwidth]{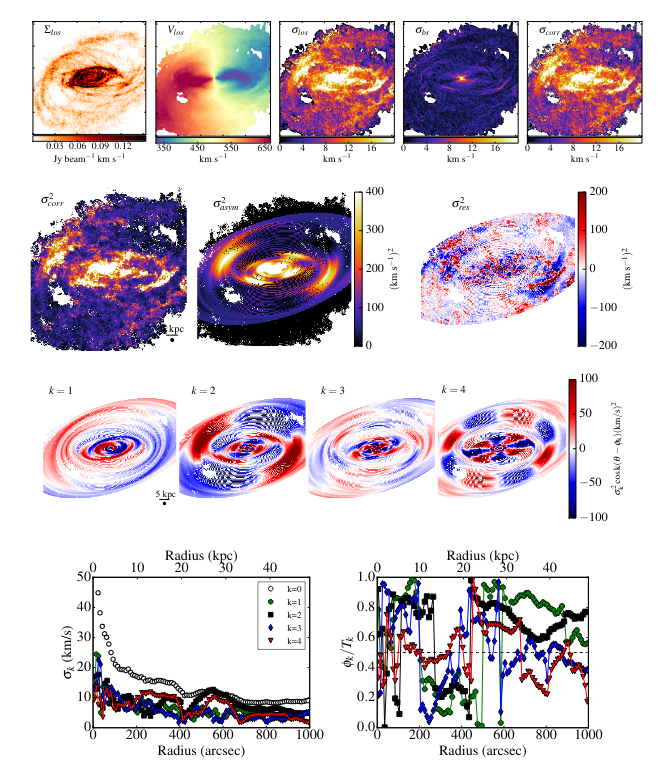}
    \caption{Same as Fig. \ref{fig:n925-exfft} for NGC5055.}
    \label{fig:n5055-exfft}
\end{figure*}  

\begin{figure*}[t]
    \includegraphics[width=\textwidth]{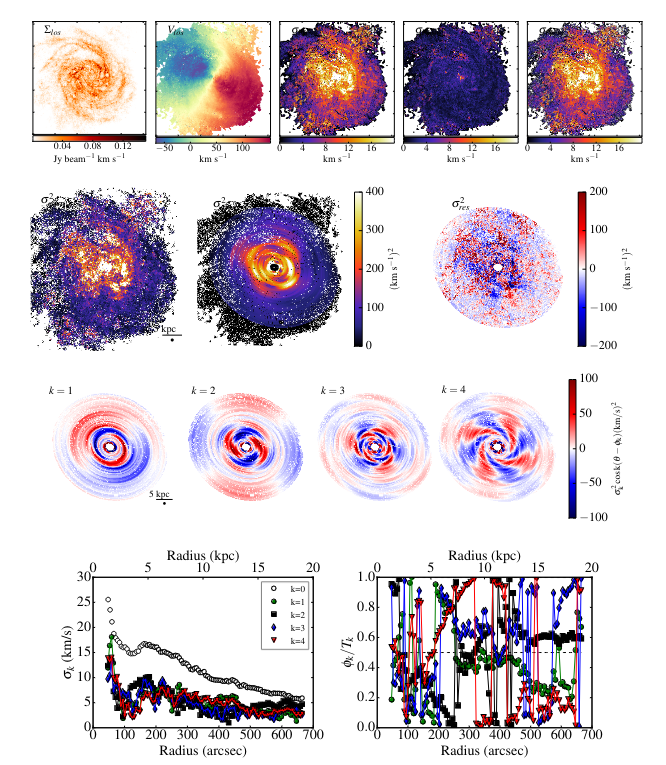}
    \caption{Same as Fig. \ref{fig:n925-exfft} for NGC6946.}
    \label{fig:n6946-exfft}
\end{figure*}  

\begin{figure*}[t]
    \includegraphics[width=\textwidth]{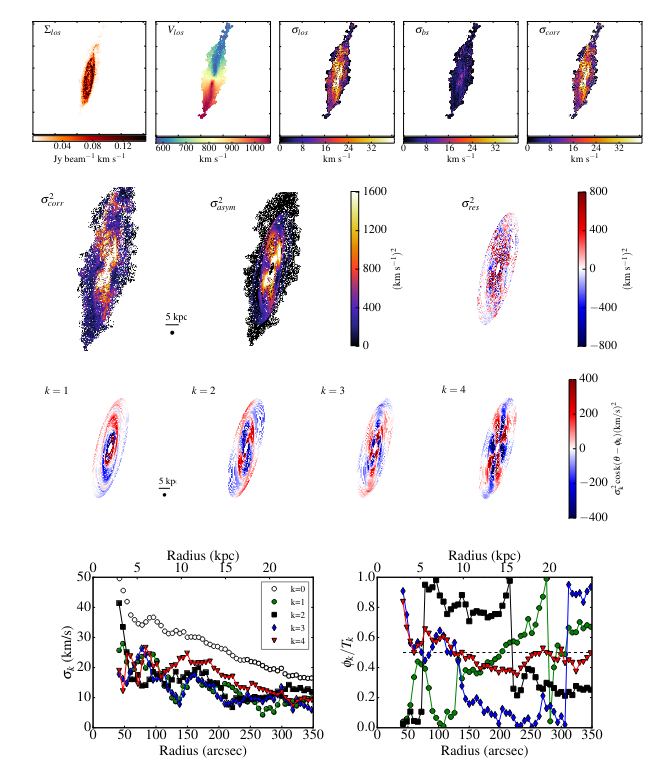}
    \caption{Same as Fig. \ref{fig:n925-exfft} for NGC7331.}
    \label{fig:n7331-exfft}
\end{figure*}  

\begin{figure*}[t]
    \includegraphics[width=\textwidth]{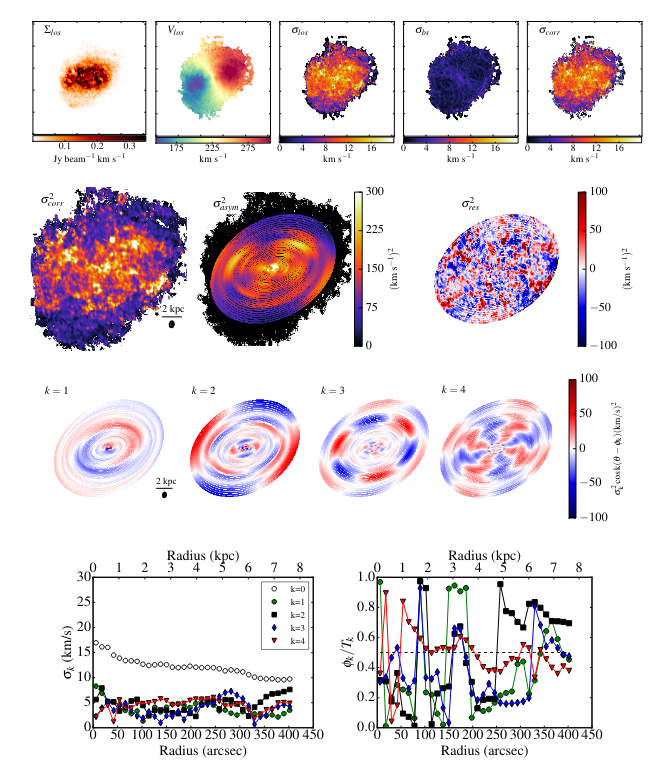}
    \caption{Same as Fig. \ref{fig:n925-exfft} for NGC7793.}
    \label{fig:n7793-exfft}
\end{figure*}

\end{appendix}
\end{document}